\documentclass[aps,pra,twocolumn]{revtex4-2}

\usepackage{graphicx}
\usepackage{dcolumn}
\usepackage{bm}
\usepackage{lineno}
\usepackage{amsmath}
\usepackage{float} 
\usepackage{placeins}
\usepackage{xcolor}
\usepackage{multirow}
\usepackage{makecell}
\usepackage{amssymb}

\begin{document}

\preprint{APS/123-QED}

\title{\textbf{Optimized ancillary drive for fast Rydberg entangling gates} 
}%

\author{Rui Li$^{1,2}$}\thanks{First Author and Second Author contribute equally to this work.\\} 
\author{Min-Hua Zhang$^{1}$}\thanks{First Author and Second Author contribute equally to this work.\\} 
\author{Jing Qian$^{1,3,4}$} 
 \email{Corresponding author: jqian1982@gmail.com}

\affiliation{$^{1}$State Key Laboratory of Precision Spectroscopy, School of Physics, East China Normal University, Shanghai, 200062, China
}
\affiliation{$^{2}$School of Physics and Astronomy, Shanghai Jiao Tong University, Shanghai, 200240, China}
\affiliation{$^{3}$Chongqing Institute of East China Normal University, Chongqing, 401120, China}
\affiliation{$^{4}$Collaborative Innovation Center of Extreme Optics, Shanxi University, Taiyuan, 030006, China}

\date{\today}

\begin{abstract}
Reaching fast and robust two-qubit gates with low infidelities has been an outstanding challenge for the long-term goal of useful quantum computers. Typically, optimizing the pulse shapes can minimize the gate infidelity and improve its robustness to certain types of errors; yet it remains incapable of speeding up the gate execution time which is fundamentally restricted by the attainable Rabi frequency in a realistic setup. In this work, we develop a fast implementation of two-qubit CZ gates using optimized ancillary drive to enhance the two-photon Rabi frequency between the ground and Rydberg states.
This ancillary drive can work in an error-robustness framework 
without increasing the original gate infidelity in the absence of the drive.
Considering the experimentally feasible parameters for $^{87}$Rb atoms,
we demonstrate that the execution time required for such CZ gates can be shortened by more than 30$\%$ as compared to standard two-photon protocols \textcolor{black}{arising the gate fidelity above 0.9954 by taking account of all relevant error sources}. Our results reduce the high-power laser requirement and \textcolor{black}{unlock the potential toward fast, high-fidelity quantum operations for large-scale quantum computation with neutral atoms.}
\end{abstract}

\maketitle


\section{Introduction}

Qubits based on arrays of trapped neutral atoms have emerged as a promising platform for realizing quantum computation, in which high-fidelity quantum operations are crucially required \textcolor{black}{\cite{Henriet2020quantumcomputing,doi:10.1116/5.0036562,PhysRevX.12.021049,Shi_2022,10.1063/5.0211071}}. However, errors due to the physical qubit itself or a noisy environment, essentially limit the accuracy of operations and must be made sufficiently low to permit efficient error correction for logical qubit performance \textcolor{black}{\cite{PhysRevA.86.032324,Wu2022,bluvstein2025}}. Recently, the optimally-modulated pulses are widely utilized to minimize the quantum gate error by increasing its robustness against various technical imperfections, but severely at the expense of longer gate duration \textcolor{black}{\cite{PhysRevA.90.032329,PhysRevLett.132.193801,PRXQuantum.4.020335}}. Because, achieving high-fidelity gates with strong robustness inherently contradicts with the requirement of fast operation, which in turn arises a big challenge toward realizing fast and high-fidelity gates for large-scale, error-tolerant quantum computation \textcolor{black}{\cite{Noiri2022}}.

Typically, faster and higher-fidelity gates are possible when the coupling laser amplitudes are greater, which is then restricted by the available laser-power system and waist diameter
imposing a critical trade-off between the gate speed and the practical feasibility \textcolor{black}{\cite{PhysRevA.109.012615}}.
Among developed techniques, the time-optimal (TO) scheme proposed by Jandura and Pupillo is particularly relevant that uses a numerically optimized phase profile for a single pulse while keeping the amplitude continuously at its maximum \textcolor{black}{\cite{Jandura2022timeoptimaltwothree}}. 
\textcolor{black}{It identifies the shortest possible global pulse with smooth time evolutions
for a CZ gate that could mitigate most type of errors. While the predicted minimal gate duration in TO protocols remains uneasy to attain practically because improving the gate robustness in a noisy environment requires the significant elongation of gate time
\cite{PhysRevResearch.5.033052,PRXQuantum.4.020336}, especially for adiabatic evolution systems \cite{PhysRevA.91.012337,PhysRevX.10.021054,PhysRevA.101.062309,robust_gate_Mitra2020}}. More recently, a different approach towards fast entangling gates is proposed \textcolor{black}{\cite{5d8p-3hm1}}, which relies on an optimized microwave modulation to facilitate the population dynamics 
mediated by the dipole-dipole interaction between pairs of Rydberg states \textcolor{black}{\cite{Chew2022}}, representing up to $20\%$ faster than the original TO schemes with van der Waals (vdWs) interaction. Notably, 
since the optimized phase and amplitude pulses required to execute two-qubit gates are not unique, providing a degree of freedom to minimize the gate duration in optimization \textcolor{black}{\cite{Goerz_2011,PhysRevA.109.022613,PhysRevResearch.6.023026,PhysRevResearch.7.023246}}.

In this work, we present an optimal-control based acceleration strategy to realize two-qubit CZ gates featuring both fast and high-fidelity merits. This method depends on a newly-introduced ancillary laser that couples the intermediate scattering state to a third hyperfine ground state (except two qubit states) serving as a control knob. Via optimizing the amplitude profile and the detuning that effectively increase the two-photon Rabi coupling strength between ground and Rydberg states, the gate operation can be significantly faster than the timescale set by native two-photon transition \cite{PhysRevA.105.042430}. In comparison to existing gates, our protocol reduces the gate execution time by more than 30$\%$ to the submicrosecond regime
without compromising high gate fidelities over a wide range of pulse parameters.

 To evaluate the performance of fast gates impacted by the ancillary drive, we present a detailed error budget, particularly quantifying the extra gate infidelity due to the imperfections of the ancillary laser: uncertainties in the applied laser amplitude and frequency \textcolor{black}{\cite{PhysRevLett.123.100501}}, as well as the laser phase noise \cite{PhysRevA.107.042611} \textcolor{black}{first extensively studied for Rydberg atoms in \cite{PhysRevA.97.053803}}. Surprisingly, our fast gates with shortened time-spent in Rydberg state, can benefit from stronger robustness to all these errors. Numerical results show that the overall infidelity purely caused by the ancillary laser drive is suppressed to a negligible level of $\sim 10^{-4}$.
 This work strongly eliminates the conventional requirement for higher laser power or specialized interaction in multiple Rydberg levels toward fast gates, paving an exceptional balance among the gate speed, the operational fidelity as well as the robustness against experimental imperfections.




\section{Acceleration Mechanism} \label{accmech}

\begin{figure}
\includegraphics[width=2.8in]{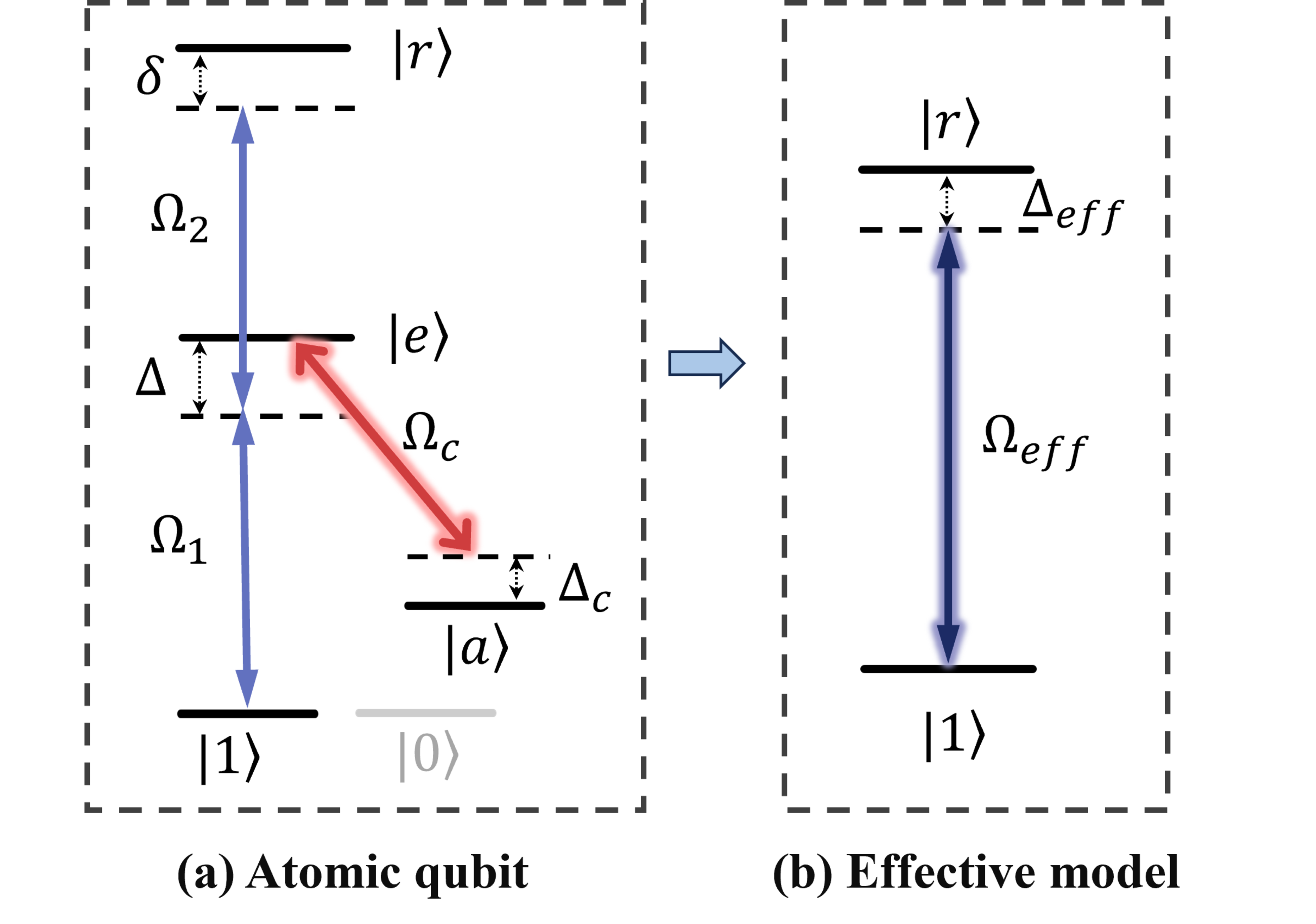}
\caption{\label{Mechanism} Acceleration mechanism. (a) Schematic representation of a general atomic qubit defined in a five-level space $\{|0\rangle$, $|1\rangle$,  $|e\rangle$, $|r\rangle$, $|a\rangle$$\}$ where $|0\rangle$ is idle. $|a\rangle$ serves an ancillary state individually coupling the intermediate state $|e\rangle$, which provides an efficient control for accelerating the standard two-photon Rabi oscillations between the ground $|1\rangle$ and the Rydberg $|r\rangle$ states. (b) The original qubit system can be mapped into an effective two-level model $\{|1\rangle,|r\rangle \}$ with Rabi frequency $\Omega_{eff}$ and detuning $\Delta_{eff}$, in the limit of large intermediate detuning conditions.}
\end{figure}

 \textcolor{black}{We start by introducing an ancillary-drive induced acceleration mechanism based on single-qubit framework, which can be used for implementing fast entangling gates later (see Sec. III).}
 Consider a general hyperfine qubit system $\{|1\rangle,|e\rangle,|r\rangle\}$ where $|0\rangle$ is idle (see Fig.\ref{Mechanism}(a)) analogous to that of Ref. \textcolor{black}{\cite{Evered2023}}, the conventional two-photon transition between $|1\rangle$ and $|r\rangle$ is driven by using two external laser fields $\Omega_1$ and $\Omega_2$, respectively detuned by $\Delta$ and $\delta$. Notably, we require that the intermediate state $|e\rangle$ here is additionally coupled by an ancillary field $\Omega_c$ to the fifth state $|a\rangle$, detuned by $\Delta+\Delta_c$, which acts as an acceleration control knob \textcolor{black}{\cite{PhysRevApplied.23.024072}}. 
 
 In the rotating frame the Hamiltonian that describes such atomic qubit depicted in Fig.\ref{Mechanism}(a), reads
\begin{eqnarray} \label{Hamn}
{H_0} =&& \frac{1}{2}(\Omega_{1} {\vert e \rangle}{\langle 1 \vert}  
 + \Omega_{2} {\vert r \rangle}{\langle e \vert} + \Omega_c|a\rangle\langle e| +\text{H.c.}) \nonumber
 \\&& -\Delta|e\rangle\langle e| - (\Delta+\Delta_c)|a\rangle\langle a|-\delta |r\rangle\langle r|
\end{eqnarray}
To understand this acceleration we first deduce the effective model. At large intermediate detuning $\Delta\gg\Omega_{1},\Omega_2$, after adiabatically eliminating the intermediate state $|e\rangle$, first the system can be approximately reduced to a three-level model, described by
\begin{eqnarray} 
{H^{\prime}_0} =&& \frac{1}{2}(\Omega_{m} {\vert a \rangle}{\langle 1 \vert}  
 + \Omega_{n} {\vert r \rangle}{\langle 1 \vert} + \Omega_k|r\rangle\langle a| +\text{H.c.}) \nonumber
 \\&& \textcolor{black}{-\Delta_1|1\rangle\langle 1|}-\Delta_a|a\rangle\langle a| - \Delta_r|r\rangle\langle r|
\end{eqnarray}
with
\begin{eqnarray}
\Omega_m&=&\frac{\Omega_1\Omega_c}{2\Delta}, \Omega_n=\frac{\Omega_1\Omega_2}{2\Delta}, \Omega_k=\frac{\Omega_2\Omega_c}{2\Delta} \nonumber\\
\textcolor{black}{\Delta_1}&=&-\frac{\Omega^2_1}{4\Delta},
\Delta_a=\Delta+\Delta_c-\frac{\Omega_c^2}{4\Delta}, \Delta_r=\delta-\frac{\Omega_2^2}{4\Delta} 
\end{eqnarray}
representing effective Rabi frequencies $\{\Omega_m,\Omega_n,\Omega_k\}$ and effective detunings $\{\Delta_1,\Delta_a,\Delta_r\}$ of the three-level system respectively. And then, we continue to perform second adiabatic elimination for state $|a\rangle$ under the condition of $\Delta_a\gg\Omega_m,\Omega_k$. Eventually, the system is conveniently described by a two-level Hamiltonian merely in the $\{|1\rangle,|r\rangle\}$ basis as shown in Fig.\ref{Mechanism}(b),
\begin{eqnarray}\label{Ut}
  \small
  &&\small
  {H}_{eff} = \left[
  \begin{array}{ccc}
   0
  &   \frac{\Omega_{eff}}{2}
  \\
  \frac{\Omega_{eff}}{2}
  &  -\Delta_{eff}
  \end{array}
  \right] \label{ef}
\end{eqnarray}
where the effective Rabi frequency and detuning are
\begin{eqnarray}
   \Omega_{eff}&=&\Omega_n+\frac{\Omega_{m}\Omega_{k}}{ 2\Delta_a}=\frac{\Omega_{1}\Omega_{2}}{ 2\Delta-\frac{\Omega_c^2}{2(\Delta+\Delta_c)}} \nonumber\\
   \textcolor{black}{\Delta_{eff}}&=&\frac{\Omega_{m}^{2}-\Omega_{k}^{2}}{4\Delta_a}+\Delta_r-\Delta_1=\frac{\Omega_{1}^{2}-\Omega_{2}^{2}}{4\Delta-\frac{\Omega_c^2}{\Delta+\Delta_c}}+\delta.
\end{eqnarray}
In this reduced picture denoted by $H_{eff}$, the diagonal term $\Delta_{eff}$ is given by the ac Stark shift which may be mitigated using a tunable detuning $\delta$ and when $\Omega_1=\Omega_2$ and $\delta=0$ this shift vanishes \textcolor{black}{\cite{PhysRevLett.129.200501}}. More crucially, we see the effective off-diagonal Rabi coupling $\Omega_{eff}$ has been modified. 
With an appropriate ancillary drive characterized by $\Omega_c$ and $\Delta_c$, an enhanced $\Omega_{eff}$ can be obtained in comparison to the general two-photon coupling rate $\frac{\Omega_1\Omega_2}{2\Delta}$ in the absence of $|a\rangle$, i.e. $\Omega_{eff}> \frac{\Omega_1\Omega_2}{2\Delta}$. For example, when $\Delta_c=-0.5\Delta$(assume $\Delta$ is positive) having a negative sign to achieve a significant acceleration and meanwhile a large $\Omega_c=\Delta$ is preferred, the modified Rabi frequency $\Omega_{eff}$ can increase to $\frac{\Omega_1\Omega_2}{\Delta}$ becoming twice the general Rabi value without ancillary drive, promising for faster and higher-fidelity quantum gates \textcolor{black}{\cite{PRXQuantum.6.010331}}.

\begin{figure}
\includegraphics[width=2.8in]{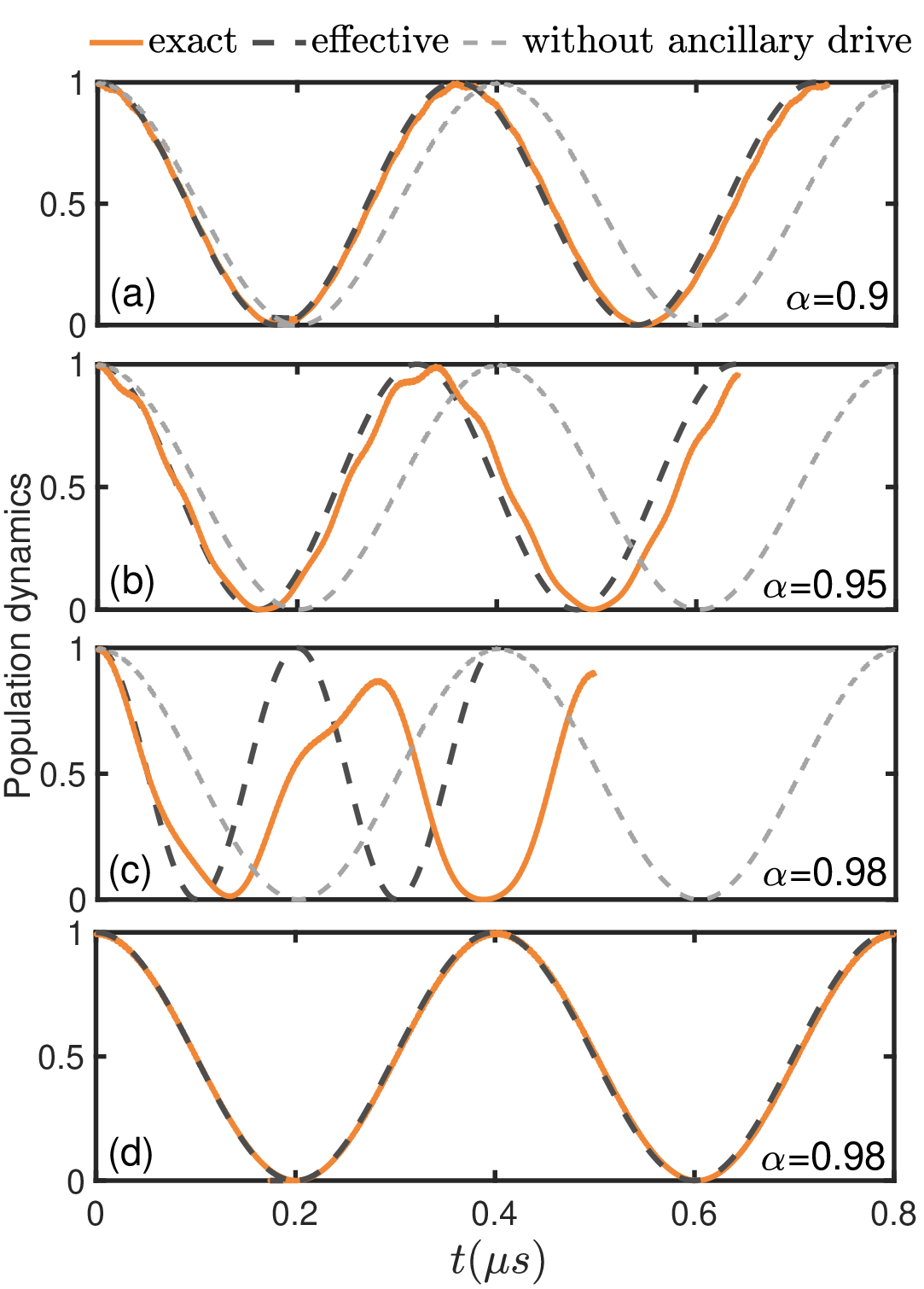}
\caption{\label{validate} Numerical verification of the ancillary-field acceleration scheme. Time-dependent population dynamics in two periods on state $|1\rangle$ are comparably shown, resolved from the exact Hamiltonian (\ref{Hamn}) with ancillary-field, the effective Hamiltonian (\ref{ef}) and the exact dynamics without ancillary-field. We set $\alpha=(0.9,0.95,0.98)$ for panels (a)-(c). Common parameters are $\Delta/2\pi=500$ MHz,  $\Omega_1=\Omega_2=0.1\Delta$, $\Omega_c=0.2\Delta$, $\Delta_c = –\alpha \Delta$ and $\delta=0$. Panel (d) separately shows the case of weak ancillary drive where $\Omega_c=0.01\Delta, \alpha=0.98$ and others are same as in (a)-(c).}
\end{figure}

To validate the acceleration mechanism, in Fig.\ref{validate}(a-c) we numerically solve the exact time-dependent population on state $|1\rangle$ \textcolor{black}{(orange-solid lines)} from the Schr\"{o}dinger equation without any decay, under different $\Delta_c$ values. \textcolor{black}{Here we define $\Delta_c = -\alpha \Delta$ with $\alpha$ a coefficient quantifying the ratio between $|\Delta_c|$ and $\Delta$.} The theoretical prediction \textcolor{black}{(black-dashed lines)} using the effective two-level model is comparably shown exhibiting a good agreement for $\alpha=0.9$ and $0.95$. \textcolor{black}{Specifically, for $\alpha=0.9$ there exists a nearly perfect agreement between the exact and effective solutions because the second adiabatic condition $\Delta_a\gg \Omega_{m,k}$ for state $|a\rangle$ fulfills. As $\alpha$ increases to 0.95 it gives rise to a small difference because of the partial participation of state $|a\rangle$ that disagrees with the effective two-level model.} Moreover, if $\alpha=0.98$, due to the break of large detuning condition for state $|a\rangle$ there exists a relatively large discrepancy between the effective and exact models.

In order to quantify the degree of acceleration we introduce a ratio $p$ taking form of
\begin{equation}
    p = \frac{T_{0}-T}{T_0}
\end{equation}
where $T_0 = 2\pi/(\frac{\Omega_1\Omega_2}{2\Delta}) = \textcolor{black}{0.4} $ $\mu$s sets the fixed timescale of one-period oscillation without ancillary drive \textcolor{black}{(gray-dashed lines)} and $T$ characterizes the realistic one-period duration estimated by the exact population dynamics. 
Remarkably,  
as $\alpha$ increases the acceleration becomes more pronounced. For $\alpha=0.9$ we obtain $p \approx \textcolor{black}{0.0991}$ corresponding to a $9.91\%$ faster for the population evolution. However, when $\alpha$ is increased to 0.95 sustaining the accuracy of effective two-level model the acceleration degree for population evolution has been improved to 15.32$\%$. If $\alpha$ grows even to $0.98$, although the effective theoretical prediction becomes slightly inconsistent with the exact simulation because of the breakdown of large detuning condition for state $|a\rangle$, the acceleration degree can reach $p \approx 25.11\%$ which means it only requires $T = 0.2795$ $\mu$s to accomplish an one-period population oscillation in an atomic qubit. Note that this case although offers the fastest dynamics, yet needs a near-resonant $|a\rangle$ state detuned by $\Delta+\Delta_c\approx 0.02\Delta$ that will strongly impact the two-photon Rydberg transition, consequently arising a larger leakage error because of a longer time spent on $|a\rangle$ (see the FAIL case). 
In Sec. III, we show that this additional acceleration knob with a moderate $\alpha$ can enable the realization of fast two-qubit CZ gates with high fidelities.

\textcolor{black}{In contrast, Fig.\ref{validate}(d) presents a weak ancillary drive case where $\Omega_c$ is tuned to be smaller than $\Omega_{1,2}$ by one order of magnitude. As expected, even if $\alpha=0.98$ corresponding to the fastest dynamics (see Fig.\ref{validate}(c)) a small $\Omega_c$ is still unable to provide efficient acceleration because the role of the ancillary drive is negligible here.}

\section{Fast and high-fidelity CZ Gates}

\subsection{Accelerated smooth-amplitude gate construction}

\begin{figure}
\includegraphics[width=3in]{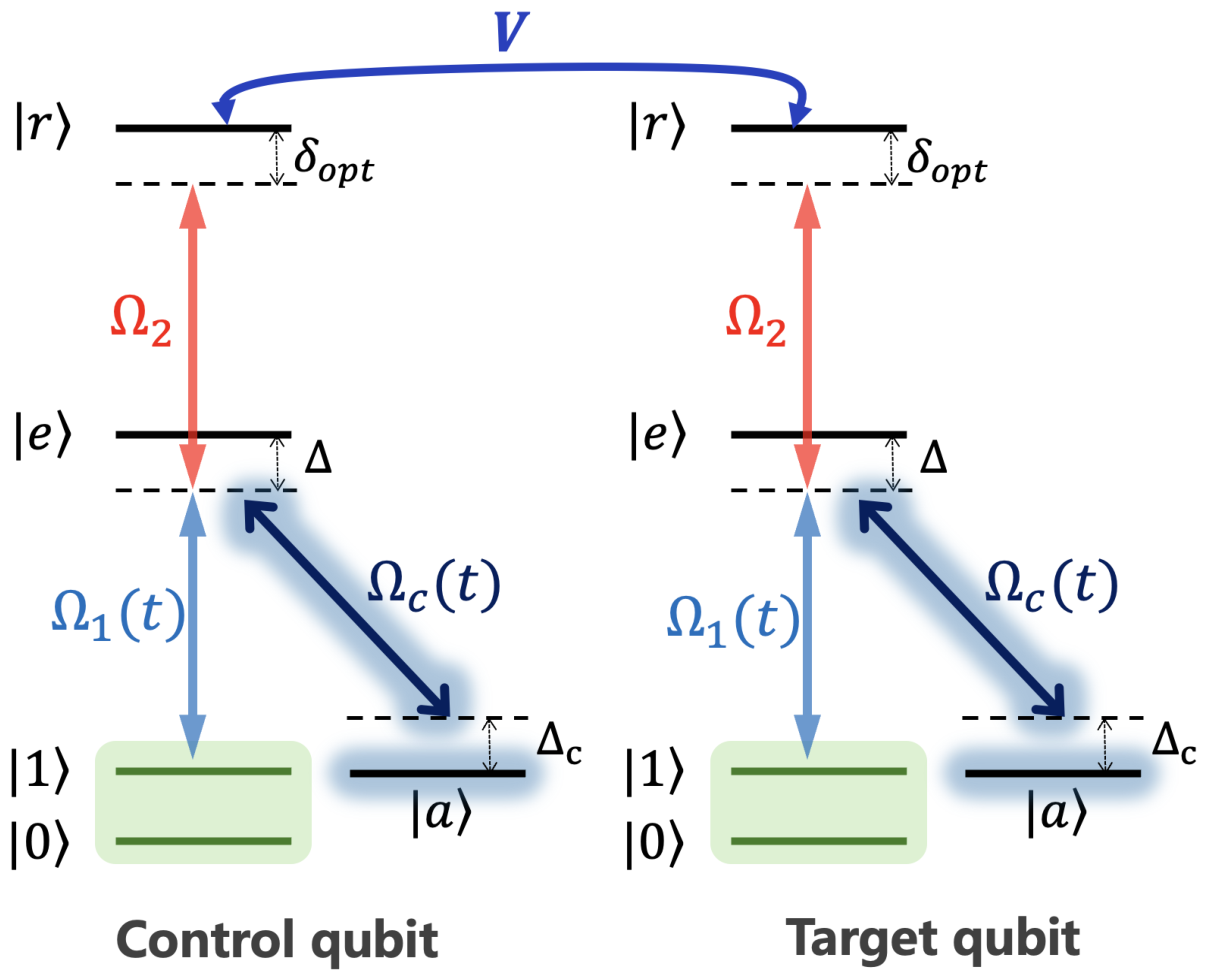}
\caption{\label{gate} Fast and high-fidelity CZ gates with Rydberg atomic qubits. Schematic diagram and atomic level structure utilized in this work. The qubit state $|1\rangle$ couples to the Rydberg state $|r\rangle$ via a native two-photon transition mediated by state $|e\rangle$, with respective Rabi frequencies $\Omega_1,\Omega_2$ and detunings $\Delta,\delta_{opt}$. Specifically, another ancillary field near-resonantly couples the intermediate state $|e\rangle$ to another hyperfine (stable) ground state $|a\rangle$ with detuning $\Delta+\Delta_c\approx 0$ due to $\Delta\Delta_c<0$ and Rabi frequency $\Omega_c$, which can serve as a robust control knob for gate acceleration. Time-dependent pulse shaping is carried out via optimal control over two laser amplitudes $\Omega_1(t)$ and $\Omega_c(t)$ while $\Omega_2$ is kept fixed. A natural vdWs interaction with strength $V$ is assumed between atoms in double excited state $|rr\rangle$ implementing the Rydberg blockade effect \textcolor{black}{\cite{Gaetan2009,PhysRevLett.119.160502}}.
 }
\end{figure}

To identify the acceleration mechanism that can be applied for realizing a faster two-qubit entangling gate, we restrict our focus to the gate scheme as shown in Fig. \ref{gate}, consisting of two five-level atomic qubits initialized in alkali $^{87}$Rb atoms, which can be described by the Hamiltonian
\begin{eqnarray} \label{Ham1}
{H(t)} = {H_0}(t) \otimes {I} + {I} \otimes {H_0}(t) + V {\vert rr\rangle} {\langle rr\vert}
\end{eqnarray}
where $I$ is the identity operator and $H_0(t)$ describes the single atomic qubit taking form of
\begin{eqnarray} \label{Ham}
{H_0}(t) =&& \frac{1}{2}(\Omega_{1}(t) {\vert e \rangle}{\langle 1 \vert}  
 + \Omega_{2} {\vert r \rangle}{\langle e \vert} + \Omega_c(t)|e\rangle\langle a| +\text{H.c.}) \nonumber
 \\&& -\Delta|e\rangle\langle e| - (\Delta+\Delta_c)|a\rangle\langle a|-\delta_{opt} |r\rangle\langle r|
\end{eqnarray}
  equivalent to Eq.(\ref{Hamn}). We first discuss a standard two-photon protocol without ancillary drive $\Omega_c(t)=0$, i.e. a native two-photon excitation scheme that is commonly-used by experiments \textcolor{black}{\cite{PhysRevLett.100.253001,PhysRevLett.110.123001,PhysRevLett.114.113002}}. \textcolor{black}{We implement the gate with a varying Rabi frequency $\Omega_1(t)$ while keeping $\Omega_2$ fixed}
  so as to realize the smooth-amplitude (SA) gates \textcolor{black}{\cite{Evered2023}}, which can help achieve higher fidelities even when the laser amplitude is limited and strongly decrease the spontaneous scatterings compared to the TO gate configurations \textcolor{black}{\cite{Baler2023}}. To minimize the set of tunable gate parameters we adopt an optimal constant detuning $\delta_{opt}$ possibly mapped by the laser phase with a simply linear modulation because the slope of phase modulation corresponds to a two-photon detuning. Note that the optimization of $\delta_{opt}$ can also partially compensate for the ac Stark shift (the first term of $\Delta_{eff}$, see Eq.(\ref{eff})) that evolves dynamically during the gate duration \textcolor{black}{\cite{Li_2023}}. 

Instead of the general two-photon SA model discussed above, we now introduce the acceleration by coupling the intermediate state $|e\rangle$ to another hyperfine ground level $|a\rangle$ which is absolutely stable avoiding extra decay. In the single-qubit framework (see Sec.II) we have verified that, the combination of auxiliary-field $\Omega_c(t)$ and its detuning $\Delta_c$ operation leads to a modified two-photon Rabi frequency and detuning between the ground-Rydberg excitation, as
\begin{equation}
\Omega_{{eff}}=\frac{\Omega_{1}(t)\Omega_{2}}{ 2\Delta-\frac{\Omega_c^2(t)}{2(\Delta+\Delta_c)}}, \Delta_{eff}=\frac{\Omega_{1}^{2}(t)-\Omega_{2}^{2}}{4\Delta-\frac{\Omega_c^2(t)}{\Delta+\Delta_c}}+\delta_{opt}.
\label{eff}
\end{equation}
We expect an appropriate optimization for a larger $\Omega_{eff}$ to speed up the population exchange between $|1\rangle$ and $|r\rangle$ in a single qubit, \textcolor{black}{giving rise to an accelerated smooth-amplitude (ASA) gate}. However, accounting for that a realistic CZ gate is also restricted by the need for the overall phase change $\phi_{11}-\phi_{10}-\phi_{01}=(2n+1)\pi$ for an integer $n$ (here $n=0$) based on four computational basis states $\{|00\rangle,|01\rangle,|10\rangle,|11\rangle\}$ \textcolor{black}{\cite{Levine2019,tmr4-gtnl}}, not a single-qubit phase shift, we further resort to robust optimization of CZ gate parameters toward the physical realization of the SA gate and the ASA gate.

\begin{table*} 
\caption{The optimized control parameters for the Rabi frequencies $\Omega_1(t)$ (given by $\beta_v^1$), $\Omega_c(t)$ (given by $\beta_v^c$), the detunings $\delta_{opt}$, $\Delta_c$ (given by $\alpha$), the gate duration $T$ as well as
the average time spent $T_e,T_r,T_a$ during the pulse.  In all these cases, some parameters are kept constant including $\Delta/2\pi=1.0$ GHz, $\Omega_2/2\pi=50$ MHz, $V/2\pi=300$ MHz and the initial search ranges for parameters are $\beta_v^{1,c}\in [0,200]$ MHz that ensures the maximal Rabi frequency $\Omega_{1,c}(t)/2\pi\lesssim 200$ MHz available in a practical setup,
$\delta_{opt}/2\pi\in[-10,10]$ MHz, $\alpha\in[0.9,1.0]$. The search ranges of minimal gate time are set by $T\in [0,1.0]$ $\mu$s, $[0,0.6]$ $\mu$s, $[0,0.3]$ $\mu$s pertaining to Cases I-III respectively.
Here $F$ presents the realistic Bell-state fidelity in the presence of spontaneous decays with rates $\gamma_r= 3.0$ kHz and $\gamma_e/2\pi= 1.0$ MHz.}
\label{tab:laser_coefficients}
\renewcommand{\arraystretch}{1.3} 
\footnotesize
\setlength{\tabcolsep}{3.3mm}{ 
\begin{tabular}{c|c|c|c|c|c|c|c|c|c}
\hline\hline 
 Case & Gate time($\mu$s)&  $\beta_v^1$(MHz) & $\frac{\delta_{opt}}{2\pi}$(MHz) & $\beta_v^c$(MHz) & $\alpha$ & $F$ & $T_{e}$(ns) & $T_{r}$($\mu$s) & $T_a$(ns)\\
\hline
\multirow{6}{*}{I} 
& \multirow{2}{*}{$T_0=1.0$} 
& \multirow{6}{*}{\makecell{$\beta^1_1=32.58$\\[2ex]$\beta^1_2=49.19$\\[2ex]$\beta^1_3=52.10$\\[2ex]$\beta^1_4=61.16$}} 
& \multirow{6}{*}{-3.794} 
& \multirow{2}{*}{/} & \multirow{2}{*}{/} & \multirow{2}{*}{0.9969} & \multirow{2}{*}{$0.311$} & \multirow{2}{*}{$0.096$} & \multirow{2}{*}{/}\\ 
&&&&&&&&& \\ 
\cline{2-2} \cline{5-10} 
& \multirow{4}{*}{$T=0.5551$} 
&&& $\beta^{c}_1=78.58$ & \multirow{4}{*}{$0.9286$} & \multirow{4}{*}{0.9973} & \multirow{4}{*}{$0.367$} & \multirow{4}{*}{$0.058$} & \multirow{4}{*}{$1.652$}\\ 
&&&& $\beta^{c}_2=65.33$ &&&&& \\ 
&&&& $\beta^{c}_3=172.7$ &&&&& \\ 
&&&& $\beta^{c}_4=125.5$ &&&&& \\ 
\hline
\multirow{6}{*}{II} 
& \multirow{2}{*}{$T_0=0.50$} 
& \multirow{6}{*}{\makecell{$\beta^1_1=17.78$\\[2ex]$\beta^1_2=5.840$\\[2ex]$\beta^1_3=27.54$\\[2ex]$\beta^1_4=157.4$}} 
& \multirow{6}{*}{-5.595} 
& \multirow{2}{*}{/} & \multirow{2}{*}{/} & \multirow{2}{*}{0.9973} & \multirow{2}{*}{$0.298$} & \multirow{2}{*}{$0.052$} & \multirow{2}{*}{/}\\ 
&&&&&&&&& \\ 
\cline{2-2} \cline{5-10}
& \multirow{4}{*}{$T=0.3338$} 
&&& $\beta^{c}_1=113.1$ & \multirow{4}{*}{$0.9479$} & \multirow{4}{*}{0.9974} & \multirow{4}{*}{$0.315$} & \multirow{4}{*}{$0.038$} & \multirow{4}{*}{$1.328$}\\ 
&&&& $\beta^{c}_2=112.6$ &&&&& \\ 
&&&& $\beta^{c}_3=64.68$ &&&&& \\ 
&&&& $\beta^{c}_4=94.88$ &&&&& \\ 
\hline
\multirow{6}{*}{III} 
& \multirow{2}{*}{$T_0=0.25$} 
& \multirow{6}{*}{\makecell{$\beta^1_1=10.19$\\[2ex]$\beta^1_2=25.47$\\[2ex]$\beta^1_3=6.044$\\[2ex]$\beta^1_4=200.0$}} 
& \multirow{6}{*}{-3.992} 
& \multirow{2}{*}{/} & \multirow{2}{*}{/} & \multirow{2}{*}{0.9981} & \multirow{2}{*}{$0.206$} & \multirow{2}{*}{$0.038$} & \multirow{2}{*}{/}\\ 
&&&&&&&&& \\ 
\cline{2-2} \cline{5-10}
& \multirow{4}{*}{$T=0.1709$} 
&&& $\beta^{c}_1=0.260$ & \multirow{4}{*}{$0.9337$} & \multirow{4}{*}{0.9980} & \multirow{4}{*}{$0.251$} & \multirow{4}{*}{$0.028$} & \multirow{4}{*}{$0.723$}\\ 
&&&& $\beta^{c}_2=75.20$ &&&&& \\ 
&&&& $\beta^{c}_3=196.7$ &&&&& \\ 
&&&& $\beta^{c}_4=60.40$ &&&&& \\ 
\hline\hline 
\multirow{10}{*}{LIM} 
& \multirow{5}{*}{$T_0=0.1$} 
& \multirow{5}{*}{\makecell{$\beta^1_1=108.5$\\[1ex]$\beta^1_2=48.11$\\[1ex]$\beta^1_3=142.5$\\[1ex]$\beta^1_4=147.1$}} 
& \multirow{5}{*}{-10.00} 
& \multirow{5}{*}{/} & \multirow{5}{*}{/} & \multirow{5}{*}{0.9872} & \multirow{5}{*}{$0.354$} & \multirow{5}{*}{$0.026$} & \multirow{5}{*}{/}\\ 
&&&&&&&&& \\ 
&&&&&&&&& \\ 
&&&&&&&&& \\
&&&&&&&&& \\
\cline{2-10}
 
& \multirow{5}{*}{$T=0.1$} 
& \multirow{5}{*}{\makecell{$\beta^1_1=80.91$\\[1ex]$\beta^1_2=18.61$\\[1ex]$\beta^1_3=143.6$\\[1ex]$\beta^1_4=132.9$}} 
& \multirow{5}{*}{-9.221} 
& \multirow{5}{*}{\makecell{$\beta^c_1=183.8$\\[1ex]$\beta^c_2=168.4$\\[1ex]$\beta^c_3=103.4$\\[1ex]$\beta^c_4=117.7$}} & \multirow{5}{*}{0.9003} & \multirow{5}{*}{0.9952} & \multirow{5}{*}{$0.350$} & \multirow{5}{*}{$0.023$} & \multirow{5}{*}{$0.634$}\\ 
&&&&&&&&& \\ 
&&&&&&&&& \\ 
&&&&&&&&& \\ 
&&&&&&&&& \\ 

\hline

\multirow{5}{*}{FAIL}& \multirow{5}{*}{$T=0.0999$} 
& \multirow{5}{*}{\makecell{$\beta^1_1=80.91$\\[1ex]$\beta^1_2=18.61$\\[1ex]$\beta^1_3=143.6$\\[1ex]$\beta^1_4=132.9$}} 
& \multirow{5}{*}{-9.221} 
& \multirow{5}{*}{\makecell{$\beta^c_1=64.39$\\[1ex]$\beta^c_2=4.81$\\[1ex]$\beta^c_3=2.13$\\[1ex]$\beta^c_4=14.60$}} & \multirow{5}{*}{0.9800} & \multirow{5}{*}{0.9041} & \multirow{5}{*}{$0.318$} & \multirow{5}{*}{$0.024$} & \multirow{5}{*}{$3.585$}\\ 
&&&&&&&&& \\ 
&&&&&&&&& \\ 
&&&&&&&&& \\ 
&&&&&&&&& \\ 

\hline\hline

\end{tabular}
}
\end{table*}

\subsection{Time-acceleration gate optimization}
\label{optimization method}

Our primary target is to determine all gate parameters that realize a high-fidelity ASA-type CZ gate in which the total gate duration $T$ is shorter than that of a standard SA gate, so we apply a commonly-used Bell-state fidelity $\mathcal{F}$ to measure the gate performance in an idealized decay-free case, defined as \textcolor{black}{\cite{PRXQuantum.4.020336}}
\begin{eqnarray} 
\mathcal{F} = \frac{1}{16} \left| \sum_{q}e^{-i\phi_{q}}\langle q|\psi_{q}\rangle\right|^2.
\end{eqnarray}
This metric accurately captures the replication between the realistic operation (final state $|\psi_q\rangle$) after gate duration $T$ and the desired output $e^{i\phi_q}|q\rangle$ for a CZ gate. The ideal final state $|\psi_q\rangle$ can be found by solving the Schr\"{o}dinger equation \textcolor{black}{$i\dot \psi_q(t) = H \psi_q(t)$} for different input states $ q \in\{00,10,01,11\}$
and the system stays unaffected for the input of $|00\rangle$. Note that we require the gate to achieve the best fidelity under intrinsic error sources, during which one significant error source arises from the spontaneous decays due to finite lifetime of the intermediate $|e\rangle$ and the Rydberg $|r\rangle$ states \textcolor{black}{\cite{PhysRevLett.123.230501}}. To minimize this intrinsic error we modify the cost function in optimization, as
\begin{equation}
   \mathcal{J}=(1-\mathcal{F})+(\gamma_e T_e+\gamma_rT_r )
   \label{cost}
\end{equation}
with $\gamma_{e},\gamma_r$ being the decay rates and $T_{e},T_r$ quantifying the time integration of intermediate and Rydberg population average over all computational basis states, given by
\begin{equation}
    T_{e,r} =  \int_0^T \bar{n}_{e,r}(t) dt
\end{equation}
Here, $\bar{n}_{e,r}(t) = \frac{1}{4}\sum_q \langle\psi_q(t)|\Pi_{e,r}|\psi_q(t)\rangle$ denotes the average population dynamics with $\Pi_{e,r}$ the projector onto the subspace where every atom occupies the lossy $|e\rangle$ or $|r\rangle$ state. The first term in Eq.(\ref{cost}) suggests minimizing the ideal infidelity $1-\mathcal{F}$ without accounting for any decay. While the minimization of the second term means that the gate is less affected by the spontaneous decay from two intermediate leakage states $|e\rangle$ and $|r\rangle$, \textcolor{black}{because $\gamma_e T_e$ and $\gamma_r T_r$ express the primary intrinsic decay errors in the gate which includes the overall decaying channels from states $|e\rangle$ and $|r\rangle$ within the Hilbert space
\textcolor{black}{\cite{56qk-rmsz}}}. 
Remarkably, the latter term is particularly important because a fast gate with shortened duration $T$ in principle enables the decrease of the time-spent $T_{e},T_r$, providing a more realistic gate performance under a noisy environment \textcolor{black}{\cite{PhysRevLett.127.050501}}.



Optimization of pulses is performed by utilizing the Genetic algorithm (GA) following our recent work \textcolor{black}{\cite{PhysRevApplied.17.024014}} which offers a global search within a wide range for all control parameters on two atomic qubits, mitigating the risk of getting a local minimum of $\mathcal{J}$. \textcolor{black}{While executing GA, one individual represents a candidate solution encoded by a full set of control parameters. Starting from the population of 100 randomly initialized individuals, the GA algorithm iteratively improves the population over 20 generations. In each generation, individuals are selected according to their fitness.
Subsequently, their parameters undergo recombination (crossover) and random perturbation (mutation) to produce the population for the next generation. 
This complete evolutionary cycle, from random initialization to convergence on a final set of optimized parameters, constitutes one optimization run. The individual with the highest fitness in the final generation is output as the optimal solution \cite{Katoch2020}. The computational time for one such run depends on the problem scale and available resources, typically averaging 1–2 hours under our practical settings.} The general SA gate, being the reference point, is provided by a fixed duration $T_0\in\{1.0,0.5,0.25\}\mu$s and a smooth-varying laser amplitude $\Omega_1(t)$. Note that, $T_0$ is defined by the total gate duration for general SA gate, not the single-qubit dynamics. Without loss of generality we choose the time-modulated waveform of lasers $\{\Omega_1(t),\Omega_c(t)\}$
in terms of \textcolor{black}{a linear combination of basis polynomials},
\begin{equation}
 \Omega_{{j\in(1,c)}}(t)/2\pi=\sum_{v=1}^{4} \beta^{j}_v \left[b_{v,n}(t/T)+b_{n-v,n}(t/T)\right]    \label{ome1}
\end{equation}
where $b_{v,n}$ is the $v$th Bernstein basis polynomial \textcolor{black}{that features advantages of inherent smoothness, symmetry and the absence of a long tail, ensuring both compatibility with numerical optimization and a straightforward experimental implementation. \textcolor{black}{In addition, we truncate at $n=8$ here \cite{PhysRevApplied.13.024059}, as a further increase in $n$ does not yield a noticeable improvement in the fidelity $F$ (e.g. the variation in $|F|$ remains $\lesssim 10^{-3}$ for $n=10$, not shown). This truncation order is sufficient for the subsequent optimization refinement}}. $\beta_{1\sim 4}^j$ is the coefficient having the unit of MHz. Note that we set the two-photon detuning $\delta_{opt}$ as another important parameter in SA gate to be optimized because it can reduce the impact of ac Stark shift during the pulse evolution.

The optimization proceeds by combining with the ancillary-state parameters for an ASA gate where we also precisely control the waveform of laser amplitude $\Omega_c(t)$(see Eq.\ref{ome1}) and leave the extra detuning $\Delta_c$ as a tunable parameter. 
The choice of detuning $\Delta_c$ is particularly important that primarily determines the acceleration degree $p$ and should be given an opposite sign with respect to the intermediate detuning $\Delta$ as verified in Sec II. Therefore, we also apply $\Delta_c = -\alpha \Delta$ with its search range $\alpha\in[0.9.1.0]$  being close to 1.0, in order to achieve an efficient acceleration.



\subsection{Implementation of accelerated CZ gates}
\label{accelerated gate}

\begin{figure*}
\includegraphics[width=7in]{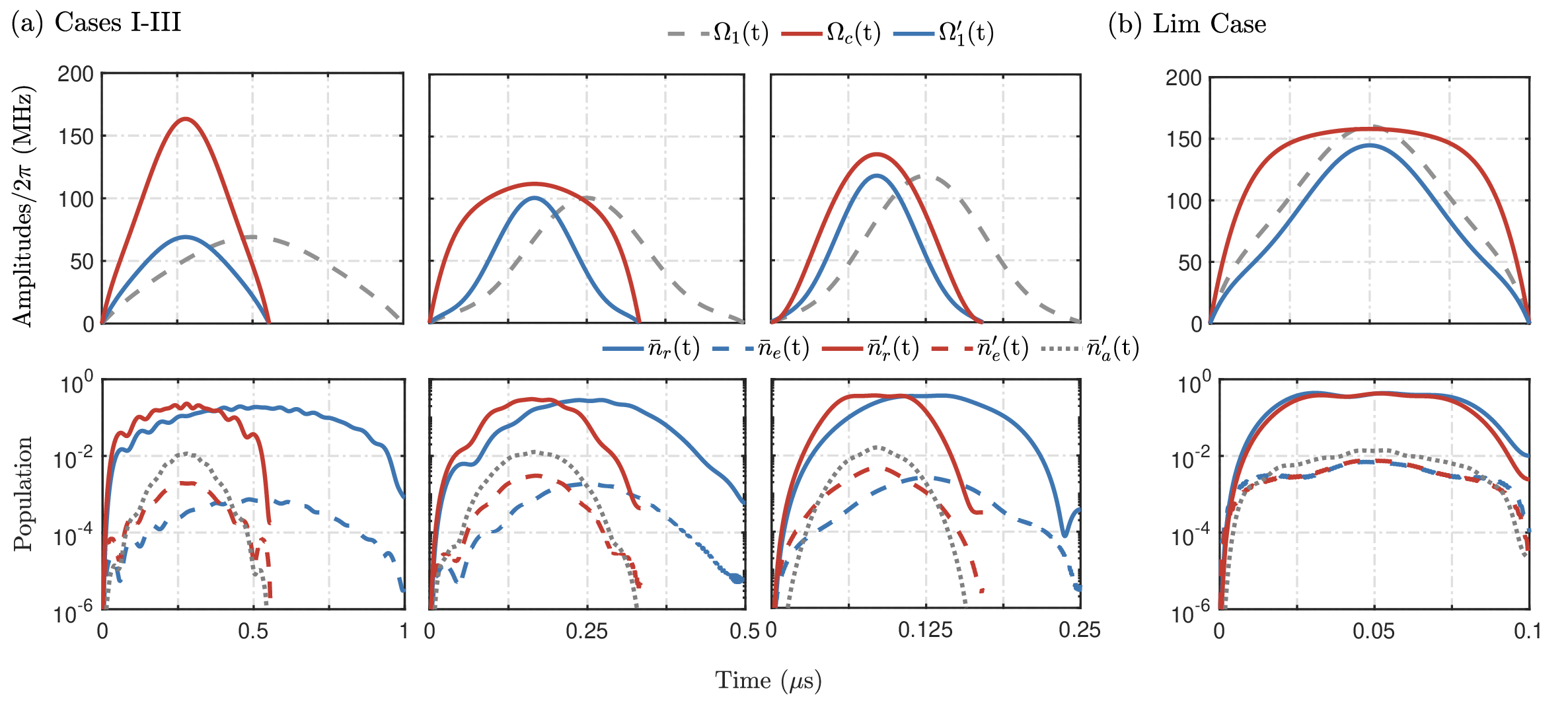}
\caption{\label{light} (a) Upper panels: Optimized pulse amplitudes as a function of time for $\Omega_1(t)$ of the SA gates and $(\Omega_1^\prime(t),\Omega_c(t))$ of the ASA gates, corresponding to Cases I-III in Table I. Note that $\Omega_1^\prime(t)$ takes a same shape as $\Omega_1(t)$ except for a reduced gate time.
Lower panels: Resolved time-dependent population on the intermediate $|e\rangle$ (dashed, $\bar{n}_e$ and $\bar{n}_e^\prime$), the Rydberg $|r\rangle$ (solid, $\bar{n}_r$ and $\bar{n}_r^\prime$) and the auxiliary $|a\rangle$ (dotted, $\bar{n}_a^\prime$) states during the pulse duration. Clearly, using the ASA gate strategy we can implement the gates with faster execution of laser pulses while keeping the intrinsic decay errors at a same level. (b) Same parameters resolved for two LIM Cases in which the pulse durations $T_0$ and $T$ are fixed to be $0.1$ $\mu$s.}
\end{figure*}

With the optimization design established we now demonstrate the implementation of an accelerated CZ gate with this approach. All results are summarized in Table I in which we compare three cases with different reference points for $T_0=\{1.0,0.5,0.25\}\mu$s in standard SA gates. Numerical plots are presented in Fig.\ref{light} for a visible supplement. In Case I (Fig.\ref{light}(a), left column), we first consider the performance of $T_0=1.0\mu$s with a general two-photon excitation model
and obtain a high fidelity of $F\approx 0.9969$ after sufficient optimization, which attributes to the simultaneous
minimization of the intermediate and Rydberg decay errors $\gamma_eT_e$ and $\gamma_rT_r$ in the cost function $\mathcal{J}$ (see Eq.\ref{cost}). The optimal Rydberg-decay induced error is $\gamma_rT_r\sim 2.888\times 10^{-4} $, 
typically smaller than that from the intermediate state 
$\gamma_eT_e\sim 1.954\times 10^{-3}$ by 1 order of magnitude, due to the long lifetime of Rydberg state. In addition we find the two-photon detuning satisfies $\Delta\delta_{opt}<0$ natively even with a symmetric search range of $\delta_{opt}/2\pi\in [-10,10]$ MHz initially, in order to improve fidelities by reducing the intermediate-state scattering \textcolor{black}{\cite{PhysRevA.107.062609}}.

As turning to the case with ancillary drive, we emphasize that the native laser $\Omega_1^\prime(t)$ takes the same modulated waveform as $\Omega_1(t)$ which is only rescaled in a shortened gate duration ($<1.0\mu$s) via optimization. By implementing an efficient control of $\Omega_c(t)$ having a higher amplitude along with an optimal detuning $\Delta_c$ characterized by $\alpha$, we re-optimize the gate by introducing the duration $T$ as a newly optimal variable.
We find that the shortest gate time needed for a comparably high fidelity $F\approx 0.9973$ can be reduced to $T=0.5551\mu$s achieving a substantial acceleration degree by $p\approx44.5\%$. We note that, such acceleration clearly arises from the presence of a suitable ancillary drive ($\Omega_c(t),\Delta_c$) that substantially speeds up the ground-Rydberg Rabi rotation with greater laser intensity
in single-qubit operation frame. To understand why the acceleration strategy does not lead to a higher fidelity (just comparable),
as plotted in Fig.\ref{light}(a) (left column) we show the average population dynamics on the intermediate and Rydberg states quantifying the impact of decay errors. \textcolor{black}{We observe that, although the time-spent in the Rydberg state can be remarkably decreased due to the integration of a shorter duration allowing the Rydberg decay error to be $\gamma_rT_r\sim 1.747\times 10^{-4}$, yet the auxiliary field $\Omega_c(t)$ will continuously employ a strong coupling between $|e\rangle$ and $|a\rangle$ that slightly increases the population on state $|e\rangle$ arising $\gamma_eT_e\sim 2.304\times 10^{-3}$. Therefore, our accelerated CZ gates have no explicit indication that they could contribute a higher fidelity.}

We further explore the acceleration strategy by optimizing the gate protocol for shorter reference points $T_0=0.5$ $\mu$s and 0.25 $\mu$s. We also show the results for the SA and ASA gates respectively, with the target of obtaining a faster gate operation. In Case II and III we note that the required laser amplitude for $\Omega_1(t)$ in SA gates becomes a little stronger due to the decrease of $T_0$. However, as long as the auxiliary field $\Omega_c(t)$ presents, an even faster gate with comparably high-fidelity can be obtained. In particular, in Case III we show the realization of an ASA gate with duration $T=0.1709\mu$s (accelerated by $p\approx31.64\%$) meanwhile contributing for a high-fidelity performance $F\approx 0.9980$. 
Hence our ASA gates can benefit from not only a faster quantum gate operation for scalable systems but also a high gate fidelity, promising for precise and large-scale quantum computing.

\section{Fundamental superiority for the ASA gates}

So far, we have demonstrated the implementation of accelerated CZ gates with an optimized ancillary drive. There raises a critical question that where is the limitation for the gate acceleration? To fundamentally validate the inherent superiority of our proposal, we turn to 
systematically explore the SA and ASA gate performance under an even shorter time constraint $T_0=0.1\mu$s, known as the LIM case (see Table \ref{tab:laser_coefficients}). During the optimization strategy, the SA gate employs the same method and we find a lower fidelity value $F=0.9872$ failing to surpass 0.99 within the given time limit, at the expense of stronger peak Rabi frequency of $\Omega_1(t)$ and a largely negative detuning $\delta_{opt}/2\pi=-10$MHz. This trend is consistent with the expected physical intuition that a fast gate is essentially limited by the attainable laser power \textcolor{black}{\cite{Schfer2018}}. \textcolor{black}{The reason for a slightly poor fidelity here is primarily due to the incomplete return of the $|11\rangle$ state population by the end of the allowed gate duration because of the imperfect blockade condition \cite{4pzb-9nlg}. As the peak Rabi frequency becomes larger it would arise a blockade error that is proportional to $\Omega^2/V^2$ acting on state $|11\rangle$ \textcolor{black}{\cite{PhysRevA.96.042306}}, thereby lowering the gate fidelity.}

 While for the ASA gate, we adopt a different optimization strategy by globally searching for all parameters including $\Omega_1^\prime(t)$. With optimized ancillary drive, it is clear that the ASA gate can achieve a higher fidelity of $F=0.9952$ under same short duration of 0.1 $\mu$s significantly outperforming the general SA gate. In this LIM case, owing to  the participation of ancillary-field coupling, the population evolution for state $|11\rangle$ can return near-completely to its initial state by the end of gate operation. Particularly, it is noteworthy that the ASA protocol effectively circumvents the speed limit inherent to traditional adiabatic pathway by incorporating non-adiabatic couplings, thereby substantially increasing the gate speed while maintaining operation accuracy \textcolor{black}{\cite{PhysRevA.110.032619,Beterov2020}}. Also, we fail to find a better ASA gate when $\alpha$ is increased to 0.98, which was initially expected to offer the fastest population evolution based on a single qubit system (see Fig.\ref{validate}(c)). This is because the time-spent in a near-resonantly ancillary state $|a\rangle$ leads to a larger residual atomic population lowering the gate fidelity.

\begin{figure}
\includegraphics[width=3.5in]{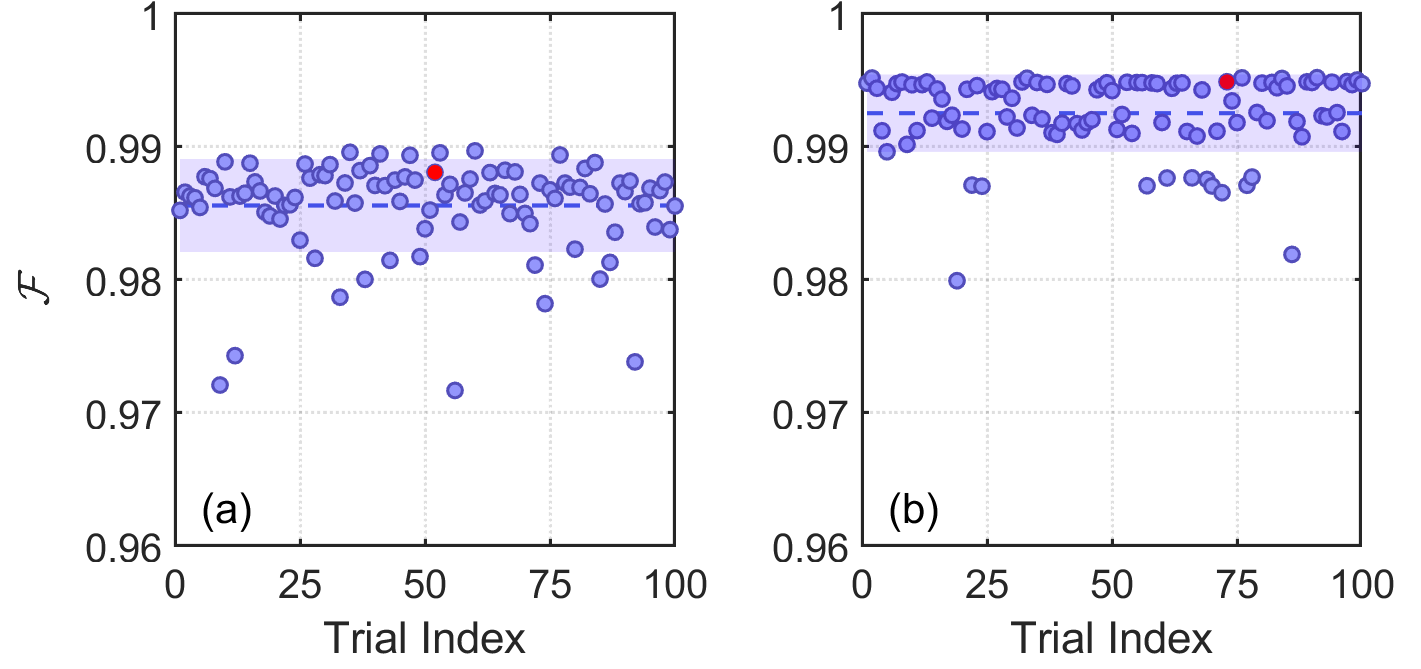}
\caption{\label{Trial} The gate fidelity $F$ with the shortest gate time $T_0=T=0.1\mu$s for (a) the SA gate and (b) the ASA gate, depending on 100 independent optimization runs. Each point represents the realistic gate fidelity of single optimization run. The dashed line and the shaded region denote the average number and the standard deviation, respectively. The \textcolor{black}{red dots} represent the LIM case specified in Table I.}
\end{figure}

In addition, to account for the stochastic nature of optimization process and ensure the reliability of our results, we perform 100 independent optimization runs to evaluate the robustness of optimized results for SA and ASA protocols in the LIM case. The statistical distribution of all optimization outcomes are displayed in Fig. \ref{Trial}. Figure \ref{Trial}(a) reveals that the best fidelity achieved by the SA protocol remains below 0.99, where all optimal solutions are concentrated within a narrow range of $F=0.9856\pm 0.0035$. This fact confirms that the observed bottleneck for the SA strategy is not incidental suggesting that the standard SA protocol fails to sustain the ultimate operation accuracy of state evolution with extreme temporal constraints. In contrast, Fig.\ref{Trial}(b) shows the distribution of fidelity values within 100 optimization trials via the ASA protocol when $T$ is also set to 0.1 $\mu$s. As expected, the ASA protocol can mostly achieve a high-fidelity ($F=0.9925\pm 0.0030$) gate
significantly outperforming the SA protocol with a same short duration.

\section{Error budget for the ancillary drive}


The ancillary drive, although significantly improves the speed of gate operation, would simultaneously arise new experimental imperfections that are not present in a general SA gate.
Consequently, developing an error budget individually due to the ancillary drive
becomes more crucial for revealing the practical feasibility of our protocol \textcolor{black}{\cite{PhysRevResearch.4.033019}}. In this section, we parametrize these imperfections that come from the ancillary drive for the ASA gates in three aspects \textcolor{black}{\cite{PhysRevA.85.042310}}, which are the laser amplitude deviations, the unknown detuning error due to uncertain laser frequency and the laser phase noise leading to the dephasing error.

\begin{figure}
\includegraphics[width=3.45in]{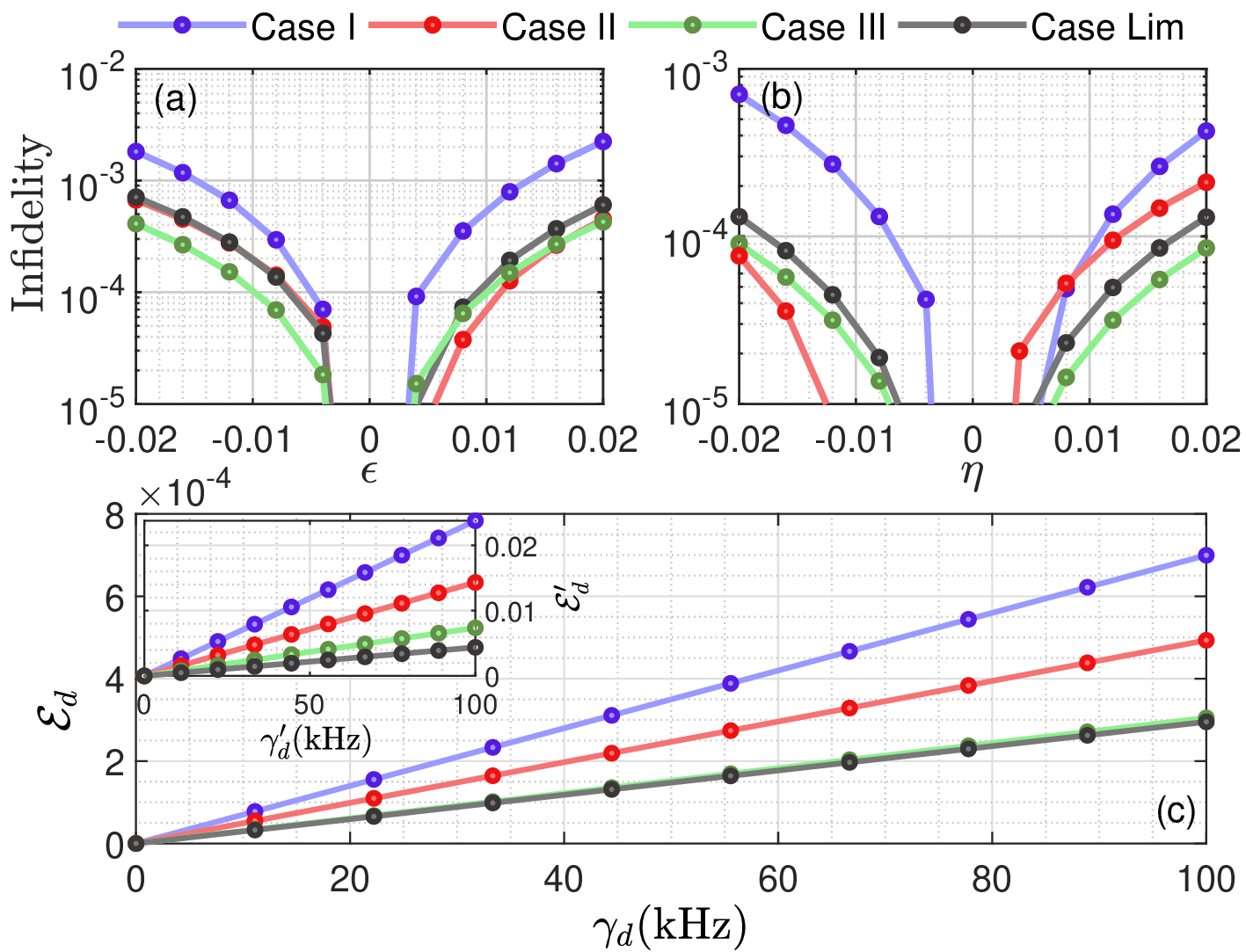}
\caption{\label{error}  The ASA gate errors arising from the ancillary drive, including (a) the amplitude deviation $\epsilon$ in $\Omega_c(t)$ while 
$\eta=0$, (b) the deviation of ancillary detuning denoted as $\eta$ in $\Delta_c$ while $\epsilon=0$, as well as (c) the laser phase noise quantified by the dephasing error with rate $\gamma_d$ between the $|e\rangle$ and $|a\rangle$ states. \textcolor{black}{Inset: another dephasing error (with rate $\gamma_d^\prime$) between the $|e\rangle$ and $|1\rangle$ states which comes from the probe drive $\Omega_1(t)$.} }
\end{figure}

\textit{Infidelities from the amplitude derivations -} Laser amplitude fluctuation is one of the common technical noise sources in neutral-atom systems especially by employing multiple laser fields \textcolor{black}{\cite{PhysRevA.101.043421, PhysRevA.102.042607}}, whose impact has been extensively explored in previous SA gate studies \textcolor{black}{\cite{PhysRevA.109.062610}}. Here, we mainly focus on the amplitude deviations to the optimized auxiliary field $\Omega_c(t)$ by adopting $\Omega_c(t)\to (1+\epsilon)\Omega_c(t)$ where $\epsilon$ treats as unknown but constant during the ASA gate duration. As displayed in Fig. \ref{error}(a) we compare the gate infidelities for four ASA gates (as summarized in Table I),
solely due to the variation of $\epsilon\in[-0.02,0.02]$ while ignoring other errors. We find that, an exponential sensitivity commonly exists leading to an explicit increase as the amplitude deviation $\epsilon$ grows. However, all cases except Case I can achieve the infidelity persistently below $10^{-3}$ even for large values of $|\epsilon|$ up to 0.02, indicating a superior amplitude noise robustness. The worst noise sensitivity in Case I primarily stems from the use of a larger laser Rabi frequency $\Omega_c(t)$ that would amplify the impact of amplitude fluctuations on the population dynamics, thereby exacerbating the phase accumulation and the population leakage \textcolor{black}{\cite{PhysRevA.109.012619}}. In contrast, other schemes employing lower peak Rabi frequencies or smoother pulse envelopes can effectively suppress this amplitude noise-induced infidelity, maintaining the infidelities at an acceptable level of $10^{-4}$ or below.
This result suggests that the laser amplitude deviation due to the ancillary drive in ASA gates brings a negligible impact.

\textit{Infidelities from the detuning derivations -} It should be noted that the laser frequency fluctuations or the ac Stark shift can result in an unknown detuning deviation which makes the actual detuning given by the ancillary drive deviate from the expected value $\Delta_c$, thereby influencing the gate fidelity \textcolor{black}{\cite{PhysRevA.72.022347}}. We analyze the effect of this error through the introduction of a deviation coefficient $\eta$ that modifies the original auxiliary frequency by $\Delta_c\to(1+\eta)\Delta_c$, where $\eta$ is also considered as unknown yet constant incorporating various frequency error sources \textcolor{black}{\cite{PhysRevA.92.022336,PhysRevApplied.17.034015}}.
In Fig. \ref{error}(b) we calculate the gate infidelity as a function of $\eta\in[-0.02,0.02]$ for all ASA gate schemes while $\epsilon=0$. We can identify the best case III that is most robust against the detuning error achieving $<10^{-4}$ for all $\eta$ values.
Notably, as compared to the infidelity caused by amplitude fluctuations (see Fig.\ref{error}(a)), this detuning error is approximately 1 order of magnitude smaller. This improved robustness can be attributed to the fact that $\Delta_c$ preserves a relatively large value instead of a time-varying one, so a small disturbance does not significantly impact the system.


\textit{Laser phase noise -} With the motivation for an ancillary-field error budget established we next proceed by dealing with the laser phase noise \textcolor{black}{\cite{PhysRevA.99.043404}}, modeled as a separately dephasing process with the corresponding the jump operator $L_d=|a\rangle\langle a|-|e\rangle\langle e|$ between $|a\rangle$ and $|e\rangle$. Here we set the dephasing rate as $\gamma_d \leq 100$kHz and satisfy the weak dissipation condition $\gamma_dT\ll 1$ (with $T$ the gate duration) \textcolor{black}{\cite{PhysRevApplied.18.044042}}. This assumption allows us to treat the dephasing effect in a perturbative expansion to the first order of $\gamma_dT$, and the resulting gate error arising from the ancillary laser phase noise can be expressed as an integral of its instantaneous contribution over the total gate duration \textcolor{black}{\cite{56qk-rmsz}}
\begin{equation}
    \mathcal{E}_d\approx\gamma_d \int_0^Tdt \delta\mathcal{E}(t,L_d)
    \label{dee}
\end{equation}
Here, the integrand $\delta\mathcal{E}(t,L_d)$ captures the infidelity strength at each moment, having an explicit form as
\begin{align}
    \delta\mathcal{E}(t,L_d) = & \frac{1}{\kappa}\mathrm{Tr}[L_d^{\dagger}(t)L_d(t)\mathbb{I}_{\mathrm{cmp}}] \nonumber \\
    & - \frac{1}{\kappa(\kappa+1)}\mathrm{Tr}[L_d^{\dagger}(t)\mathbb{I}_{\mathrm{cmp}}L_d(t)\mathbb{I}_{\mathrm{cmp}}] \nonumber \\
    & - \frac{1}{\kappa(\kappa+1)}\mathrm{Tr}[L_d^{\dagger}(t)\mathbb{I}_{\mathrm{cmp}}]\mathrm{Tr}[L_d(t)\mathbb{I}_{\mathrm{cmp}}]
\end{align}
where the space dimension is $\kappa=2^N$ with $N=2$ the qubit number, $\mathbb{I}_{\text{cmp}}$ denotes the projector onto the computational subspace. Here, $L_d(t)=U^{\dagger}(t)L_dU(t)$ is the noise operator in the interaction picture, with $U(t)=\text{exp}[-i\int_0^t H(t^{\prime})dt^{\prime}]$ being the unitary time-evolution operator of the system. In this case the dephasing error $\mathcal{E}_d$ caused by the laser phase noise is directly proportional to the rate $\gamma_d$.

Figure \ref{error}(c) shows the behavior of $\mathcal{E}_d$ as a function of the dephasing rate $\gamma_d$ over a wide range of $[0,100]$ kHz. The results indicate that our ASA gates are notably robust to the laser phase noise from auxiliary laser field. The gate error is merely on the order of $10^{-4}$ for all four cases, even under a very large dephasing rate up to $\gamma_d=100$ kHz. Specifically, Cases III and LIM benefiting from much shorter gate times (0.1709 $\mu$s and 0.1 $\mu$s), achieve a competitive value of $\mathcal{E}_d\approx 3\times 10^{-4}$ for $\gamma_d=100$ kHz.
This observation is consistent with the small differences between the time spent $T_e$ and $T_a$ on states $|e\rangle$ and $|a\rangle$ during the gate duration (see Table I).
\textcolor{black}{Analogously, we also evaluate another dephasing error $\mathcal{E}_{d}^\prime$ arising from the phase noise of the probe laser $\Omega_1(t)$. This is modeled by replacing the jump operator with $L_{d}^\prime=|1\rangle\langle 1| - |e\rangle\langle e|$, quantified by the same rate range $\gamma_d^\prime \in [0,100]$ kHz (see inset of Fig.\ref{error}c). As expected, this error is more pronounced because of the large population difference between $|1\rangle$ and $|e\rangle$ states. For shorter-duration gates (Cases III and LIM), $\mathcal{E}_{d}^\prime$ remains at the level of $10^{-3}$ (the same error scale as in a normal two-photon model \cite{PhysRevApplied.18.044042}), and exceeds the $|a\rangle-|e\rangle$ dephasing error by one order of magnitude.}

\section{Scheme Feasibility}

Finally, it is worthwhile to point out that, the ancillary drive introduced in our scheme is able to provide an effective control for the gate acceleration, and meanwhile it will not bring significant gate infidelity due to the presence of state $|a\rangle$, which is confirmed by our error budget above. For a practical Rydberg excitation, we consider parameters proposed for $^{87}$Rb qubits using a native two-photon transition mediated by \textcolor{black}{$|e\rangle=|5P_{3/2},F=3,m_F=-1\rangle$} to the Rydberg state $|r\rangle=|70S_{1/2},J=1/2,m_J=-1/2 \rangle$ \textcolor{black}{\cite{ReetzLamour2008}}. The initial atomic qubits are prepared in $|0\rangle=|5S_{1/2},F=1,m_F=0 \rangle$, $|1\rangle=|5S_{1/2},F=2,m_F=0 \rangle$ forming the computational basis states, and $|a\rangle=|5S_{1/2},F=2,m_F=-1\rangle$ is chosen as the ancillary state. This choice of $|a\rangle$ not only ensures its stability avoiding the decay error but also arises a direct single-photon transition between $|a\rangle$ and $|e\rangle$ implementing the effective acceleration control.
\textcolor{black}{In addition, under realistic conditions we quantify the population escaping outside the computational subspace, e.g. due to the presence of a leakage state $|k\rangle=|5S_{1/2},F=2,m_F=-2\rangle $, that also contributes the gate infidelity. To simulate this leakage error we define the jump operator $L_{k} = |k\rangle\langle e| $ and take the leakage rate to be {$\gamma_k = b \gamma_e$} with $b=0.0667$
the branching ratio \cite{SIBALIC2017319}. As demonstrated in Ref.\cite{PhysRevA.97.032306} the leakage error can be estimated as }
\begin{equation}
    \textcolor{black}{\mathcal{E}_k \approx \gamma_k \int_0^T dt \delta \mathcal{E}(t,L_{k})} 
\end{equation}
\textcolor{black}{with} 
\begin{eqnarray}
    \color{black}\delta \mathcal{E}(t,L_{k}) &\color{black}=& \color{black}\frac{1}{\kappa}\text{Tr}[L_{k}^\dagger(t)L_{k} \mathbb{I}_{\mathrm{cmp}}] \nonumber \\
    &\color{black}-&\color{black}\frac{1}{\kappa}\text{Tr}[L_{k}^\dagger(t)\mathbb{I}_{\mathrm{cmp}}L_{k}(t)\mathbb{I}_{\mathrm{cmp}}]
\end{eqnarray}
\textcolor{black}{Via resolving the noise operator $L_k(t) = U^\dagger L_k U(t)$ in the complete Hilbert space (including $|k\rangle$) we find that this error source only contributes a minor gate infidelity of $\mathcal{E}_k\approx (3.956,4.016)\times 10^{-4}$ for Cases III and LIM.}

As an illustrative example for the practical implementation of fast and high-fidelity gates, we assume
the blockade strength is $V=2\pi\times 300$ MHz corresponding to the vdWs coefficient $C_6 = 2\pi \times 863 $ GHz$\cdot\mu m^6$ with a distance of $r\approx3.77$ $\mu$m  \textcolor{black}{\cite{SIBALIC2017319}}. Other specific parameters are $\Omega_2/2\pi=50$ MHz, $\Omega_1(t)/2\pi\leq 200$ MHz, $\Delta/2\pi = 1.0$ GHz leading to 
the native two-photon Rabi frequency given by $\frac{\Omega_1\Omega_2}{2\Delta}\leq 2\pi\times5.0$ MHz. Note that we can extract an enhanced Rabi frequency as
\begin{equation}
    \Omega_{eff} = \frac{\Omega_1\Omega_2}{2\Delta - \frac{\Omega_c^2}{2(\Delta+\Delta_c)}}\leq 2\pi\times 5.46 \text{ MHz}
\end{equation}
by taking $\Omega_c(t)/2\pi\leq 150$ MHz and $\Delta_c/2\pi  = -0.9337$ GHz in Case III because of the ancillary drive. 
Therefore, we provide a conservative lower bound for the overall gate fidelity in Case III that leads to $F\approx 0.9973$ within the gate time of 0.1709 $\mu$s
by taking account of $(\epsilon,\eta,\gamma_d)$ =$(0.02,0.02,50$ kHz), demonstrating that the error contribution from the ancillary field has been decreased to be negligible $\sim \textcolor{black}{6.62}\times 10^{-4}$. Even for the LIM case, 
by considering all ancillary-field errors the practical gate fidelity will decrease to $F\approx 0.9943$ suggesting the same level of errors $\sim \textcolor{black}{8.83}\times 10^{-4}$.

Besides errors from the ancillary drive, our proposed gates will be affected by error sources inherent to the native two-photon transition (i.e. general SA gate), such as the probe dephasing error $\mathcal{E}_d^\prime$ serving as a significant error source as well as the state leakage $\mathcal{E}_k$. 
\textcolor{black}{After considering $\mathcal{E}_d^\prime$ (estimated by $\gamma_d^\prime = 20$ kHz) and $\mathcal{E}_k$
we obtain a more conservative estimation for the predicted gate fidelity which is $F> 0.9954$ for the best Case III}. Therefore suppressing these imperfections in the design of ancillary fields will be the subject of future investigation.


\section{Conclusion and Outlook} 

We have developed an optimal-control method with acceleration strategy for implementing the fast two-qubit entangling gates in neutral Rydberg atom platforms. Such acceleration relies on an ancillary-field drive with its amplitude and detuning both optimized giving rise to an enhanced two-photon Rabi coupling strength between the ground and Rydberg states \textcolor{black}{\cite{Chang2023}}. We identify that these pulses can strongly shorten the gate execution time by more than 30$\%$ as compared to the conventional two-photon protocols, while nearly completely eliminating the extra technical errors such as the amplitude deviations, the detuning deviations and the laser phase noise coming from the ancillary laser. Thanks to the shortened operation time, our gates can maintain a high Bell state fidelity, for example, \textcolor{black}{above $\sim 0.9973$} with a 170.9-nanosecond duration even in the presence of all decay errors and ancillary-field imperfections. \textcolor{black}{After
taking account of other relevant error sources a more conservative lower bound for the ASA gate fidelity remains 0.9954}.
This improvement has significantly outperformed the existing two-photon protocols in both speed and error robustness, promising for a near-term implementation in a realistic setup \textcolor{black}{\cite{Evered2023}}.

Notably, our approach can relax the conventional requirement for high-power lasers which has traditionally been deemed essential for achieving fast quantum gates \textcolor{black}{\cite{Tang2022}}. By leveraging optimal control in an auxiliary-driven setting, we are able to realize high-fidelity quantum operations at submicrosecond timescales, while still operating under moderate laser power conditions. Our findings highlight the potential of integrating optimal control strategy with ancillary drive to facilitate large-scale, high-speed quantum computing in neutral-atom platforms \textcolor{black}{\cite{Ebadi2022,PhysRevLett.131.170601}}, opening up new avenues toward more efficient quantum gate synthesis with reduced reliance on extreme experimental parameters.


\begin{acknowledgments}

We acknowledge financial support from the National Natural Science Foundation of China under Grants Nos. 12174106, 11474094 and 11104076, the Natural Science Foundation of Chongqing under Grant No. CSTB2024NSCQ-MSX1117, the Shanghai Science and Technology Innovation Project under Grant No. 24LZ1400600, and the Science and Technology Commission of Shanghai Municipality under Grant No.18ZR1412800.

\end{acknowledgments}

\appendix

\textcolor{black}{\section{Improving the SA gates for LIM case}}


\begin{figure}
\includegraphics[width=3.3in]{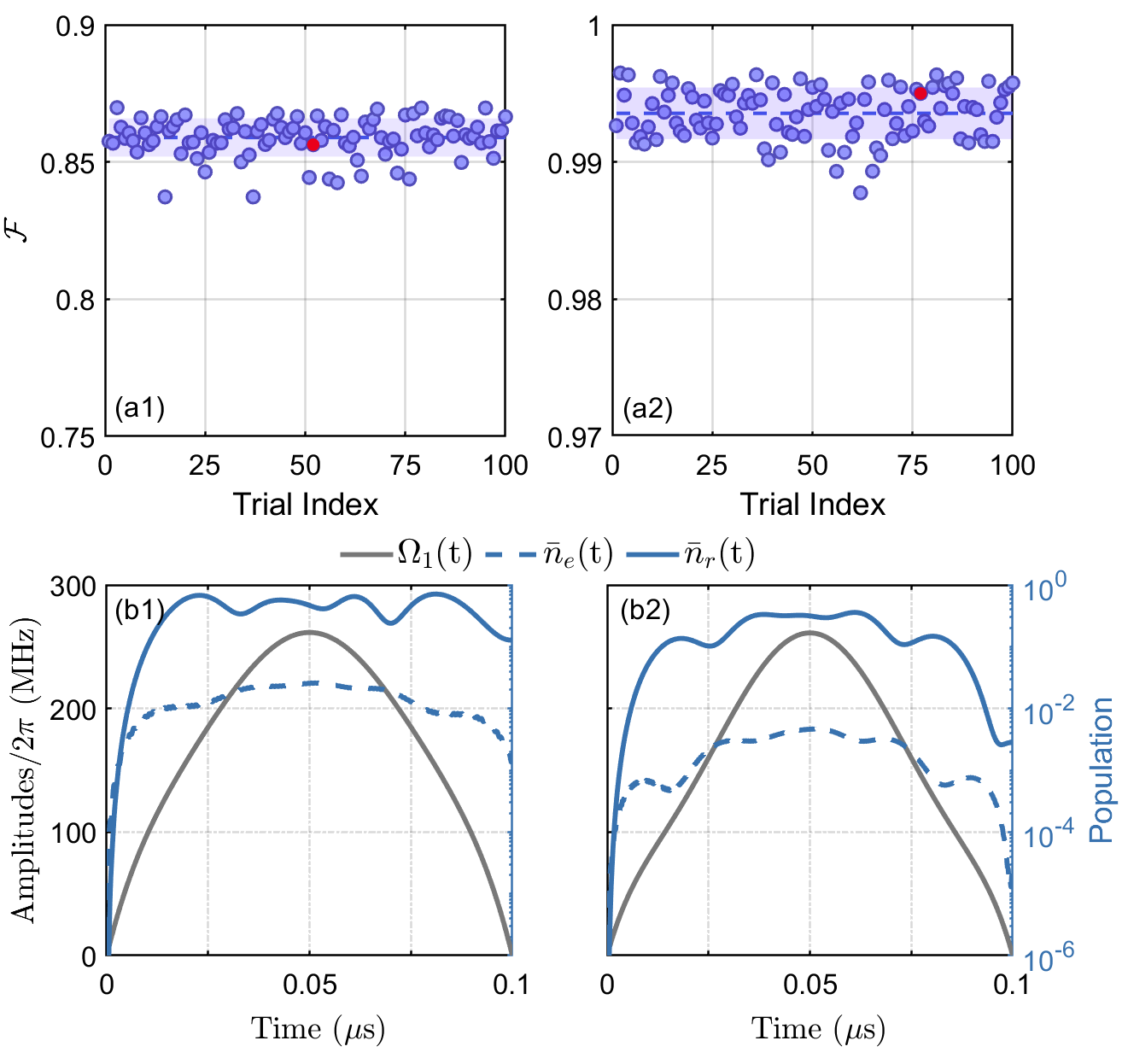}
\caption{\label{appendix} \textcolor{black}{The gate fidelity $F$ over 100 independent optimization runs for two additional SA gates at the LIM case ($T_0 = 0.1 \mu s$), where (a1) the maximum of $\Omega_1(t)/2\pi$ is optimized within the range of $[200,300]$ MHz and (a2) alongside with the increase of $\Delta/2\pi$ to 2.0 GHz. The dashed line and the shaded region respectively denote the average number and the standard deviation.
(b1-b2) Optimized pulse amplitudes of $\Omega_1(t)$ and the resolved time-dependent population $(\bar{n}_e(t),\bar{n}_r(t))$ on the intermediate and Rydberg states, as specifically highlighted by red dots in (a1-a2). Here, the optimized parameters are ($\delta_{opt}/2\pi$, $\beta^{1}_{1\sim4}$) =(-9.219, 167.71, 167.70, 210.47, 223.60)MHz and (-10.00, 150.49, 59.879, 274.54, 216.76) MHz, respectively.}}
\end{figure}

\textcolor{black}{This appendix will provide a detailed analysis of the standard two-photon SA gates which implement in the LIM case with $T_0=0.1$ $\mu$s, for a fastest and high-fidelity SA gate. In the main text we have shown a lower fidelity value $F=0.9872$ in this case because of the available laser intensity set by $\Omega_1(t)/2\pi\lesssim 200$ MHz. Here, we explore two intuitive strategies for improving the SA gate performance and reveal their inherent bottlenecks as compared with the ASA gate protocols.}

\textcolor{black}{\textit{Strategy I: Increasing the upper bound of $\Omega_1(t)$ -} To make a fair comparison with the ASA protocol we first individually increase the maximum of $\Omega_1(t)$ to $2\pi\times[200,300]$ MHz for optimization search which benefits from an enhanced effective Rabi frequency for the ground-Rydberg transition due to $\Omega_{eff}\approx \frac{\Omega_1\Omega_2}{2\Delta}$, leading to an improved gate performance in principle. However, we find that simply increasing $\Omega_1$ is difficult to improve the gate fidelity at the LIM case. As displayed in Fig. \ref{appendix}(a1) which presents the fidelity outcomes depending on 100 independent optimization runs,run counterintuitively, the average fidelity value is only {$F = 0.8589 \pm 0.0070$}, even much smaller than that of the LIM case (Table \ref{tab:laser_coefficients}) where a smaller $\Omega_1$ is used. The physical interpretation of this degradation originates from the breakdown of the large-detuning condition for state $|e\rangle$. See Fig.\ref{appendix}(b1) a higher $\Omega_1$ gives rise to a substantial increase of the time integration ${T_e = \int_0^{T_0} \bar{n}_e(t)dt \approx 1.419}$ ns in the lossy state $|e\rangle$, leading to the intermediate decay error {$\gamma_eT_e\sim 8.896\times 10^{-3}$} dominating the gate infidelity. Therefore, simply using higher-intensity pulses to obtain fast SA gates is unfeasible here.}

\textcolor{black}{\textit{Strategy II: Along with the increase of intermediate detuning $\Delta$ -} To suppress the intermediate decay error we additionally use a larger $\Delta$ while keeping the maximum of $\Omega_1(t)$ within $2\pi\times[200,300]$ MHz, in order to improve the SA gates in LIM case. We still carry out 100 independent optimizations with the same set of parameters at $T_0=0.1\mu$s. As plotted by Fig. \ref{appendix}(a2), the statistical distribution of optimization outcomes shows that the average fidelity number has grown to $F = 0.9935 \pm 0.0018$, having reached a comparable level with the LIM case of ASA gates (see Fig.\ref{Trial}(b)). That mainly arises from the significant decrease of the time-spent $T_e\approx 0.205$ ns on state $|e\rangle$ due to the far off-resonance condition, which in turn leads to a much smaller decay error ${\gamma_e T_e\sim 1.290\times 10^{-3}}$.
 Nevertheless, we also find that this improvement for a fastest SA gate requires more demanding cost i.e. a much larger laser amplitude $\Omega_1(t)$ with its peak value more than $2\pi\times 250$ MHz, which lowers the feasibility in a practically experimental setup.}




\textcolor{black}{Based on the comparison of two strategies above, we highlights that the standard SA gate scheme possesses an inherent trade-off in the LIM case, i.e. accelerating the gate speed via simply increasing the laser drive inevitably exacerbates the intrinsic decay errors, while suppressing these errors via a larger detuning will impose impractical experimental demands. In contrast, our new ASA protocol relying on a moderate ancillary drive as an independent control knob, provides an alternative way to accelerate the gate without the need for even stronger laser pulses. In Sec. IV we have presented the realization of the fastest ASA gates at $T=0.1$ $\mu$s with a high fidelity of $F>0.995$ (with decay error only), in which all laser pulses are moderate and experimentally accessible outperforming the standard SA gates.}

\nocite{*}


%


\begin{thebibliography}{61}%
\makeatletter
\providecommand \@ifxundefined [1]{%
 \@ifx{#1\undefined}
}%
\providecommand \@ifnum [1]{%
 \ifnum #1\expandafter \@firstoftwo
 \else \expandafter \@secondoftwo
 \fi
}%
\providecommand \@ifx [1]{%
 \ifx #1\expandafter \@firstoftwo
 \else \expandafter \@secondoftwo
 \fi
}%
\providecommand \natexlab [1]{#1}%
\providecommand \enquote  [1]{``#1''}%
\providecommand \bibnamefont  [1]{#1}%
\providecommand \bibfnamefont [1]{#1}%
\providecommand \citenamefont [1]{#1}%
\providecommand \href@noop [0]{\@secondoftwo}%
\providecommand \href [0]{\begingroup \@sanitize@url \@href}%
\providecommand \@href[1]{\@@startlink{#1}\@@href}%
\providecommand \@@href[1]{\endgroup#1\@@endlink}%
\providecommand \@sanitize@url [0]{\catcode `\\12\catcode `\$12\catcode `\&12\catcode `\#12\catcode `\^12\catcode `\_12\catcode `\%12\relax}%
\providecommand \@@startlink[1]{}%
\providecommand \@@endlink[0]{}%
\providecommand \url  [0]{\begingroup\@sanitize@url \@url }%
\providecommand \@url [1]{\endgroup\@href {#1}{\urlprefix }}%
\providecommand \urlprefix  [0]{URL }%
\providecommand \Eprint [0]{\href }%
\providecommand \doibase [0]{https://doi.org/}%
\providecommand \selectlanguage [0]{\@gobble}%
\providecommand \bibinfo  [0]{\@secondoftwo}%
\providecommand \bibfield  [0]{\@secondoftwo}%
\providecommand \translation [1]{[#1]}%
\providecommand \BibitemOpen [0]{}%
\providecommand \bibitemStop [0]{}%
\providecommand \bibitemNoStop [0]{.\EOS\space}%
\providecommand \EOS [0]{\spacefactor3000\relax}%
\providecommand \BibitemShut  [1]{\csname bibitem#1\endcsname}%
\let\auto@bib@innerbib\@empty
\bibitem [{\citenamefont {Cong}\ \emph {et~al.}(2022)\citenamefont {Cong}, \citenamefont {Levine}, \citenamefont {Keesling}, \citenamefont {Bluvstein}, \citenamefont {Wang},\ and\ \citenamefont {Lukin}}]{robust_gate_Cong2022}%
  \BibitemOpen
  \bibfield  {author} {\bibinfo {author} {\bibfnamefont {I.}~\bibnamefont {Cong}}, \bibinfo {author} {\bibfnamefont {H.}~\bibnamefont {Levine}}, \bibinfo {author} {\bibfnamefont {A.}~\bibnamefont {Keesling}}, \bibinfo {author} {\bibfnamefont {D.}~\bibnamefont {Bluvstein}}, \bibinfo {author} {\bibfnamefont {S.-T.}\ \bibnamefont {Wang}},\ and\ \bibinfo {author} {\bibfnamefont {M.~D.}\ \bibnamefont {Lukin}},\ }\bibfield  {title} {\bibinfo {title} {Hardware-efficient, fault-tolerant quantum computation with rydberg atoms},\ }\href {https://doi.org/10.1103/PhysRevX.12.021049} {\bibfield  {journal} {\bibinfo  {journal} {Phys. Rev. X}\ }\textbf {\bibinfo {volume} {12}},\ \bibinfo {pages} {021049} (\bibinfo {year} {2022})}\BibitemShut {NoStop}%
\bibitem [{\citenamefont {Mitra}\ \emph {et~al.}(2020)\citenamefont {Mitra}, \citenamefont {Martin}, \citenamefont {Biedermann}, \citenamefont {Marino}, \citenamefont {Poggi},\ and\ \citenamefont {Deutsch}}]{robust_gate_Mitra2020}%
  \BibitemOpen
  \bibfield  {author} {\bibinfo {author} {\bibfnamefont {A.}~\bibnamefont {Mitra}}, \bibinfo {author} {\bibfnamefont {M.~J.}\ \bibnamefont {Martin}}, \bibinfo {author} {\bibfnamefont {G.~W.}\ \bibnamefont {Biedermann}}, \bibinfo {author} {\bibfnamefont {A.~M.}\ \bibnamefont {Marino}}, \bibinfo {author} {\bibfnamefont {P.~M.}\ \bibnamefont {Poggi}},\ and\ \bibinfo {author} {\bibfnamefont {I.~H.}\ \bibnamefont {Deutsch}},\ }\bibfield  {title} {\bibinfo {title} {Robust m\o{}lmer-s\o{}rensen gate for neutral atoms using rapid adiabatic rydberg dressing},\ }\href {https://doi.org/10.1103/PhysRevA.101.030301} {\bibfield  {journal} {\bibinfo  {journal} {Phys. Rev. A}\ }\textbf {\bibinfo {volume} {101}},\ \bibinfo {pages} {030301} (\bibinfo {year} {2020})}\BibitemShut {NoStop}%
\bibitem [{\citenamefont {Mohan}\ \emph {et~al.}(2023)\citenamefont {Mohan}, \citenamefont {de~Keijzer},\ and\ \citenamefont {Kokkelmans}}]{robust_gate_Mohan2023}%
  \BibitemOpen
  \bibfield  {author} {\bibinfo {author} {\bibfnamefont {M.}~\bibnamefont {Mohan}}, \bibinfo {author} {\bibfnamefont {R.}~\bibnamefont {de~Keijzer}},\ and\ \bibinfo {author} {\bibfnamefont {S.}~\bibnamefont {Kokkelmans}},\ }\bibfield  {title} {\bibinfo {title} {Robust control and optimal rydberg states for neutral atom two-qubit gates},\ }\href {https://doi.org/10.1103/PhysRevResearch.5.033052} {\bibfield  {journal} {\bibinfo  {journal} {Phys. Rev. Res.}\ }\textbf {\bibinfo {volume} {5}},\ \bibinfo {pages} {033052} (\bibinfo {year} {2023})}\BibitemShut {NoStop}%
\bibitem [{\citenamefont {Fromonteil}\ \emph {et~al.}(2023)\citenamefont {Fromonteil}, \citenamefont {Bluvstein},\ and\ \citenamefont {Pichler}}]{robust_gate_Fromonteil2023}%
  \BibitemOpen
  \bibfield  {author} {\bibinfo {author} {\bibfnamefont {C.}~\bibnamefont {Fromonteil}}, \bibinfo {author} {\bibfnamefont {D.}~\bibnamefont {Bluvstein}},\ and\ \bibinfo {author} {\bibfnamefont {H.}~\bibnamefont {Pichler}},\ }\bibfield  {title} {\bibinfo {title} {Protocols for rydberg entangling gates featuring robustness against quasistatic errors},\ }\href {https://doi.org/10.1103/PRXQuantum.4.020335} {\bibfield  {journal} {\bibinfo  {journal} {PRX Quantum}\ }\textbf {\bibinfo {volume} {4}},\ \bibinfo {pages} {020335} (\bibinfo {year} {2023})}\BibitemShut {NoStop}%
\bibitem [{\citenamefont {Shao}\ \emph {et~al.}(2024)\citenamefont {Shao}, \citenamefont {Su}, \citenamefont {Li}, \citenamefont {Nath}, \citenamefont {Wu},\ and\ \citenamefont {Li}}]{review_Shao2024}%
  \BibitemOpen
  \bibfield  {author} {\bibinfo {author} {\bibfnamefont {X.-Q.}\ \bibnamefont {Shao}}, \bibinfo {author} {\bibfnamefont {S.-L.}\ \bibnamefont {Su}}, \bibinfo {author} {\bibfnamefont {L.}~\bibnamefont {Li}}, \bibinfo {author} {\bibfnamefont {R.}~\bibnamefont {Nath}}, \bibinfo {author} {\bibfnamefont {J.-H.}\ \bibnamefont {Wu}},\ and\ \bibinfo {author} {\bibfnamefont {W.}~\bibnamefont {Li}},\ }\bibfield  {title} {\bibinfo {title} {Rydberg superatoms: An artificial quantum system for quantum information processing and quantum optics},\ }\bibfield  {journal} {\bibinfo  {journal} {Applied Physics Reviews}\ }\textbf {\bibinfo {volume} {11}},\ \href {https://doi.org/10.1063/5.0211071} {10.1063/5.0211071} (\bibinfo {year} {2024})\BibitemShut {NoStop}%
\bibitem [{\citenamefont {Shi}(2022)}]{review_Shi2022}%
  \BibitemOpen
  \bibfield  {author} {\bibinfo {author} {\bibfnamefont {X.-F.}\ \bibnamefont {Shi}},\ }\bibfield  {title} {\bibinfo {title} {Quantum logic and entanglement by neutral rydberg atoms: methods and fidelity},\ }\href {https://doi.org/10.1088/2058-9565/ac18b8} {\bibfield  {journal} {\bibinfo  {journal} {Quantum Science and Technology}\ }\textbf {\bibinfo {volume} {7}},\ \bibinfo {pages} {023002} (\bibinfo {year} {2022})}\BibitemShut {NoStop}%
\bibitem [{\citenamefont {Pagano}\ \emph {et~al.}(2022)\citenamefont {Pagano}, \citenamefont {Weber}, \citenamefont {Jaschke}, \citenamefont {Pfau}, \citenamefont {Meinert}, \citenamefont {Montangero},\ and\ \citenamefont {B\"uchler}}]{gate_error_Pagano2022}%
  \BibitemOpen
  \bibfield  {author} {\bibinfo {author} {\bibfnamefont {A.}~\bibnamefont {Pagano}}, \bibinfo {author} {\bibfnamefont {S.}~\bibnamefont {Weber}}, \bibinfo {author} {\bibfnamefont {D.}~\bibnamefont {Jaschke}}, \bibinfo {author} {\bibfnamefont {T.}~\bibnamefont {Pfau}}, \bibinfo {author} {\bibfnamefont {F.}~\bibnamefont {Meinert}}, \bibinfo {author} {\bibfnamefont {S.}~\bibnamefont {Montangero}},\ and\ \bibinfo {author} {\bibfnamefont {H.~P.}\ \bibnamefont {B\"uchler}},\ }\bibfield  {title} {\bibinfo {title} {Error budgeting for a controlled-phase gate with strontium-88 rydberg atoms},\ }\href {https://doi.org/10.1103/PhysRevResearch.4.033019} {\bibfield  {journal} {\bibinfo  {journal} {Phys. Rev. Res.}\ }\textbf {\bibinfo {volume} {4}},\ \bibinfo {pages} {033019} (\bibinfo {year} {2022})}\BibitemShut {NoStop}%
\bibitem [{\citenamefont {Guo}\ \emph {et~al.}(2020)\citenamefont {Guo}, \citenamefont {Yan}, \citenamefont {Zhang}, \citenamefont {Su},\ and\ \citenamefont {Li}}]{errors_Guo2020}%
  \BibitemOpen
  \bibfield  {author} {\bibinfo {author} {\bibfnamefont {C.-Y.}\ \bibnamefont {Guo}}, \bibinfo {author} {\bibfnamefont {L.-L.}\ \bibnamefont {Yan}}, \bibinfo {author} {\bibfnamefont {S.}~\bibnamefont {Zhang}}, \bibinfo {author} {\bibfnamefont {S.-L.}\ \bibnamefont {Su}},\ and\ \bibinfo {author} {\bibfnamefont {W.}~\bibnamefont {Li}},\ }\bibfield  {title} {\bibinfo {title} {Optimized geometric quantum computation with a mesoscopic ensemble of rydberg atoms},\ }\href {https://doi.org/10.1103/PhysRevA.102.042607} {\bibfield  {journal} {\bibinfo  {journal} {Phys. Rev. A}\ }\textbf {\bibinfo {volume} {102}},\ \bibinfo {pages} {042607} (\bibinfo {year} {2020})}\BibitemShut {NoStop}%
\bibitem [{\citenamefont {Zhang}\ \emph {et~al.}(2012)\citenamefont {Zhang}, \citenamefont {Gill}, \citenamefont {Isenhower}, \citenamefont {Walker},\ and\ \citenamefont {Saffman}}]{error_stem_Zhang2012}%
  \BibitemOpen
  \bibfield  {author} {\bibinfo {author} {\bibfnamefont {X.~L.}\ \bibnamefont {Zhang}}, \bibinfo {author} {\bibfnamefont {A.~T.}\ \bibnamefont {Gill}}, \bibinfo {author} {\bibfnamefont {L.}~\bibnamefont {Isenhower}}, \bibinfo {author} {\bibfnamefont {T.~G.}\ \bibnamefont {Walker}},\ and\ \bibinfo {author} {\bibfnamefont {M.}~\bibnamefont {Saffman}},\ }\bibfield  {title} {\bibinfo {title} {Fidelity of a rydberg-blockade quantum gate from simulated quantum process tomography},\ }\href {https://doi.org/10.1103/PhysRevA.85.042310} {\bibfield  {journal} {\bibinfo  {journal} {Phys. Rev. A}\ }\textbf {\bibinfo {volume} {85}},\ \bibinfo {pages} {042310} (\bibinfo {year} {2012})}\BibitemShut {NoStop}%
\bibitem [{\citenamefont {Levine}\ \emph {et~al.}(2019)\citenamefont {Levine}, \citenamefont {Keesling}, \citenamefont {Semeghini}, \citenamefont {Omran}, \citenamefont {Wang}, \citenamefont {Ebadi}, \citenamefont {Bernien}, \citenamefont {Greiner}, \citenamefont {Vuleti\ifmmode~\acute{c}\else \'{c}\fi{}}, \citenamefont {Pichler},\ and\ \citenamefont {Lukin}}]{Modulated_Pulses_Fidelity_Levine2019}%
  \BibitemOpen
  \bibfield  {author} {\bibinfo {author} {\bibfnamefont {H.}~\bibnamefont {Levine}}, \bibinfo {author} {\bibfnamefont {A.}~\bibnamefont {Keesling}}, \bibinfo {author} {\bibfnamefont {G.}~\bibnamefont {Semeghini}}, \bibinfo {author} {\bibfnamefont {A.}~\bibnamefont {Omran}}, \bibinfo {author} {\bibfnamefont {T.~T.}\ \bibnamefont {Wang}}, \bibinfo {author} {\bibfnamefont {S.}~\bibnamefont {Ebadi}}, \bibinfo {author} {\bibfnamefont {H.}~\bibnamefont {Bernien}}, \bibinfo {author} {\bibfnamefont {M.}~\bibnamefont {Greiner}}, \bibinfo {author} {\bibfnamefont {V.}~\bibnamefont {Vuleti\ifmmode~\acute{c}\else \'{c}\fi{}}}, \bibinfo {author} {\bibfnamefont {H.}~\bibnamefont {Pichler}},\ and\ \bibinfo {author} {\bibfnamefont {M.~D.}\ \bibnamefont {Lukin}},\ }\bibfield  {title} {\bibinfo {title} {Parallel implementation of high-fidelity multiqubit gates with neutral atoms},\ }\href {https://doi.org/10.1103/PhysRevLett.123.170503} {\bibfield  {journal} {\bibinfo  {journal} {Phys. Rev. Lett.}\ }\textbf {\bibinfo {volume}
  {123}},\ \bibinfo {pages} {170503} (\bibinfo {year} {2019})}\BibitemShut {NoStop}%
\bibitem [{\citenamefont {Fu}\ \emph {et~al.}(2022)\citenamefont {Fu}, \citenamefont {Xu}, \citenamefont {Sun}, \citenamefont {Liu}, \citenamefont {He}, \citenamefont {Li}, \citenamefont {Liu}, \citenamefont {Li}, \citenamefont {Wang}, \citenamefont {Liu},\ and\ \citenamefont {Zhan}}]{Modulated_Pulses_Fidelity_Fu2022}%
  \BibitemOpen
  \bibfield  {author} {\bibinfo {author} {\bibfnamefont {Z.}~\bibnamefont {Fu}}, \bibinfo {author} {\bibfnamefont {P.}~\bibnamefont {Xu}}, \bibinfo {author} {\bibfnamefont {Y.}~\bibnamefont {Sun}}, \bibinfo {author} {\bibfnamefont {Y.-Y.}\ \bibnamefont {Liu}}, \bibinfo {author} {\bibfnamefont {X.-D.}\ \bibnamefont {He}}, \bibinfo {author} {\bibfnamefont {X.}~\bibnamefont {Li}}, \bibinfo {author} {\bibfnamefont {M.}~\bibnamefont {Liu}}, \bibinfo {author} {\bibfnamefont {R.-B.}\ \bibnamefont {Li}}, \bibinfo {author} {\bibfnamefont {J.}~\bibnamefont {Wang}}, \bibinfo {author} {\bibfnamefont {L.}~\bibnamefont {Liu}},\ and\ \bibinfo {author} {\bibfnamefont {M.-S.}\ \bibnamefont {Zhan}},\ }\bibfield  {title} {\bibinfo {title} {High-fidelity entanglement of neutral atoms via a rydberg-mediated single-modulated-pulse controlled-phase gate},\ }\href {https://doi.org/10.1103/PhysRevA.105.042430} {\bibfield  {journal} {\bibinfo  {journal} {Phys. Rev. A}\ }\textbf {\bibinfo {volume} {105}},\ \bibinfo {pages} {042430}
  (\bibinfo {year} {2022})}\BibitemShut {NoStop}%
\bibitem [{\citenamefont {Evered}\ \emph {et~al.}(2023)\citenamefont {Evered}, \citenamefont {Bluvstein}, \citenamefont {Kalinowski}, \citenamefont {Ebadi}, \citenamefont {Manovitz}, \citenamefont {Zhou}, \citenamefont {Li}, \citenamefont {Geim}, \citenamefont {Wang}, \citenamefont {Maskara}, \citenamefont {Levine}, \citenamefont {Semeghini}, \citenamefont {Greiner}, \citenamefont {Vuletić},\ and\ \citenamefont {Lukin}}]{Nature_Evered2023}%
  \BibitemOpen
  \bibfield  {author} {\bibinfo {author} {\bibfnamefont {S.~J.}\ \bibnamefont {Evered}}, \bibinfo {author} {\bibfnamefont {D.}~\bibnamefont {Bluvstein}}, \bibinfo {author} {\bibfnamefont {M.}~\bibnamefont {Kalinowski}}, \bibinfo {author} {\bibfnamefont {S.}~\bibnamefont {Ebadi}}, \bibinfo {author} {\bibfnamefont {T.}~\bibnamefont {Manovitz}}, \bibinfo {author} {\bibfnamefont {H.}~\bibnamefont {Zhou}}, \bibinfo {author} {\bibfnamefont {S.~H.}\ \bibnamefont {Li}}, \bibinfo {author} {\bibfnamefont {A.~A.}\ \bibnamefont {Geim}}, \bibinfo {author} {\bibfnamefont {T.~T.}\ \bibnamefont {Wang}}, \bibinfo {author} {\bibfnamefont {N.}~\bibnamefont {Maskara}}, \bibinfo {author} {\bibfnamefont {H.}~\bibnamefont {Levine}}, \bibinfo {author} {\bibfnamefont {G.}~\bibnamefont {Semeghini}}, \bibinfo {author} {\bibfnamefont {M.}~\bibnamefont {Greiner}}, \bibinfo {author} {\bibfnamefont {V.}~\bibnamefont {Vuletić}},\ and\ \bibinfo {author} {\bibfnamefont {M.~D.}\ \bibnamefont {Lukin}},\ }\bibfield  {title} {\bibinfo {title}
  {High-fidelity parallel entangling gates on a neutral-atom quantum computer},\ }\href {https://doi.org/10.1038/s41586-023-06481-y} {\bibfield  {journal} {\bibinfo  {journal} {Nature}\ }\textbf {\bibinfo {volume} {622}},\ \bibinfo {pages} {268–272} (\bibinfo {year} {2023})}\BibitemShut {NoStop}%
\bibitem [{\citenamefont {Daems}\ \emph {et~al.}(2013)\citenamefont {Daems}, \citenamefont {Ruschhaupt}, \citenamefont {Sugny},\ and\ \citenamefont {Gu\'erin}}]{Modulated_Pulses_Robust_Daems2013}%
  \BibitemOpen
  \bibfield  {author} {\bibinfo {author} {\bibfnamefont {D.}~\bibnamefont {Daems}}, \bibinfo {author} {\bibfnamefont {A.}~\bibnamefont {Ruschhaupt}}, \bibinfo {author} {\bibfnamefont {D.}~\bibnamefont {Sugny}},\ and\ \bibinfo {author} {\bibfnamefont {S.}~\bibnamefont {Gu\'erin}},\ }\bibfield  {title} {\bibinfo {title} {Robust quantum control by a single-shot shaped pulse},\ }\href {https://doi.org/10.1103/PhysRevLett.111.050404} {\bibfield  {journal} {\bibinfo  {journal} {Phys. Rev. Lett.}\ }\textbf {\bibinfo {volume} {111}},\ \bibinfo {pages} {050404} (\bibinfo {year} {2013})}\BibitemShut {NoStop}%
\bibitem [{\citenamefont {Goerz}\ \emph {et~al.}(2014)\citenamefont {Goerz}, \citenamefont {Halperin}, \citenamefont {Aytac}, \citenamefont {Koch},\ and\ \citenamefont {Whaley}}]{Modulated_Pulses_Robust_Goerz2014}%
  \BibitemOpen
  \bibfield  {author} {\bibinfo {author} {\bibfnamefont {M.~H.}\ \bibnamefont {Goerz}}, \bibinfo {author} {\bibfnamefont {E.~J.}\ \bibnamefont {Halperin}}, \bibinfo {author} {\bibfnamefont {J.~M.}\ \bibnamefont {Aytac}}, \bibinfo {author} {\bibfnamefont {C.~P.}\ \bibnamefont {Koch}},\ and\ \bibinfo {author} {\bibfnamefont {K.~B.}\ \bibnamefont {Whaley}},\ }\bibfield  {title} {\bibinfo {title} {Robustness of high-fidelity rydberg gates with single-site addressability},\ }\href {https://doi.org/10.1103/PhysRevA.90.032329} {\bibfield  {journal} {\bibinfo  {journal} {Phys. Rev. A}\ }\textbf {\bibinfo {volume} {90}},\ \bibinfo {pages} {032329} (\bibinfo {year} {2014})}\BibitemShut {NoStop}%
\bibitem [{\citenamefont {Poggi}\ \emph {et~al.}(2024)\citenamefont {Poggi}, \citenamefont {De~Chiara}, \citenamefont {Campbell},\ and\ \citenamefont {Kiely}}]{Modulated_Pulses_Robust_Poggi2024}%
  \BibitemOpen
  \bibfield  {author} {\bibinfo {author} {\bibfnamefont {P.~M.}\ \bibnamefont {Poggi}}, \bibinfo {author} {\bibfnamefont {G.}~\bibnamefont {De~Chiara}}, \bibinfo {author} {\bibfnamefont {S.}~\bibnamefont {Campbell}},\ and\ \bibinfo {author} {\bibfnamefont {A.}~\bibnamefont {Kiely}},\ }\bibfield  {title} {\bibinfo {title} {Universally robust quantum control},\ }\href {https://doi.org/10.1103/PhysRevLett.132.193801} {\bibfield  {journal} {\bibinfo  {journal} {Phys. Rev. Lett.}\ }\textbf {\bibinfo {volume} {132}},\ \bibinfo {pages} {193801} (\bibinfo {year} {2024})}\BibitemShut {NoStop}%
\bibitem [{\citenamefont {Zhang}\ \emph {et~al.}(2024)\citenamefont {Zhang}, \citenamefont {Liu}, \citenamefont {Song}, \citenamefont {Xia},\ and\ \citenamefont {Shi}}]{Modulated_Pulses_Robust_Zhang2024}%
  \BibitemOpen
  \bibfield  {author} {\bibinfo {author} {\bibfnamefont {C.}~\bibnamefont {Zhang}}, \bibinfo {author} {\bibfnamefont {Y.}~\bibnamefont {Liu}}, \bibinfo {author} {\bibfnamefont {J.}~\bibnamefont {Song}}, \bibinfo {author} {\bibfnamefont {Y.}~\bibnamefont {Xia}},\ and\ \bibinfo {author} {\bibfnamefont {Z.-C.}\ \bibnamefont {Shi}},\ }\bibfield  {title} {\bibinfo {title} {High-fidelity quantum gates via optimizing short pulse sequences in three-level systems},\ }\href {https://doi.org/10.1088/1367-2630/ad1a2a} {\bibfield  {journal} {\bibinfo  {journal} {New Journal of Physics}\ }\textbf {\bibinfo {volume} {26}},\ \bibinfo {pages} {013024} (\bibinfo {year} {2024})}\BibitemShut {NoStop}%
\bibitem [{\citenamefont {G\"ung\"ord\"u}\ and\ \citenamefont {Kestner}(2022)}]{Modulated_Pulses_Robust_Goerz2014_PRR}%
  \BibitemOpen
  \bibfield  {author} {\bibinfo {author} {\bibfnamefont {U.}~\bibnamefont {G\"ung\"ord\"u}}\ and\ \bibinfo {author} {\bibfnamefont {J.~P.}\ \bibnamefont {Kestner}},\ }\bibfield  {title} {\bibinfo {title} {Robust quantum gates using smooth pulses and physics-informed neural networks},\ }\href {https://doi.org/10.1103/PhysRevResearch.4.023155} {\bibfield  {journal} {\bibinfo  {journal} {Phys. Rev. Res.}\ }\textbf {\bibinfo {volume} {4}},\ \bibinfo {pages} {023155} (\bibinfo {year} {2022})}\BibitemShut {NoStop}%
\bibitem [{\citenamefont {Theis}\ \emph {et~al.}(2016)\citenamefont {Theis}, \citenamefont {Motzoi}, \citenamefont {Wilhelm},\ and\ \citenamefont {Saffman}}]{Modulated_Pulses_Fidelity_Theis2016}%
  \BibitemOpen
  \bibfield  {author} {\bibinfo {author} {\bibfnamefont {L.~S.}\ \bibnamefont {Theis}}, \bibinfo {author} {\bibfnamefont {F.}~\bibnamefont {Motzoi}}, \bibinfo {author} {\bibfnamefont {F.~K.}\ \bibnamefont {Wilhelm}},\ and\ \bibinfo {author} {\bibfnamefont {M.}~\bibnamefont {Saffman}},\ }\bibfield  {title} {\bibinfo {title} {High-fidelity rydberg-blockade entangling gate using shaped, analytic pulses},\ }\href {https://doi.org/10.1103/PhysRevA.94.032306} {\bibfield  {journal} {\bibinfo  {journal} {Phys. Rev. A}\ }\textbf {\bibinfo {volume} {94}},\ \bibinfo {pages} {032306} (\bibinfo {year} {2016})}\BibitemShut {NoStop}%
\bibitem [{\citenamefont {Fauseweh}\ \emph {et~al.}(2012)\citenamefont {Fauseweh}, \citenamefont {Pasini},\ and\ \citenamefont {Uhrig}}]{One_objective_Robust_Fauseweh2012}%
  \BibitemOpen
  \bibfield  {author} {\bibinfo {author} {\bibfnamefont {B.}~\bibnamefont {Fauseweh}}, \bibinfo {author} {\bibfnamefont {S.}~\bibnamefont {Pasini}},\ and\ \bibinfo {author} {\bibfnamefont {G.~S.}\ \bibnamefont {Uhrig}},\ }\bibfield  {title} {\bibinfo {title} {Frequency-modulated pulses for quantum bits coupled to time-dependent baths},\ }\href {https://doi.org/10.1103/PhysRevA.85.022310} {\bibfield  {journal} {\bibinfo  {journal} {Phys. Rev. A}\ }\textbf {\bibinfo {volume} {85}},\ \bibinfo {pages} {022310} (\bibinfo {year} {2012})}\BibitemShut {NoStop}%
\bibitem [{\citenamefont {Hou}\ \emph {et~al.}(2024)\citenamefont {Hou}, \citenamefont {Wang},\ and\ \citenamefont {Qian}}]{One_objective_Robust_Hou2024}%
  \BibitemOpen
  \bibfield  {author} {\bibinfo {author} {\bibfnamefont {Q.-L.}\ \bibnamefont {Hou}}, \bibinfo {author} {\bibfnamefont {H.}~\bibnamefont {Wang}},\ and\ \bibinfo {author} {\bibfnamefont {J.}~\bibnamefont {Qian}},\ }\bibfield  {title} {\bibinfo {title} {Active robustness against detuning error for rydberg quantum gates},\ }\href {https://doi.org/10.1103/PhysRevApplied.22.034054} {\bibfield  {journal} {\bibinfo  {journal} {Phys. Rev. Appl.}\ }\textbf {\bibinfo {volume} {22}},\ \bibinfo {pages} {034054} (\bibinfo {year} {2024})}\BibitemShut {NoStop}%
\bibitem [{\citenamefont {Xiao}\ \emph {et~al.}(2024)\citenamefont {Xiao}, \citenamefont {Kang}, \citenamefont {Zheng}, \citenamefont {Song}, \citenamefont {Chen},\ and\ \citenamefont {Xia}}]{One_objective_Robust_Xiao2024}%
  \BibitemOpen
  \bibfield  {author} {\bibinfo {author} {\bibfnamefont {Y.}~\bibnamefont {Xiao}}, \bibinfo {author} {\bibfnamefont {Y.-H.}\ \bibnamefont {Kang}}, \bibinfo {author} {\bibfnamefont {R.-H.}\ \bibnamefont {Zheng}}, \bibinfo {author} {\bibfnamefont {J.}~\bibnamefont {Song}}, \bibinfo {author} {\bibfnamefont {Y.-H.}\ \bibnamefont {Chen}},\ and\ \bibinfo {author} {\bibfnamefont {Y.}~\bibnamefont {Xia}},\ }\bibfield  {title} {\bibinfo {title} {Effective nonadiabatic holonomic swap gate with rydberg atoms using invariant-based reverse engineering},\ }\href {https://doi.org/10.1103/PhysRevA.109.062610} {\bibfield  {journal} {\bibinfo  {journal} {Phys. Rev. A}\ }\textbf {\bibinfo {volume} {109}},\ \bibinfo {pages} {062610} (\bibinfo {year} {2024})}\BibitemShut {NoStop}%
\bibitem [{\citenamefont {Liang}\ \emph {et~al.}(2023)\citenamefont {Liang}, \citenamefont {Shen}, \citenamefont {Ji},\ and\ \citenamefont {Xue}}]{decline_other_objective_Liang2023}%
  \BibitemOpen
  \bibfield  {author} {\bibinfo {author} {\bibfnamefont {Y.}~\bibnamefont {Liang}}, \bibinfo {author} {\bibfnamefont {P.}~\bibnamefont {Shen}}, \bibinfo {author} {\bibfnamefont {L.-N.}\ \bibnamefont {Ji}},\ and\ \bibinfo {author} {\bibfnamefont {Z.-Y.}\ \bibnamefont {Xue}},\ }\bibfield  {title} {\bibinfo {title} {State-independent nonadiabatic geometric quantum gates},\ }\href {https://doi.org/10.1103/PhysRevApplied.19.024051} {\bibfield  {journal} {\bibinfo  {journal} {Phys. Rev. Appl.}\ }\textbf {\bibinfo {volume} {19}},\ \bibinfo {pages} {024051} (\bibinfo {year} {2023})}\BibitemShut {NoStop}%
\bibitem [{\citenamefont {Jandura}\ \emph {et~al.}(2023)\citenamefont {Jandura}, \citenamefont {Thompson},\ and\ \citenamefont {Pupillo}}]{Two_Cost_Function_into_One_PRX_Jandura2023}%
  \BibitemOpen
  \bibfield  {author} {\bibinfo {author} {\bibfnamefont {S.}~\bibnamefont {Jandura}}, \bibinfo {author} {\bibfnamefont {J.~D.}\ \bibnamefont {Thompson}},\ and\ \bibinfo {author} {\bibfnamefont {G.}~\bibnamefont {Pupillo}},\ }\bibfield  {title} {\bibinfo {title} {Optimizing rydberg gates for logical-qubit performance},\ }\href {https://doi.org/10.1103/PRXQuantum.4.020336} {\bibfield  {journal} {\bibinfo  {journal} {PRX Quantum}\ }\textbf {\bibinfo {volume} {4}},\ \bibinfo {pages} {020336} (\bibinfo {year} {2023})}\BibitemShut {NoStop}%
\bibitem [{\citenamefont {Sharma}\ and\ \citenamefont {Kumar}(2022)}]{Multi_objective_optimization_Sharma2022}%
  \BibitemOpen
  \bibfield  {author} {\bibinfo {author} {\bibfnamefont {S.}~\bibnamefont {Sharma}}\ and\ \bibinfo {author} {\bibfnamefont {V.}~\bibnamefont {Kumar}},\ }\bibfield  {title} {\bibinfo {title} {A comprehensive review on multi-objective optimization techniques: Past, present and future},\ }\href {https://doi.org/10.1007/s11831-022-09778-9} {\bibfield  {journal} {\bibinfo  {journal} {Archives of Computational Methods in Engineering}\ }\textbf {\bibinfo {volume} {29}},\ \bibinfo {pages} {5605–5633} (\bibinfo {year} {2022})}\BibitemShut {NoStop}%
\bibitem [{\citenamefont {Gollub}\ and\ \citenamefont {de~Vivie-Riedle}(2009)}]{Multi_objective_optimization_Gollub2009}%
  \BibitemOpen
  \bibfield  {author} {\bibinfo {author} {\bibfnamefont {C.}~\bibnamefont {Gollub}}\ and\ \bibinfo {author} {\bibfnamefont {R.}~\bibnamefont {de~Vivie-Riedle}},\ }\bibfield  {title} {\bibinfo {title} {Multi-objective genetic algorithm optimization of 2d- and 3d-pareto fronts for vibrational quantum processes},\ }\href {https://doi.org/10.1088/1367-2630/11/1/013019} {\bibfield  {journal} {\bibinfo  {journal} {New Journal of Physics}\ }\textbf {\bibinfo {volume} {11}},\ \bibinfo {pages} {013019} (\bibinfo {year} {2009})}\BibitemShut {NoStop}%
\bibitem [{\citenamefont {Chatterjee}\ and\ \citenamefont {Jacobs}(2025)}]{Multi_objective_optimization_Chatterjee2025}%
  \BibitemOpen
  \bibfield  {author} {\bibinfo {author} {\bibfnamefont {S.}~\bibnamefont {Chatterjee}}\ and\ \bibinfo {author} {\bibfnamefont {W.~M.}\ \bibnamefont {Jacobs}},\ }\bibfield  {title} {\bibinfo {title} {Multiobjective optimization for targeted self-assembly among competing polymorphs},\ }\href {https://doi.org/10.1103/PhysRevX.15.011075} {\bibfield  {journal} {\bibinfo  {journal} {Phys. Rev. X}\ }\textbf {\bibinfo {volume} {15}},\ \bibinfo {pages} {011075} (\bibinfo {year} {2025})}\BibitemShut {NoStop}%
\bibitem [{\citenamefont {Chen}\ \emph {et~al.}(2025)\citenamefont {Chen}, \citenamefont {Zhang}, \citenamefont {Lin}, \citenamefont {Lin}, \citenamefont {Zhao}, \citenamefont {Zhang},\ and\ \citenamefont {Kwok}}]{Multi_objective_optimization_chen2025modl}%
  \BibitemOpen
  \bibfield  {author} {\bibinfo {author} {\bibfnamefont {W.}~\bibnamefont {Chen}}, \bibinfo {author} {\bibfnamefont {X.}~\bibnamefont {Zhang}}, \bibinfo {author} {\bibfnamefont {B.}~\bibnamefont {Lin}}, \bibinfo {author} {\bibfnamefont {X.}~\bibnamefont {Lin}}, \bibinfo {author} {\bibfnamefont {H.}~\bibnamefont {Zhao}}, \bibinfo {author} {\bibfnamefont {Q.}~\bibnamefont {Zhang}},\ and\ \bibinfo {author} {\bibfnamefont {J.~T.}\ \bibnamefont {Kwok}},\ }\bibfield  {title} {\bibinfo {title} {Gradient-based multi-objective deep learning: Algorithms, theories, applications, and beyond},\ }\href@noop {} {\bibfield  {journal} {\bibinfo  {journal} {arXiv preprint arXiv:2501.10945}\ } (\bibinfo {year} {2025})}\BibitemShut {NoStop}%
\bibitem [{\citenamefont {Abdumalikov~Jr}\ \emph {et~al.}(2013)\citenamefont {Abdumalikov~Jr}, \citenamefont {Fink}, \citenamefont {Juliusson}, \citenamefont {Pechal}, \citenamefont {Berger}, \citenamefont {Wallraff},\ and\ \citenamefont {Filipp}}]{NHQC_AbdumalikovJr2013}%
  \BibitemOpen
  \bibfield  {author} {\bibinfo {author} {\bibfnamefont {A.~A.}\ \bibnamefont {Abdumalikov~Jr}}, \bibinfo {author} {\bibfnamefont {J.~M.}\ \bibnamefont {Fink}}, \bibinfo {author} {\bibfnamefont {K.}~\bibnamefont {Juliusson}}, \bibinfo {author} {\bibfnamefont {M.}~\bibnamefont {Pechal}}, \bibinfo {author} {\bibfnamefont {S.}~\bibnamefont {Berger}}, \bibinfo {author} {\bibfnamefont {A.}~\bibnamefont {Wallraff}},\ and\ \bibinfo {author} {\bibfnamefont {S.}~\bibnamefont {Filipp}},\ }\bibfield  {title} {\bibinfo {title} {Experimental realization of non-abelian non-adiabatic geometric gates},\ }\href {https://doi.org/10.1038/nature12010} {\bibfield  {journal} {\bibinfo  {journal} {Nature}\ }\textbf {\bibinfo {volume} {496}},\ \bibinfo {pages} {482–485} (\bibinfo {year} {2013})}\BibitemShut {NoStop}%
\bibitem [{\citenamefont {Feng}\ \emph {et~al.}(2013)\citenamefont {Feng}, \citenamefont {Xu},\ and\ \citenamefont {Long}}]{NHQC_Feng2013}%
  \BibitemOpen
  \bibfield  {author} {\bibinfo {author} {\bibfnamefont {G.}~\bibnamefont {Feng}}, \bibinfo {author} {\bibfnamefont {G.}~\bibnamefont {Xu}},\ and\ \bibinfo {author} {\bibfnamefont {G.}~\bibnamefont {Long}},\ }\bibfield  {title} {\bibinfo {title} {Experimental realization of nonadiabatic holonomic quantum computation},\ }\href {https://doi.org/10.1103/PhysRevLett.110.190501} {\bibfield  {journal} {\bibinfo  {journal} {Phys. Rev. Lett.}\ }\textbf {\bibinfo {volume} {110}},\ \bibinfo {pages} {190501} (\bibinfo {year} {2013})}\BibitemShut {NoStop}%
\bibitem [{\citenamefont {Zhou}\ \emph {et~al.}(2017)\citenamefont {Zhou}, \citenamefont {Jerger}, \citenamefont {Shkolnikov}, \citenamefont {Heremans}, \citenamefont {Burkard},\ and\ \citenamefont {Awschalom}}]{NHQC_Zhou2017}%
  \BibitemOpen
  \bibfield  {author} {\bibinfo {author} {\bibfnamefont {B.~B.}\ \bibnamefont {Zhou}}, \bibinfo {author} {\bibfnamefont {P.~C.}\ \bibnamefont {Jerger}}, \bibinfo {author} {\bibfnamefont {V.~O.}\ \bibnamefont {Shkolnikov}}, \bibinfo {author} {\bibfnamefont {F.~J.}\ \bibnamefont {Heremans}}, \bibinfo {author} {\bibfnamefont {G.}~\bibnamefont {Burkard}},\ and\ \bibinfo {author} {\bibfnamefont {D.~D.}\ \bibnamefont {Awschalom}},\ }\bibfield  {title} {\bibinfo {title} {Holonomic quantum control by coherent optical excitation in diamond},\ }\href {https://doi.org/10.1103/PhysRevLett.119.140503} {\bibfield  {journal} {\bibinfo  {journal} {Phys. Rev. Lett.}\ }\textbf {\bibinfo {volume} {119}},\ \bibinfo {pages} {140503} (\bibinfo {year} {2017})}\BibitemShut {NoStop}%
\bibitem [{\citenamefont {Zhao}\ \emph {et~al.}(2020)\citenamefont {Zhao}, \citenamefont {Li}, \citenamefont {Xu},\ and\ \citenamefont {Tong}}]{NHQC_Zhao2020}%
  \BibitemOpen
  \bibfield  {author} {\bibinfo {author} {\bibfnamefont {P.~Z.}\ \bibnamefont {Zhao}}, \bibinfo {author} {\bibfnamefont {K.~Z.}\ \bibnamefont {Li}}, \bibinfo {author} {\bibfnamefont {G.~F.}\ \bibnamefont {Xu}},\ and\ \bibinfo {author} {\bibfnamefont {D.~M.}\ \bibnamefont {Tong}},\ }\bibfield  {title} {\bibinfo {title} {General approach for constructing hamiltonians for nonadiabatic holonomic quantum computation},\ }\href {https://doi.org/10.1103/PhysRevA.101.062306} {\bibfield  {journal} {\bibinfo  {journal} {Phys. Rev. A}\ }\textbf {\bibinfo {volume} {101}},\ \bibinfo {pages} {062306} (\bibinfo {year} {2020})}\BibitemShut {NoStop}%
\bibitem [{\citenamefont {Jin}\ and\ \citenamefont {Jing}(2024)}]{NHQC_Jin2024}%
  \BibitemOpen
  \bibfield  {author} {\bibinfo {author} {\bibfnamefont {Z.-y.}\ \bibnamefont {Jin}}\ and\ \bibinfo {author} {\bibfnamefont {J.}~\bibnamefont {Jing}},\ }\bibfield  {title} {\bibinfo {title} {Geometric quantum gates via dark paths in rydberg atoms},\ }\href {https://doi.org/10.1103/PhysRevA.109.012619} {\bibfield  {journal} {\bibinfo  {journal} {Phys. Rev. A}\ }\textbf {\bibinfo {volume} {109}},\ \bibinfo {pages} {012619} (\bibinfo {year} {2024})}\BibitemShut {NoStop}%
\bibitem [{\citenamefont {Jing}\ \emph {et~al.}(2017)\citenamefont {Jing}, \citenamefont {Lam},\ and\ \citenamefont {Wu}}]{Control_error_PhysRevA.95.012334}%
  \BibitemOpen
  \bibfield  {author} {\bibinfo {author} {\bibfnamefont {J.}~\bibnamefont {Jing}}, \bibinfo {author} {\bibfnamefont {C.-H.}\ \bibnamefont {Lam}},\ and\ \bibinfo {author} {\bibfnamefont {L.-A.}\ \bibnamefont {Wu}},\ }\bibfield  {title} {\bibinfo {title} {Non-abelian holonomic transformation in the presence of classical noise},\ }\href {https://doi.org/10.1103/PhysRevA.95.012334} {\bibfield  {journal} {\bibinfo  {journal} {Phys. Rev. A}\ }\textbf {\bibinfo {volume} {95}},\ \bibinfo {pages} {012334} (\bibinfo {year} {2017})}\BibitemShut {NoStop}%
\bibitem [{\citenamefont {Thomas}\ \emph {et~al.}(2011)\citenamefont {Thomas}, \citenamefont {Lababidi},\ and\ \citenamefont {Tian}}]{Control_error_Thomas2011}%
  \BibitemOpen
  \bibfield  {author} {\bibinfo {author} {\bibfnamefont {J.~T.}\ \bibnamefont {Thomas}}, \bibinfo {author} {\bibfnamefont {M.}~\bibnamefont {Lababidi}},\ and\ \bibinfo {author} {\bibfnamefont {M.}~\bibnamefont {Tian}},\ }\bibfield  {title} {\bibinfo {title} {Robustness of single-qubit geometric gate against systematic error},\ }\href {https://doi.org/10.1103/PhysRevA.84.042335} {\bibfield  {journal} {\bibinfo  {journal} {Phys. Rev. A}\ }\textbf {\bibinfo {volume} {84}},\ \bibinfo {pages} {042335} (\bibinfo {year} {2011})}\BibitemShut {NoStop}%
\bibitem [{\citenamefont {Zheng}\ \emph {et~al.}(2016)\citenamefont {Zheng}, \citenamefont {Yang},\ and\ \citenamefont {Nori}}]{Control_error_Zheng2016}%
  \BibitemOpen
  \bibfield  {author} {\bibinfo {author} {\bibfnamefont {S.-B.}\ \bibnamefont {Zheng}}, \bibinfo {author} {\bibfnamefont {C.-P.}\ \bibnamefont {Yang}},\ and\ \bibinfo {author} {\bibfnamefont {F.}~\bibnamefont {Nori}},\ }\bibfield  {title} {\bibinfo {title} {Comparison of the sensitivity to systematic errors between nonadiabatic non-abelian geometric gates and their dynamical counterparts},\ }\href {https://doi.org/10.1103/PhysRevA.93.032313} {\bibfield  {journal} {\bibinfo  {journal} {Phys. Rev. A}\ }\textbf {\bibinfo {volume} {93}},\ \bibinfo {pages} {032313} (\bibinfo {year} {2016})}\BibitemShut {NoStop}%
\bibitem [{\citenamefont {Liang}\ \emph {et~al.}(2024)\citenamefont {Liang}, \citenamefont {Wu},\ and\ \citenamefont {Xue}}]{Precisely_Modify_the_Path_Parameters_Liang2024}%
  \BibitemOpen
  \bibfield  {author} {\bibinfo {author} {\bibfnamefont {Y.}~\bibnamefont {Liang}}, \bibinfo {author} {\bibfnamefont {Y.-X.}\ \bibnamefont {Wu}},\ and\ \bibinfo {author} {\bibfnamefont {Z.-Y.}\ \bibnamefont {Xue}},\ }\bibfield  {title} {\bibinfo {title} {Nonadiabatic geometric quantum gates that are robust against systematic errors},\ }\href {https://doi.org/10.1103/PhysRevApplied.22.024061} {\bibfield  {journal} {\bibinfo  {journal} {Phys. Rev. Appl.}\ }\textbf {\bibinfo {volume} {22}},\ \bibinfo {pages} {024061} (\bibinfo {year} {2024})}\BibitemShut {NoStop}%
\bibitem [{\citenamefont {Liang}\ \emph {et~al.}(2022)\citenamefont {Liang}, \citenamefont {Shen}, \citenamefont {Chen},\ and\ \citenamefont {Xue}}]{Composite_Dynamical_Decoupling_pulses_Liang2022}%
  \BibitemOpen
  \bibfield  {author} {\bibinfo {author} {\bibfnamefont {Y.}~\bibnamefont {Liang}}, \bibinfo {author} {\bibfnamefont {P.}~\bibnamefont {Shen}}, \bibinfo {author} {\bibfnamefont {T.}~\bibnamefont {Chen}},\ and\ \bibinfo {author} {\bibfnamefont {Z.-Y.}\ \bibnamefont {Xue}},\ }\bibfield  {title} {\bibinfo {title} {Composite short-path nonadiabatic holonomic quantum gates},\ }\href {https://doi.org/10.1103/PhysRevApplied.17.034015} {\bibfield  {journal} {\bibinfo  {journal} {Phys. Rev. Appl.}\ }\textbf {\bibinfo {volume} {17}},\ \bibinfo {pages} {034015} (\bibinfo {year} {2022})}\BibitemShut {NoStop}%
\bibitem [{\citenamefont {Haxhiraj}\ \emph {et~al.}(2025)\citenamefont {Haxhiraj}, \citenamefont {Shahu},\ and\ \citenamefont {Agastra}}]{Pareto_front_Haxhiraj2025}%
  \BibitemOpen
  \bibfield  {author} {\bibinfo {author} {\bibfnamefont {E.}~\bibnamefont {Haxhiraj}}, \bibinfo {author} {\bibfnamefont {D.}~\bibnamefont {Shahu}},\ and\ \bibinfo {author} {\bibfnamefont {E.}~\bibnamefont {Agastra}},\ }\bibfield  {title} {\bibinfo {title} {Pareto front transformation in the decision-making process for spectral and energy efficiency trade-off in massive mimo systems},\ }\href {https://doi.org/10.3390/s25051451} {\bibfield  {journal} {\bibinfo  {journal} {Sensors}\ }\textbf {\bibinfo {volume} {25}},\ \bibinfo {pages} {1451} (\bibinfo {year} {2025})}\BibitemShut {NoStop}%
\bibitem [{\citenamefont {Kumar}\ \emph {et~al.}(2021)\citenamefont {Kumar}, \citenamefont {Singh}, \citenamefont {Bilga}, \citenamefont {Jatin}, \citenamefont {Singh}, \citenamefont {Singh}, \citenamefont {Scutaru},\ and\ \citenamefont {Pruncu}}]{EWM_Kumar2021}%
  \BibitemOpen
  \bibfield  {author} {\bibinfo {author} {\bibfnamefont {R.}~\bibnamefont {Kumar}}, \bibinfo {author} {\bibfnamefont {S.}~\bibnamefont {Singh}}, \bibinfo {author} {\bibfnamefont {P.~S.}\ \bibnamefont {Bilga}}, \bibinfo {author} {\bibnamefont {Jatin}}, \bibinfo {author} {\bibfnamefont {J.}~\bibnamefont {Singh}}, \bibinfo {author} {\bibfnamefont {S.}~\bibnamefont {Singh}}, \bibinfo {author} {\bibfnamefont {M.-L.}\ \bibnamefont {Scutaru}},\ and\ \bibinfo {author} {\bibfnamefont {C.~I.}\ \bibnamefont {Pruncu}},\ }\bibfield  {title} {\bibinfo {title} {Revealing the benefits of entropy weights method for multi-objective optimization in machining operations: A critical review},\ }\href {https://doi.org/10.1016/j.jmrt.2020.12.114} {\bibfield  {journal} {\bibinfo  {journal} {Journal of Materials Research and Technology}\ }\textbf {\bibinfo {volume} {10}},\ \bibinfo {pages} {1471–1492} (\bibinfo {year} {2021})}\BibitemShut {NoStop}%
\bibitem [{\citenamefont {Sj\"{o}qvist}\ \emph {et~al.}(2012)\citenamefont {Sj\"{o}qvist}, \citenamefont {Tong}, \citenamefont {Mauritz~Andersson}, \citenamefont {Hessmo}, \citenamefont {Johansson},\ and\ \citenamefont {Singh}}]{geometric_condition_Sjqvist2012}%
  \BibitemOpen
  \bibfield  {author} {\bibinfo {author} {\bibfnamefont {E.}~\bibnamefont {Sj\"{o}qvist}}, \bibinfo {author} {\bibfnamefont {D.~M.}\ \bibnamefont {Tong}}, \bibinfo {author} {\bibfnamefont {L.}~\bibnamefont {Mauritz~Andersson}}, \bibinfo {author} {\bibfnamefont {B.}~\bibnamefont {Hessmo}}, \bibinfo {author} {\bibfnamefont {M.}~\bibnamefont {Johansson}},\ and\ \bibinfo {author} {\bibfnamefont {K.}~\bibnamefont {Singh}},\ }\bibfield  {title} {\bibinfo {title} {Non-adiabatic holonomic quantum computation},\ }\href {https://doi.org/10.1088/1367-2630/14/10/103035} {\bibfield  {journal} {\bibinfo  {journal} {New Journal of Physics}\ }\textbf {\bibinfo {volume} {14}},\ \bibinfo {pages} {103035} (\bibinfo {year} {2012})}\BibitemShut {NoStop}%
\bibitem [{\citenamefont {Ai}\ \emph {et~al.}(2022)\citenamefont {Ai}, \citenamefont {Li}, \citenamefont {He}, \citenamefont {Xue}, \citenamefont {Cui}, \citenamefont {Huang}, \citenamefont {Li},\ and\ \citenamefont {Guo}}]{Hamiltonian_Ai2022}%
  \BibitemOpen
  \bibfield  {author} {\bibinfo {author} {\bibfnamefont {M.-Z.}\ \bibnamefont {Ai}}, \bibinfo {author} {\bibfnamefont {S.}~\bibnamefont {Li}}, \bibinfo {author} {\bibfnamefont {R.}~\bibnamefont {He}}, \bibinfo {author} {\bibfnamefont {Z.-Y.}\ \bibnamefont {Xue}}, \bibinfo {author} {\bibfnamefont {J.-M.}\ \bibnamefont {Cui}}, \bibinfo {author} {\bibfnamefont {Y.-F.}\ \bibnamefont {Huang}}, \bibinfo {author} {\bibfnamefont {C.-F.}\ \bibnamefont {Li}},\ and\ \bibinfo {author} {\bibfnamefont {G.-C.}\ \bibnamefont {Guo}},\ }\bibfield  {title} {\bibinfo {title} {Experimental realization of nonadiabatic holonomic single‐qubit quantum gates with two dark paths in a trapped ion},\ }\href {https://doi.org/10.1016/j.fmre.2021.11.031} {\bibfield  {journal} {\bibinfo  {journal} {Fundamental Research}\ }\textbf {\bibinfo {volume} {2}},\ \bibinfo {pages} {661–666} (\bibinfo {year} {2022})}\BibitemShut {NoStop}%
\bibitem [{\citenamefont {Isenhower}\ \emph {et~al.}(2010)\citenamefont {Isenhower}, \citenamefont {Urban}, \citenamefont {Zhang}, \citenamefont {Gill}, \citenamefont {Henage}, \citenamefont {Johnson}, \citenamefont {Walker},\ and\ \citenamefont {Saffman}}]{CNOT_gate_Isenhower2010}%
  \BibitemOpen
  \bibfield  {author} {\bibinfo {author} {\bibfnamefont {L.}~\bibnamefont {Isenhower}}, \bibinfo {author} {\bibfnamefont {E.}~\bibnamefont {Urban}}, \bibinfo {author} {\bibfnamefont {X.~L.}\ \bibnamefont {Zhang}}, \bibinfo {author} {\bibfnamefont {A.~T.}\ \bibnamefont {Gill}}, \bibinfo {author} {\bibfnamefont {T.}~\bibnamefont {Henage}}, \bibinfo {author} {\bibfnamefont {T.~A.}\ \bibnamefont {Johnson}}, \bibinfo {author} {\bibfnamefont {T.~G.}\ \bibnamefont {Walker}},\ and\ \bibinfo {author} {\bibfnamefont {M.}~\bibnamefont {Saffman}},\ }\bibfield  {title} {\bibinfo {title} {Demonstration of a neutral atom controlled-not quantum gate},\ }\href {https://doi.org/10.1103/PhysRevLett.104.010503} {\bibfield  {journal} {\bibinfo  {journal} {Phys. Rev. Lett.}\ }\textbf {\bibinfo {volume} {104}},\ \bibinfo {pages} {010503} (\bibinfo {year} {2010})}\BibitemShut {NoStop}%
\bibitem [{\citenamefont {Liu}\ \emph {et~al.}(2019)\citenamefont {Liu}, \citenamefont {Song}, \citenamefont {Xue}, \citenamefont {Wang},\ and\ \citenamefont {Yung}}]{von_Neumann_equation_Liu2019}%
  \BibitemOpen
  \bibfield  {author} {\bibinfo {author} {\bibfnamefont {B.-J.}\ \bibnamefont {Liu}}, \bibinfo {author} {\bibfnamefont {X.-K.}\ \bibnamefont {Song}}, \bibinfo {author} {\bibfnamefont {Z.-Y.}\ \bibnamefont {Xue}}, \bibinfo {author} {\bibfnamefont {X.}~\bibnamefont {Wang}},\ and\ \bibinfo {author} {\bibfnamefont {M.-H.}\ \bibnamefont {Yung}},\ }\bibfield  {title} {\bibinfo {title} {Plug-and-play approach to nonadiabatic geometric quantum gates},\ }\href {https://doi.org/10.1103/PhysRevLett.123.100501} {\bibfield  {journal} {\bibinfo  {journal} {Phys. Rev. Lett.}\ }\textbf {\bibinfo {volume} {123}},\ \bibinfo {pages} {100501} (\bibinfo {year} {2019})}\BibitemShut {NoStop}%
\bibitem [{\citenamefont {Braaten}\ \emph {et~al.}(2017)\citenamefont {Braaten}, \citenamefont {Hammer},\ and\ \citenamefont {Lepage}}]{Lindblad_master_equationBraaten2017}%
  \BibitemOpen
  \bibfield  {author} {\bibinfo {author} {\bibfnamefont {E.}~\bibnamefont {Braaten}}, \bibinfo {author} {\bibfnamefont {H.-W.}\ \bibnamefont {Hammer}},\ and\ \bibinfo {author} {\bibfnamefont {G.~P.}\ \bibnamefont {Lepage}},\ }\bibfield  {title} {\bibinfo {title} {Lindblad equation for the inelastic loss of ultracold atoms},\ }\href {https://doi.org/10.1103/PhysRevA.95.012708} {\bibfield  {journal} {\bibinfo  {journal} {Phys. Rev. A}\ }\textbf {\bibinfo {volume} {95}},\ \bibinfo {pages} {012708} (\bibinfo {year} {2017})}\BibitemShut {NoStop}%
\bibitem [{\citenamefont {Tamura}\ \emph {et~al.}(2020)\citenamefont {Tamura}, \citenamefont {Yamakoshi},\ and\ \citenamefont {Nakagawa}}]{decay_and_dephasing_Tamura2020}%
  \BibitemOpen
  \bibfield  {author} {\bibinfo {author} {\bibfnamefont {H.}~\bibnamefont {Tamura}}, \bibinfo {author} {\bibfnamefont {T.}~\bibnamefont {Yamakoshi}},\ and\ \bibinfo {author} {\bibfnamefont {K.}~\bibnamefont {Nakagawa}},\ }\bibfield  {title} {\bibinfo {title} {Analysis of coherent dynamics of a rydberg-atom quantum simulator},\ }\href {https://doi.org/10.1103/PhysRevA.101.043421} {\bibfield  {journal} {\bibinfo  {journal} {Phys. Rev. A}\ }\textbf {\bibinfo {volume} {101}},\ \bibinfo {pages} {043421} (\bibinfo {year} {2020})}\BibitemShut {NoStop}%
\bibitem [{\citenamefont {Nielsen}\ and\ \citenamefont {Chuang}(2012)}]{Fidelity_Nielsen2012}%
  \BibitemOpen
  \bibfield  {author} {\bibinfo {author} {\bibfnamefont {M.~A.}\ \bibnamefont {Nielsen}}\ and\ \bibinfo {author} {\bibfnamefont {I.~L.}\ \bibnamefont {Chuang}},\ }\href {https://doi.org/10.1017/cbo9780511976667} {\emph {\bibinfo {title} {Quantum Computation and Quantum Information: 10th Anniversary Edition}}}\ (\bibinfo  {publisher} {Cambridge University Press},\ \bibinfo {year} {2012})\BibitemShut {NoStop}%
\bibitem [{\citenamefont {Liang}\ and\ \citenamefont {Xue}(2024)}]{Light_Liang2024}%
  \BibitemOpen
  \bibfield  {author} {\bibinfo {author} {\bibfnamefont {Y.}~\bibnamefont {Liang}}\ and\ \bibinfo {author} {\bibfnamefont {Z.-Y.}\ \bibnamefont {Xue}},\ }\bibfield  {title} {\bibinfo {title} {Nonadiabatic geometric quantum gates with on-demand trajectories},\ }\href {https://doi.org/10.1103/PhysRevApplied.21.064048} {\bibfield  {journal} {\bibinfo  {journal} {Phys. Rev. Appl.}\ }\textbf {\bibinfo {volume} {21}},\ \bibinfo {pages} {064048} (\bibinfo {year} {2024})}\BibitemShut {NoStop}%
\bibitem [{\citenamefont {Liu}\ \emph {et~al.}(2021)\citenamefont {Liu}, \citenamefont {Wang},\ and\ \citenamefont {Yung}}]{change_designed_path_Liu2021}%
  \BibitemOpen
  \bibfield  {author} {\bibinfo {author} {\bibfnamefont {B.-J.}\ \bibnamefont {Liu}}, \bibinfo {author} {\bibfnamefont {Y.-S.}\ \bibnamefont {Wang}},\ and\ \bibinfo {author} {\bibfnamefont {M.-H.}\ \bibnamefont {Yung}},\ }\bibfield  {title} {\bibinfo {title} {Super-robust nonadiabatic geometric quantum control},\ }\href {https://doi.org/10.1103/PhysRevResearch.3.L032066} {\bibfield  {journal} {\bibinfo  {journal} {Phys. Rev. Res.}\ }\textbf {\bibinfo {volume} {3}},\ \bibinfo {pages} {L032066} (\bibinfo {year} {2021})}\BibitemShut {NoStop}%
\bibitem [{\citenamefont {McDonnell}\ \emph {et~al.}(2022)\citenamefont {McDonnell}, \citenamefont {Keary},\ and\ \citenamefont {Pritchard}}]{Same_amplitude_error_McDonnell2022}%
  \BibitemOpen
  \bibfield  {author} {\bibinfo {author} {\bibfnamefont {K.}~\bibnamefont {McDonnell}}, \bibinfo {author} {\bibfnamefont {L.~F.}\ \bibnamefont {Keary}},\ and\ \bibinfo {author} {\bibfnamefont {J.~D.}\ \bibnamefont {Pritchard}},\ }\bibfield  {title} {\bibinfo {title} {Demonstration of a quantum gate using electromagnetically induced transparency},\ }\href {https://doi.org/10.1103/PhysRevLett.129.200501} {\bibfield  {journal} {\bibinfo  {journal} {Phys. Rev. Lett.}\ }\textbf {\bibinfo {volume} {129}},\ \bibinfo {pages} {200501} (\bibinfo {year} {2022})}\BibitemShut {NoStop}%
\bibitem [{\citenamefont {Li}\ \emph {et~al.}(2024)\citenamefont {Li}, \citenamefont {Wu}, \citenamefont {Su},\ and\ \citenamefont {Qian}}]{previous_work_Li2024}%
  \BibitemOpen
  \bibfield  {author} {\bibinfo {author} {\bibfnamefont {W.-X.}\ \bibnamefont {Li}}, \bibinfo {author} {\bibfnamefont {J.-L.}\ \bibnamefont {Wu}}, \bibinfo {author} {\bibfnamefont {S.-L.}\ \bibnamefont {Su}},\ and\ \bibinfo {author} {\bibfnamefont {J.}~\bibnamefont {Qian}},\ }\bibfield  {title} {\bibinfo {title} {High-tolerance antiblockade $\mathrm{SWAP}$ gates using optimal pulse drivings},\ }\href {https://doi.org/10.1103/PhysRevA.109.012608} {\bibfield  {journal} {\bibinfo  {journal} {Phys. Rev. A}\ }\textbf {\bibinfo {volume} {109}},\ \bibinfo {pages} {012608} (\bibinfo {year} {2024})}\BibitemShut {NoStop}%
\bibitem [{\citenamefont {Cantu}\ \emph {et~al.}(2020)\citenamefont {Cantu}, \citenamefont {Venkatramani}, \citenamefont {Xu}, \citenamefont {Zhou}, \citenamefont {Jelenković}, \citenamefont {Lukin},\ and\ \citenamefont {Vuletić}}]{two_photon_coupling_Cantu2020}%
  \BibitemOpen
  \bibfield  {author} {\bibinfo {author} {\bibfnamefont {S.~H.}\ \bibnamefont {Cantu}}, \bibinfo {author} {\bibfnamefont {A.~V.}\ \bibnamefont {Venkatramani}}, \bibinfo {author} {\bibfnamefont {W.}~\bibnamefont {Xu}}, \bibinfo {author} {\bibfnamefont {L.}~\bibnamefont {Zhou}}, \bibinfo {author} {\bibfnamefont {B.}~\bibnamefont {Jelenković}}, \bibinfo {author} {\bibfnamefont {M.~D.}\ \bibnamefont {Lukin}},\ and\ \bibinfo {author} {\bibfnamefont {V.}~\bibnamefont {Vuletić}},\ }\bibfield  {title} {\bibinfo {title} {Repulsive photons in a quantum nonlinear medium},\ }\href {https://doi.org/10.1038/s41567-020-0917-6} {\bibfield  {journal} {\bibinfo  {journal} {Nature Physics}\ }\textbf {\bibinfo {volume} {16}},\ \bibinfo {pages} {921–925} (\bibinfo {year} {2020})}\BibitemShut {NoStop}%
\bibitem [{\citenamefont {Levine}\ \emph {et~al.}(2018)\citenamefont {Levine}, \citenamefont {Keesling}, \citenamefont {Omran}, \citenamefont {Bernien}, \citenamefont {Schwartz}, \citenamefont {Zibrov}, \citenamefont {Endres}, \citenamefont {Greiner}, \citenamefont {Vuleti\ifmmode~\acute{c}\else \'{c}\fi{}},\ and\ \citenamefont {Lukin}}]{errors_Levine2018}%
  \BibitemOpen
  \bibfield  {author} {\bibinfo {author} {\bibfnamefont {H.}~\bibnamefont {Levine}}, \bibinfo {author} {\bibfnamefont {A.}~\bibnamefont {Keesling}}, \bibinfo {author} {\bibfnamefont {A.}~\bibnamefont {Omran}}, \bibinfo {author} {\bibfnamefont {H.}~\bibnamefont {Bernien}}, \bibinfo {author} {\bibfnamefont {S.}~\bibnamefont {Schwartz}}, \bibinfo {author} {\bibfnamefont {A.~S.}\ \bibnamefont {Zibrov}}, \bibinfo {author} {\bibfnamefont {M.}~\bibnamefont {Endres}}, \bibinfo {author} {\bibfnamefont {M.}~\bibnamefont {Greiner}}, \bibinfo {author} {\bibfnamefont {V.}~\bibnamefont {Vuleti\ifmmode~\acute{c}\else \'{c}\fi{}}},\ and\ \bibinfo {author} {\bibfnamefont {M.~D.}\ \bibnamefont {Lukin}},\ }\bibfield  {title} {\bibinfo {title} {High-fidelity control and entanglement of rydberg-atom qubits},\ }\href {https://doi.org/10.1103/PhysRevLett.121.123603} {\bibfield  {journal} {\bibinfo  {journal} {Phys. Rev. Lett.}\ }\textbf {\bibinfo {volume} {121}},\ \bibinfo {pages} {123603} (\bibinfo {year} {2018})}\BibitemShut
  {NoStop}%
\bibitem [{\citenamefont {Kang}\ \emph {et~al.}(2024)\citenamefont {Kang}, \citenamefont {Li},\ and\ \citenamefont {Wang}}]{Pareto_front_Kang2024}%
  \BibitemOpen
  \bibfield  {author} {\bibinfo {author} {\bibfnamefont {S.}~\bibnamefont {Kang}}, \bibinfo {author} {\bibfnamefont {K.}~\bibnamefont {Li}},\ and\ \bibinfo {author} {\bibfnamefont {R.}~\bibnamefont {Wang}},\ }\bibfield  {title} {\bibinfo {title} {A survey on pareto front learning for multi-objective optimization},\ }\bibfield  {journal} {\bibinfo  {journal} {Journal of Membrane Computing}\ }\href {https://doi.org/10.1007/s41965-024-00170-z} {10.1007/s41965-024-00170-z} (\bibinfo {year} {2024})\BibitemShut {NoStop}%
\bibitem [{\citenamefont {Li}\ \emph {et~al.}(2022{\natexlab{a}})\citenamefont {Li}, \citenamefont {Luo}, \citenamefont {Wang}, \citenamefont {Fan},\ and\ \citenamefont {Zhang}}]{EWM_Li2022}%
  \BibitemOpen
  \bibfield  {author} {\bibinfo {author} {\bibfnamefont {Z.}~\bibnamefont {Li}}, \bibinfo {author} {\bibfnamefont {Z.}~\bibnamefont {Luo}}, \bibinfo {author} {\bibfnamefont {Y.}~\bibnamefont {Wang}}, \bibinfo {author} {\bibfnamefont {G.}~\bibnamefont {Fan}},\ and\ \bibinfo {author} {\bibfnamefont {J.}~\bibnamefont {Zhang}},\ }\bibfield  {title} {\bibinfo {title} {Suitability evaluation system for the shallow geothermal energy implementation in region by entropy weight method and topsis method},\ }\href {https://doi.org/10.1016/j.renene.2021.11.112} {\bibfield  {journal} {\bibinfo  {journal} {Renewable Energy}\ }\textbf {\bibinfo {volume} {184}},\ \bibinfo {pages} {564–576} (\bibinfo {year} {2022}{\natexlab{a}})}\BibitemShut {NoStop}%
\bibitem [{\citenamefont {Li}\ \emph {et~al.}(2022{\natexlab{b}})\citenamefont {Li}, \citenamefont {You}, \citenamefont {Shao},\ and\ \citenamefont {Li}}]{Energy_level_Li2022}%
  \BibitemOpen
  \bibfield  {author} {\bibinfo {author} {\bibfnamefont {X.~X.}\ \bibnamefont {Li}}, \bibinfo {author} {\bibfnamefont {J.~B.}\ \bibnamefont {You}}, \bibinfo {author} {\bibfnamefont {X.~Q.}\ \bibnamefont {Shao}},\ and\ \bibinfo {author} {\bibfnamefont {W.}~\bibnamefont {Li}},\ }\bibfield  {title} {\bibinfo {title} {Coherent ground-state transport of neutral atoms},\ }\href {https://doi.org/10.1103/PhysRevA.105.032417} {\bibfield  {journal} {\bibinfo  {journal} {Phys. Rev. A}\ }\textbf {\bibinfo {volume} {105}},\ \bibinfo {pages} {032417} (\bibinfo {year} {2022}{\natexlab{b}})}\BibitemShut {NoStop}%
\bibitem [{\citenamefont {Maller}\ \emph {et~al.}(2015)\citenamefont {Maller}, \citenamefont {Lichtman}, \citenamefont {Xia}, \citenamefont {Sun}, \citenamefont {Piotrowicz}, \citenamefont {Carr}, \citenamefont {Isenhower},\ and\ \citenamefont {Saffman}}]{CNOT_Maller2015}%
  \BibitemOpen
  \bibfield  {author} {\bibinfo {author} {\bibfnamefont {K.~M.}\ \bibnamefont {Maller}}, \bibinfo {author} {\bibfnamefont {M.~T.}\ \bibnamefont {Lichtman}}, \bibinfo {author} {\bibfnamefont {T.}~\bibnamefont {Xia}}, \bibinfo {author} {\bibfnamefont {Y.}~\bibnamefont {Sun}}, \bibinfo {author} {\bibfnamefont {M.~J.}\ \bibnamefont {Piotrowicz}}, \bibinfo {author} {\bibfnamefont {A.~W.}\ \bibnamefont {Carr}}, \bibinfo {author} {\bibfnamefont {L.}~\bibnamefont {Isenhower}},\ and\ \bibinfo {author} {\bibfnamefont {M.}~\bibnamefont {Saffman}},\ }\bibfield  {title} {\bibinfo {title} {Rydberg-blockade controlled-not gate and entanglement in a two-dimensional array of neutral-atom qubits},\ }\href {https://doi.org/10.1103/PhysRevA.92.022336} {\bibfield  {journal} {\bibinfo  {journal} {Phys. Rev. A}\ }\textbf {\bibinfo {volume} {92}},\ \bibinfo {pages} {022336} (\bibinfo {year} {2015})}\BibitemShut {NoStop}%
\bibitem [{\citenamefont {Šibalić}\ \emph {et~al.}(2017)\citenamefont {Šibalić}, \citenamefont {Pritchard}, \citenamefont {Adams},\ and\ \citenamefont {Weatherill}}]{ARC_ibali2017}%
  \BibitemOpen
  \bibfield  {author} {\bibinfo {author} {\bibfnamefont {N.}~\bibnamefont {Šibalić}}, \bibinfo {author} {\bibfnamefont {J.}~\bibnamefont {Pritchard}}, \bibinfo {author} {\bibfnamefont {C.}~\bibnamefont {Adams}},\ and\ \bibinfo {author} {\bibfnamefont {K.}~\bibnamefont {Weatherill}},\ }\bibfield  {title} {\bibinfo {title} {Arc: An open-source library for calculating properties of alkali rydberg atoms},\ }\href {https://doi.org/10.1016/j.cpc.2017.06.015} {\bibfield  {journal} {\bibinfo  {journal} {Computer Physics Communications}\ }\textbf {\bibinfo {volume} {220}},\ \bibinfo {pages} {319–331} (\bibinfo {year} {2017})}\BibitemShut {NoStop}%
\bibitem [{\citenamefont {Shi}(2017)}]{Blockade_error_Shi2017}%
  \BibitemOpen
  \bibfield  {author} {\bibinfo {author} {\bibfnamefont {X.-F.}\ \bibnamefont {Shi}},\ }\bibfield  {title} {\bibinfo {title} {Rydberg quantum gates free from blockade error},\ }\href {https://doi.org/10.1103/PhysRevApplied.7.064017} {\bibfield  {journal} {\bibinfo  {journal} {Phys. Rev. Appl.}\ }\textbf {\bibinfo {volume} {7}},\ \bibinfo {pages} {064017} (\bibinfo {year} {2017})}\BibitemShut {NoStop}%
\bibitem [{\citenamefont {Chang}\ \emph {et~al.}(2023)\citenamefont {Chang}, \citenamefont {Wang}, \citenamefont {Jen},\ and\ \citenamefont {Chen}}]{Modulated_Pulses_Fidelity_Chang2023}%
  \BibitemOpen
  \bibfield  {author} {\bibinfo {author} {\bibfnamefont {T.~H.}\ \bibnamefont {Chang}}, \bibinfo {author} {\bibfnamefont {T.~N.}\ \bibnamefont {Wang}}, \bibinfo {author} {\bibfnamefont {H.~H.}\ \bibnamefont {Jen}},\ and\ \bibinfo {author} {\bibfnamefont {Y.-C.}\ \bibnamefont {Chen}},\ }\bibfield  {title} {\bibinfo {title} {High-fidelity rydberg controlled-z gates with optimized pulses},\ }\href {https://doi.org/10.1088/1367-2630/ad0fa9} {\bibfield  {journal} {\bibinfo  {journal} {New Journal of Physics}\ }\textbf {\bibinfo {volume} {25}},\ \bibinfo {pages} {123007} (\bibinfo {year} {2023})}\BibitemShut {NoStop}%
\bibitem [{\citenamefont {Xu}\ \emph {et~al.}(2017)\citenamefont {Xu}, \citenamefont {Zhao}, \citenamefont {Tong},\ and\ \citenamefont {Sj\"oqvist}}]{One_objective_Robust_Xu2017}%
  \BibitemOpen
  \bibfield  {author} {\bibinfo {author} {\bibfnamefont {G.~F.}\ \bibnamefont {Xu}}, \bibinfo {author} {\bibfnamefont {P.~Z.}\ \bibnamefont {Zhao}}, \bibinfo {author} {\bibfnamefont {D.~M.}\ \bibnamefont {Tong}},\ and\ \bibinfo {author} {\bibfnamefont {E.}~\bibnamefont {Sj\"oqvist}},\ }\bibfield  {title} {\bibinfo {title} {Robust paths to realize nonadiabatic holonomic gates},\ }\href {https://doi.org/10.1103/PhysRevA.95.052349} {\bibfield  {journal} {\bibinfo  {journal} {Phys. Rev. A}\ }\textbf {\bibinfo {volume} {95}},\ \bibinfo {pages} {052349} (\bibinfo {year} {2017})}\BibitemShut {NoStop}%
\bibitem [{\citenamefont {Khazali}\ and\ \citenamefont {M\o{}lmer}(2020)}]{Multiqubit_gate_Khazali2020}%
  \BibitemOpen
  \bibfield  {author} {\bibinfo {author} {\bibfnamefont {M.}~\bibnamefont {Khazali}}\ and\ \bibinfo {author} {\bibfnamefont {K.}~\bibnamefont {M\o{}lmer}},\ }\bibfield  {title} {\bibinfo {title} {Fast multiqubit gates by adiabatic evolution in interacting excited-state manifolds of rydberg atoms and superconducting circuits},\ }\href {https://doi.org/10.1103/PhysRevX.10.021054} {\bibfield  {journal} {\bibinfo  {journal} {Phys. Rev. X}\ }\textbf {\bibinfo {volume} {10}},\ \bibinfo {pages} {021054} (\bibinfo {year} {2020})}\BibitemShut {NoStop}%
\end{thebibliography}%


\begin{thebibliography}{73}%
\makeatletter
\providecommand \@ifxundefined [1]{%
 \@ifx{#1\undefined}
}%
\providecommand \@ifnum [1]{%
 \ifnum #1\expandafter \@firstoftwo
 \else \expandafter \@secondoftwo
 \fi
}%
\providecommand \@ifx [1]{%
 \ifx #1\expandafter \@firstoftwo
 \else \expandafter \@secondoftwo
 \fi
}%
\providecommand \natexlab [1]{#1}%
\providecommand \enquote  [1]{``#1''}%
\providecommand \bibnamefont  [1]{#1}%
\providecommand \bibfnamefont [1]{#1}%
\providecommand \citenamefont [1]{#1}%
\providecommand \href@noop [0]{\@secondoftwo}%
\providecommand \href [0]{\begingroup \@sanitize@url \@href}%
\providecommand \@href[1]{\@@startlink{#1}\@@href}%
\providecommand \@@href[1]{\endgroup#1\@@endlink}%
\providecommand \@sanitize@url [0]{\catcode `\\12\catcode `\$12\catcode `\&12\catcode `\#12\catcode `\^12\catcode `\_12\catcode `\%12\relax}%
\providecommand \@@startlink[1]{}%
\providecommand \@@endlink[0]{}%
\providecommand \url  [0]{\begingroup\@sanitize@url \@url }%
\providecommand \@url [1]{\endgroup\@href {#1}{\urlprefix }}%
\providecommand \urlprefix  [0]{URL }%
\providecommand \Eprint [0]{\href }%
\providecommand \doibase [0]{https://doi.org/}%
\providecommand \selectlanguage [0]{\@gobble}%
\providecommand \bibinfo  [0]{\@secondoftwo}%
\providecommand \bibfield  [0]{\@secondoftwo}%
\providecommand \translation [1]{[#1]}%
\providecommand \BibitemOpen [0]{}%
\providecommand \bibitemStop [0]{}%
\providecommand \bibitemNoStop [0]{.\EOS\space}%
\providecommand \EOS [0]{\spacefactor3000\relax}%
\providecommand \BibitemShut  [1]{\csname bibitem#1\endcsname}%
\let\auto@bib@innerbib\@empty
\bibitem [{\citenamefont {Henriet}\ \emph {et~al.}(2020)\citenamefont {Henriet}, \citenamefont {Beguin}, \citenamefont {Signoles}, \citenamefont {Lahaye}, \citenamefont {Browaeys}, \citenamefont {Reymond},\ and\ \citenamefont {Jurczak}}]{Henriet2020quantumcomputing}%
  \BibitemOpen
  \bibfield  {author} {\bibinfo {author} {\bibfnamefont {L.}~\bibnamefont {Henriet}}, \bibinfo {author} {\bibfnamefont {L.}~\bibnamefont {Beguin}}, \bibinfo {author} {\bibfnamefont {A.}~\bibnamefont {Signoles}}, \bibinfo {author} {\bibfnamefont {T.}~\bibnamefont {Lahaye}}, \bibinfo {author} {\bibfnamefont {A.}~\bibnamefont {Browaeys}}, \bibinfo {author} {\bibfnamefont {G.}~\bibnamefont {Reymond}},\ and\ \bibinfo {author} {\bibfnamefont {C.}~\bibnamefont {Jurczak}},\ }\bibfield  {title} {\bibinfo {title} {Quantum computing with neutral atoms},\ }\href {https://doi.org/10.22331/q-2020-09-21-327} {\bibfield  {journal} {\bibinfo  {journal} {{Quantum}}\ }\textbf {\bibinfo {volume} {4}},\ \bibinfo {pages} {327} (\bibinfo {year} {2020})}\BibitemShut {NoStop}%
\bibitem [{\citenamefont {Morgado}\ and\ \citenamefont {Whitlock}(2021)}]{doi:10.1116/5.0036562}%
  \BibitemOpen
  \bibfield  {author} {\bibinfo {author} {\bibfnamefont {M.}~\bibnamefont {Morgado}}\ and\ \bibinfo {author} {\bibfnamefont {S.}~\bibnamefont {Whitlock}},\ }\bibfield  {title} {\bibinfo {title} {Quantum simulation and computing with rydberg-interacting qubits},\ }\href {https://doi.org/10.1116/5.0036562} {\bibfield  {journal} {\bibinfo  {journal} {AVS Quantum Science}\ }\textbf {\bibinfo {volume} {3}},\ \bibinfo {pages} {023501} (\bibinfo {year} {2021})}\BibitemShut {NoStop}%
\bibitem [{\citenamefont {Cong}\ \emph {et~al.}(2022)\citenamefont {Cong}, \citenamefont {Levine}, \citenamefont {Keesling}, \citenamefont {Bluvstein}, \citenamefont {Wang},\ and\ \citenamefont {Lukin}}]{PhysRevX.12.021049}%
  \BibitemOpen
  \bibfield  {author} {\bibinfo {author} {\bibfnamefont {I.}~\bibnamefont {Cong}}, \bibinfo {author} {\bibfnamefont {H.}~\bibnamefont {Levine}}, \bibinfo {author} {\bibfnamefont {A.}~\bibnamefont {Keesling}}, \bibinfo {author} {\bibfnamefont {D.}~\bibnamefont {Bluvstein}}, \bibinfo {author} {\bibfnamefont {S.-T.}\ \bibnamefont {Wang}},\ and\ \bibinfo {author} {\bibfnamefont {M.}~\bibnamefont {Lukin}},\ }\bibfield  {title} {\bibinfo {title} {Hardware-efficient, fault-tolerant quantum computation with rydberg atoms},\ }\href {https://doi.org/10.1103/PhysRevX.12.021049} {\bibfield  {journal} {\bibinfo  {journal} {Phys. Rev. X}\ }\textbf {\bibinfo {volume} {12}},\ \bibinfo {pages} {021049} (\bibinfo {year} {2022})}\BibitemShut {NoStop}%
\bibitem [{\citenamefont {Shi}(2022)}]{Shi_2022}%
  \BibitemOpen
  \bibfield  {author} {\bibinfo {author} {\bibfnamefont {X.}~\bibnamefont {Shi}},\ }\bibfield  {title} {\bibinfo {title} {Quantum logic and entanglement by neutral rydberg atoms: methods and fidelity},\ }\href {https://doi.org/10.1088/2058-9565/ac18b8} {\bibfield  {journal} {\bibinfo  {journal} {Quantum Science and Technology}\ }\textbf {\bibinfo {volume} {7}},\ \bibinfo {pages} {023002} (\bibinfo {year} {2022})}\BibitemShut {NoStop}%
\bibitem [{\citenamefont {Shao}\ \emph {et~al.}(2024)\citenamefont {Shao}, \citenamefont {Su}, \citenamefont {Li}, \citenamefont {Nath}, \citenamefont {Wu},\ and\ \citenamefont {Li}}]{10.1063/5.0211071}%
  \BibitemOpen
  \bibfield  {author} {\bibinfo {author} {\bibfnamefont {X.}~\bibnamefont {Shao}}, \bibinfo {author} {\bibfnamefont {S.}~\bibnamefont {Su}}, \bibinfo {author} {\bibfnamefont {L.}~\bibnamefont {Li}}, \bibinfo {author} {\bibfnamefont {R.}~\bibnamefont {Nath}}, \bibinfo {author} {\bibfnamefont {J.}~\bibnamefont {Wu}},\ and\ \bibinfo {author} {\bibfnamefont {W.}~\bibnamefont {Li}},\ }\bibfield  {title} {\bibinfo {title} {Rydberg superatoms: An artificial quantum system for quantum information processing and quantum optics},\ }\href {https://doi.org/10.1063/5.0211071} {\bibfield  {journal} {\bibinfo  {journal} {Applied Physics Reviews}\ }\textbf {\bibinfo {volume} {11}},\ \bibinfo {pages} {031320} (\bibinfo {year} {2024})}\BibitemShut {NoStop}%
\bibitem [{\citenamefont {Fowler}\ \emph {et~al.}(2012)\citenamefont {Fowler}, \citenamefont {Mariantoni}, \citenamefont {Martinis},\ and\ \citenamefont {Cleland}}]{PhysRevA.86.032324}%
  \BibitemOpen
  \bibfield  {author} {\bibinfo {author} {\bibfnamefont {A.}~\bibnamefont {Fowler}}, \bibinfo {author} {\bibfnamefont {M.}~\bibnamefont {Mariantoni}}, \bibinfo {author} {\bibfnamefont {J.}~\bibnamefont {Martinis}},\ and\ \bibinfo {author} {\bibfnamefont {A.}~\bibnamefont {Cleland}},\ }\bibfield  {title} {\bibinfo {title} {Surface codes: Towards practical large-scale quantum computation},\ }\href {https://doi.org/10.1103/PhysRevA.86.032324} {\bibfield  {journal} {\bibinfo  {journal} {Phys. Rev. A}\ }\textbf {\bibinfo {volume} {86}},\ \bibinfo {pages} {032324} (\bibinfo {year} {2012})}\BibitemShut {NoStop}%
\bibitem [{\citenamefont {Wu}\ \emph {et~al.}(2022)\citenamefont {Wu}, \citenamefont {Kolkowitz}, \citenamefont {Puri},\ and\ \citenamefont {Thompson}}]{Wu2022}%
  \BibitemOpen
  \bibfield  {author} {\bibinfo {author} {\bibfnamefont {Y.}~\bibnamefont {Wu}}, \bibinfo {author} {\bibfnamefont {S.}~\bibnamefont {Kolkowitz}}, \bibinfo {author} {\bibfnamefont {S.}~\bibnamefont {Puri}},\ and\ \bibinfo {author} {\bibfnamefont {J.}~\bibnamefont {Thompson}},\ }\bibfield  {title} {\bibinfo {title} {Erasure conversion for fault-tolerant quantum computing in alkaline earth rydberg atom arrays},\ }\href {https://doi.org/10.1038/s41467-022-32094-6} {\bibfield  {journal} {\bibinfo  {journal} {Nature Communications}\ }\textbf {\bibinfo {volume} {13}},\ \bibinfo {pages} {4657} (\bibinfo {year} {2022})}\BibitemShut {NoStop}%
\bibitem [{\citenamefont {Bluvstein}\ \emph {et~al.}(2025)\citenamefont {Bluvstein}, \citenamefont {Geim}, \citenamefont {Li}, \citenamefont {Evered}, \citenamefont {Ataides}, \citenamefont {Baranes}, \citenamefont {Gu}, \citenamefont {Manovitz}, \citenamefont {Xu}, \citenamefont {Kalinowski}, \citenamefont {Majidy}, \citenamefont {Kokail}, \citenamefont {Maskara}, \citenamefont {Trapp}, \citenamefont {Stewart}, \citenamefont {Hollerith}, \citenamefont {Zhou}, \citenamefont {Gullans}, \citenamefont {Yelin}, \citenamefont {Greiner}, \citenamefont {Vuletic}, \citenamefont {Cain},\ and\ \citenamefont {Lukin}}]{bluvstein2025}%
  \BibitemOpen
  \bibfield  {author} {\bibinfo {author} {\bibfnamefont {D.}~\bibnamefont {Bluvstein}}, \bibinfo {author} {\bibfnamefont {A.}~\bibnamefont {Geim}}, \bibinfo {author} {\bibfnamefont {S.}~\bibnamefont {Li}}, \bibinfo {author} {\bibfnamefont {S.}~\bibnamefont {Evered}}, \bibinfo {author} {\bibfnamefont {J.}~\bibnamefont {Ataides}}, \bibinfo {author} {\bibfnamefont {G.}~\bibnamefont {Baranes}}, \bibinfo {author} {\bibfnamefont {A.}~\bibnamefont {Gu}}, \bibinfo {author} {\bibfnamefont {T.}~\bibnamefont {Manovitz}}, \bibinfo {author} {\bibfnamefont {M.}~\bibnamefont {Xu}}, \bibinfo {author} {\bibfnamefont {M.}~\bibnamefont {Kalinowski}}, \bibinfo {author} {\bibfnamefont {S.}~\bibnamefont {Majidy}}, \bibinfo {author} {\bibfnamefont {C.}~\bibnamefont {Kokail}}, \bibinfo {author} {\bibfnamefont {N.}~\bibnamefont {Maskara}}, \bibinfo {author} {\bibfnamefont {E.}~\bibnamefont {Trapp}}, \bibinfo {author} {\bibfnamefont {L.}~\bibnamefont {Stewart}}, \bibinfo {author} {\bibfnamefont {S.}~\bibnamefont {Hollerith}}, \bibinfo
  {author} {\bibfnamefont {H.}~\bibnamefont {Zhou}}, \bibinfo {author} {\bibfnamefont {M.}~\bibnamefont {Gullans}}, \bibinfo {author} {\bibfnamefont {S.}~\bibnamefont {Yelin}}, \bibinfo {author} {\bibfnamefont {M.}~\bibnamefont {Greiner}}, \bibinfo {author} {\bibfnamefont {V.}~\bibnamefont {Vuletic}}, \bibinfo {author} {\bibfnamefont {M.}~\bibnamefont {Cain}},\ and\ \bibinfo {author} {\bibfnamefont {M.}~\bibnamefont {Lukin}},\ }\href@noop {} {\bibinfo {title} {Architectural mechanisms of a universal fault-tolerant quantum computer}} (\bibinfo {year} {2025}),\ \Eprint {https://arxiv.org/abs/2506.20661} {arXiv:2506.20661} \BibitemShut {NoStop}%
\bibitem [{\citenamefont {Goerz}\ \emph {et~al.}(2014)\citenamefont {Goerz}, \citenamefont {Halperin}, \citenamefont {Aytac}, \citenamefont {Koch},\ and\ \citenamefont {Whaley}}]{PhysRevA.90.032329}%
  \BibitemOpen
  \bibfield  {author} {\bibinfo {author} {\bibfnamefont {M.}~\bibnamefont {Goerz}}, \bibinfo {author} {\bibfnamefont {E.}~\bibnamefont {Halperin}}, \bibinfo {author} {\bibfnamefont {J.}~\bibnamefont {Aytac}}, \bibinfo {author} {\bibfnamefont {C.}~\bibnamefont {Koch}},\ and\ \bibinfo {author} {\bibfnamefont {K.}~\bibnamefont {Whaley}},\ }\bibfield  {title} {\bibinfo {title} {Robustness of high-fidelity rydberg gates with single-site addressability},\ }\href {https://doi.org/10.1103/PhysRevA.90.032329} {\bibfield  {journal} {\bibinfo  {journal} {Phys. Rev. A}\ }\textbf {\bibinfo {volume} {90}},\ \bibinfo {pages} {032329} (\bibinfo {year} {2014})}\BibitemShut {NoStop}%
\bibitem [{\citenamefont {Poggi}\ \emph {et~al.}(2024)\citenamefont {Poggi}, \citenamefont {De~Chiara}, \citenamefont {Campbell},\ and\ \citenamefont {Kiely}}]{PhysRevLett.132.193801}%
  \BibitemOpen
  \bibfield  {author} {\bibinfo {author} {\bibfnamefont {P.}~\bibnamefont {Poggi}}, \bibinfo {author} {\bibfnamefont {G.}~\bibnamefont {De~Chiara}}, \bibinfo {author} {\bibfnamefont {S.}~\bibnamefont {Campbell}},\ and\ \bibinfo {author} {\bibfnamefont {A.}~\bibnamefont {Kiely}},\ }\bibfield  {title} {\bibinfo {title} {Universally robust quantum control},\ }\href {https://doi.org/10.1103/PhysRevLett.132.193801} {\bibfield  {journal} {\bibinfo  {journal} {Phys. Rev. Lett.}\ }\textbf {\bibinfo {volume} {132}},\ \bibinfo {pages} {193801} (\bibinfo {year} {2024})}\BibitemShut {NoStop}%
\bibitem [{\citenamefont {Fromonteil}\ \emph {et~al.}(2023)\citenamefont {Fromonteil}, \citenamefont {Bluvstein},\ and\ \citenamefont {Pichler}}]{PRXQuantum.4.020335}%
  \BibitemOpen
  \bibfield  {author} {\bibinfo {author} {\bibfnamefont {C.}~\bibnamefont {Fromonteil}}, \bibinfo {author} {\bibfnamefont {D.}~\bibnamefont {Bluvstein}},\ and\ \bibinfo {author} {\bibfnamefont {H.}~\bibnamefont {Pichler}},\ }\bibfield  {title} {\bibinfo {title} {Protocols for rydberg entangling gates featuring robustness against quasistatic errors},\ }\href {https://doi.org/10.1103/PRXQuantum.4.020335} {\bibfield  {journal} {\bibinfo  {journal} {PRX Quantum}\ }\textbf {\bibinfo {volume} {4}},\ \bibinfo {pages} {020335} (\bibinfo {year} {2023})}\BibitemShut {NoStop}%
\bibitem [{\citenamefont {Noiri}\ \emph {et~al.}(2022)\citenamefont {Noiri}, \citenamefont {Takeda}, \citenamefont {Nakajima}, \citenamefont {Kobayashi}, \citenamefont {Sammak}, \citenamefont {Scappucci},\ and\ \citenamefont {Tarucha}}]{Noiri2022}%
  \BibitemOpen
  \bibfield  {author} {\bibinfo {author} {\bibfnamefont {A.}~\bibnamefont {Noiri}}, \bibinfo {author} {\bibfnamefont {K.}~\bibnamefont {Takeda}}, \bibinfo {author} {\bibfnamefont {T.}~\bibnamefont {Nakajima}}, \bibinfo {author} {\bibfnamefont {T.}~\bibnamefont {Kobayashi}}, \bibinfo {author} {\bibfnamefont {A.}~\bibnamefont {Sammak}}, \bibinfo {author} {\bibfnamefont {G.}~\bibnamefont {Scappucci}},\ and\ \bibinfo {author} {\bibfnamefont {S.}~\bibnamefont {Tarucha}},\ }\bibfield  {title} {\bibinfo {title} {Fast universal quantum gate above the fault-tolerance threshold in silicon},\ }\href {https://doi.org/10.1038/s41586-021-04182-y} {\bibfield  {journal} {\bibinfo  {journal} {Nature}\ }\textbf {\bibinfo {volume} {601}},\ \bibinfo {pages} {338} (\bibinfo {year} {2022})}\BibitemShut {NoStop}%
\bibitem [{\citenamefont {Buchemmavari}\ \emph {et~al.}(2024)\citenamefont {Buchemmavari}, \citenamefont {Omanakuttan}, \citenamefont {Jau},\ and\ \citenamefont {Deutsch}}]{PhysRevA.109.012615}%
  \BibitemOpen
  \bibfield  {author} {\bibinfo {author} {\bibfnamefont {V.}~\bibnamefont {Buchemmavari}}, \bibinfo {author} {\bibfnamefont {S.}~\bibnamefont {Omanakuttan}}, \bibinfo {author} {\bibfnamefont {Y.}~\bibnamefont {Jau}},\ and\ \bibinfo {author} {\bibfnamefont {I.}~\bibnamefont {Deutsch}},\ }\bibfield  {title} {\bibinfo {title} {Entangling quantum logic gates in neutral atoms via the microwave-driven spin-flip blockade},\ }\href {https://doi.org/10.1103/PhysRevA.109.012615} {\bibfield  {journal} {\bibinfo  {journal} {Phys. Rev. A}\ }\textbf {\bibinfo {volume} {109}},\ \bibinfo {pages} {012615} (\bibinfo {year} {2024})}\BibitemShut {NoStop}%
\bibitem [{\citenamefont {Jandura}\ and\ \citenamefont {Pupillo}(2022)}]{Jandura2022timeoptimaltwothree}%
  \BibitemOpen
  \bibfield  {author} {\bibinfo {author} {\bibfnamefont {S.}~\bibnamefont {Jandura}}\ and\ \bibinfo {author} {\bibfnamefont {G.}~\bibnamefont {Pupillo}},\ }\bibfield  {title} {\bibinfo {title} {Time-{O}ptimal {T}wo- and {T}hree-{Q}ubit {G}ates for {R}ydberg {A}toms},\ }\href {https://doi.org/10.22331/q-2022-05-13-712} {\bibfield  {journal} {\bibinfo  {journal} {{Quantum}}\ }\textbf {\bibinfo {volume} {6}},\ \bibinfo {pages} {712} (\bibinfo {year} {2022})}\BibitemShut {NoStop}%
\bibitem [{\citenamefont {Mohan}\ \emph {et~al.}(2023)\citenamefont {Mohan}, \citenamefont {de~Keijzer},\ and\ \citenamefont {Kokkelmans}}]{PhysRevResearch.5.033052}%
  \BibitemOpen
  \bibfield  {author} {\bibinfo {author} {\bibfnamefont {M.}~\bibnamefont {Mohan}}, \bibinfo {author} {\bibfnamefont {R.}~\bibnamefont {de~Keijzer}},\ and\ \bibinfo {author} {\bibfnamefont {S.}~\bibnamefont {Kokkelmans}},\ }\bibfield  {title} {\bibinfo {title} {Robust control and optimal rydberg states for neutral atom two-qubit gates},\ }\href {https://doi.org/10.1103/PhysRevResearch.5.033052} {\bibfield  {journal} {\bibinfo  {journal} {Phys. Rev. Res.}\ }\textbf {\bibinfo {volume} {5}},\ \bibinfo {pages} {033052} (\bibinfo {year} {2023})}\BibitemShut {NoStop}%
\bibitem [{\citenamefont {Jandura}\ \emph {et~al.}(2023)\citenamefont {Jandura}, \citenamefont {Thompson},\ and\ \citenamefont {Pupillo}}]{PRXQuantum.4.020336}%
  \BibitemOpen
  \bibfield  {author} {\bibinfo {author} {\bibfnamefont {S.}~\bibnamefont {Jandura}}, \bibinfo {author} {\bibfnamefont {J.}~\bibnamefont {Thompson}},\ and\ \bibinfo {author} {\bibfnamefont {G.}~\bibnamefont {Pupillo}},\ }\bibfield  {title} {\bibinfo {title} {Optimizing rydberg gates for logical-qubit performance},\ }\href {https://doi.org/10.1103/PRXQuantum.4.020336} {\bibfield  {journal} {\bibinfo  {journal} {PRX Quantum}\ }\textbf {\bibinfo {volume} {4}},\ \bibinfo {pages} {020336} (\bibinfo {year} {2023})}\BibitemShut {NoStop}%
\bibitem [{\citenamefont {Keating}\ \emph {et~al.}(2015)\citenamefont {Keating}, \citenamefont {Cook}, \citenamefont {Hankin}, \citenamefont {Jau}, \citenamefont {Biedermann},\ and\ \citenamefont {Deutsch}}]{PhysRevA.91.012337}%
  \BibitemOpen
  \bibfield  {author} {\bibinfo {author} {\bibfnamefont {T.}~\bibnamefont {Keating}}, \bibinfo {author} {\bibfnamefont {R.}~\bibnamefont {Cook}}, \bibinfo {author} {\bibfnamefont {A.}~\bibnamefont {Hankin}}, \bibinfo {author} {\bibfnamefont {Y.}~\bibnamefont {Jau}}, \bibinfo {author} {\bibfnamefont {G.}~\bibnamefont {Biedermann}},\ and\ \bibinfo {author} {\bibfnamefont {I.}~\bibnamefont {Deutsch}},\ }\bibfield  {title} {\bibinfo {title} {Robust quantum logic in neutral atoms via adiabatic rydberg dressing},\ }\href {https://doi.org/10.1103/PhysRevA.91.012337} {\bibfield  {journal} {\bibinfo  {journal} {Phys. Rev. A}\ }\textbf {\bibinfo {volume} {91}},\ \bibinfo {pages} {012337} (\bibinfo {year} {2015})}\BibitemShut {NoStop}%
\bibitem [{\citenamefont {Khazali}\ and\ \citenamefont {M\o{}lmer}(2020)}]{PhysRevX.10.021054}%
  \BibitemOpen
  \bibfield  {author} {\bibinfo {author} {\bibfnamefont {M.}~\bibnamefont {Khazali}}\ and\ \bibinfo {author} {\bibfnamefont {K.}~\bibnamefont {M\o{}lmer}},\ }\bibfield  {title} {\bibinfo {title} {Fast multiqubit gates by adiabatic evolution in interacting excited-state manifolds of rydberg atoms and superconducting circuits},\ }\href {https://doi.org/10.1103/PhysRevX.10.021054} {\bibfield  {journal} {\bibinfo  {journal} {Phys. Rev. X}\ }\textbf {\bibinfo {volume} {10}},\ \bibinfo {pages} {021054} (\bibinfo {year} {2020})}\BibitemShut {NoStop}%
\bibitem [{\citenamefont {Saffman}\ \emph {et~al.}(2020)\citenamefont {Saffman}, \citenamefont {Beterov}, \citenamefont {Dalal}, \citenamefont {P\'aez},\ and\ \citenamefont {Sanders}}]{PhysRevA.101.062309}%
  \BibitemOpen
  \bibfield  {author} {\bibinfo {author} {\bibfnamefont {M.}~\bibnamefont {Saffman}}, \bibinfo {author} {\bibfnamefont {I.}~\bibnamefont {Beterov}}, \bibinfo {author} {\bibfnamefont {A.}~\bibnamefont {Dalal}}, \bibinfo {author} {\bibfnamefont {E.}~\bibnamefont {P\'aez}},\ and\ \bibinfo {author} {\bibfnamefont {B.}~\bibnamefont {Sanders}},\ }\bibfield  {title} {\bibinfo {title} {Symmetric rydberg controlled-$z$ gates with adiabatic pulses},\ }\href {https://doi.org/10.1103/PhysRevA.101.062309} {\bibfield  {journal} {\bibinfo  {journal} {Phys. Rev. A}\ }\textbf {\bibinfo {volume} {101}},\ \bibinfo {pages} {062309} (\bibinfo {year} {2020})}\BibitemShut {NoStop}%
\bibitem [{\citenamefont {Mitra}\ \emph {et~al.}(2020)\citenamefont {Mitra}, \citenamefont {Martin}, \citenamefont {Biedermann}, \citenamefont {Marino}, \citenamefont {Poggi},\ and\ \citenamefont {Deutsch}}]{robust_gate_Mitra2020}%
  \BibitemOpen
  \bibfield  {author} {\bibinfo {author} {\bibfnamefont {A.}~\bibnamefont {Mitra}}, \bibinfo {author} {\bibfnamefont {M.}~\bibnamefont {Martin}}, \bibinfo {author} {\bibfnamefont {G.}~\bibnamefont {Biedermann}}, \bibinfo {author} {\bibfnamefont {A.}~\bibnamefont {Marino}}, \bibinfo {author} {\bibfnamefont {P.}~\bibnamefont {Poggi}},\ and\ \bibinfo {author} {\bibfnamefont {I.}~\bibnamefont {Deutsch}},\ }\bibfield  {title} {\bibinfo {title} {Robust m\o{}lmer-s\o{}rensen gate for neutral atoms using rapid adiabatic rydberg dressing},\ }\href {https://doi.org/10.1103/PhysRevA.101.030301} {\bibfield  {journal} {\bibinfo  {journal} {Phys. Rev. A}\ }\textbf {\bibinfo {volume} {101}},\ \bibinfo {pages} {030301} (\bibinfo {year} {2020})}\BibitemShut {NoStop}%
\bibitem [{\citenamefont {Giudici}\ \emph {et~al.}(2025)\citenamefont {Giudici}, \citenamefont {Veroni}, \citenamefont {Giudice}, \citenamefont {Pichler},\ and\ \citenamefont {Zeiher}}]{5d8p-3hm1}%
  \BibitemOpen
  \bibfield  {author} {\bibinfo {author} {\bibfnamefont {G.}~\bibnamefont {Giudici}}, \bibinfo {author} {\bibfnamefont {S.}~\bibnamefont {Veroni}}, \bibinfo {author} {\bibfnamefont {G.}~\bibnamefont {Giudice}}, \bibinfo {author} {\bibfnamefont {H.}~\bibnamefont {Pichler}},\ and\ \bibinfo {author} {\bibfnamefont {J.}~\bibnamefont {Zeiher}},\ }\bibfield  {title} {\bibinfo {title} {Fast entangling gates for rydberg atoms via resonant dipole-dipole interaction},\ }\href {https://doi.org/10.1103/5d8p-3hm1} {\bibfield  {journal} {\bibinfo  {journal} {PRX Quantum}\ }\textbf {\bibinfo {volume} {6}},\ \bibinfo {pages} {030308} (\bibinfo {year} {2025})}\BibitemShut {NoStop}%
\bibitem [{\citenamefont {Chew}\ \emph {et~al.}(2022)\citenamefont {Chew}, \citenamefont {Tomita}, \citenamefont {Mahesh}, \citenamefont {Sugawa}, \citenamefont {de~L{\'e}s{\'e}leuc},\ and\ \citenamefont {Ohmori}}]{Chew2022}%
  \BibitemOpen
  \bibfield  {author} {\bibinfo {author} {\bibfnamefont {Y.}~\bibnamefont {Chew}}, \bibinfo {author} {\bibfnamefont {T.}~\bibnamefont {Tomita}}, \bibinfo {author} {\bibfnamefont {T.}~\bibnamefont {Mahesh}}, \bibinfo {author} {\bibfnamefont {S.}~\bibnamefont {Sugawa}}, \bibinfo {author} {\bibfnamefont {S.}~\bibnamefont {de~L{\'e}s{\'e}leuc}},\ and\ \bibinfo {author} {\bibfnamefont {K.}~\bibnamefont {Ohmori}},\ }\bibfield  {title} {\bibinfo {title} {Ultrafast energy exchange between two single rydberg atoms on a nanosecond timescale},\ }\href {https://doi.org/10.1038/s41566-022-01047-2} {\bibfield  {journal} {\bibinfo  {journal} {Nature Photonics}\ }\textbf {\bibinfo {volume} {16}},\ \bibinfo {pages} {724} (\bibinfo {year} {2022})}\BibitemShut {NoStop}%
\bibitem [{\citenamefont {Goerz}\ \emph {et~al.}(2011)\citenamefont {Goerz}, \citenamefont {Calarco},\ and\ \citenamefont {Koch}}]{Goerz_2011}%
  \BibitemOpen
  \bibfield  {author} {\bibinfo {author} {\bibfnamefont {M.}~\bibnamefont {Goerz}}, \bibinfo {author} {\bibfnamefont {T.}~\bibnamefont {Calarco}},\ and\ \bibinfo {author} {\bibfnamefont {C.}~\bibnamefont {Koch}},\ }\bibfield  {title} {\bibinfo {title} {The quantum speed limit of optimal controlled phasegates for trapped neutral atoms},\ }\href {https://doi.org/10.1088/0953-4075/44/15/154011} {\bibfield  {journal} {\bibinfo  {journal} {Journal of Physics B: Atomic, Molecular and Optical Physics}\ }\textbf {\bibinfo {volume} {44}},\ \bibinfo {pages} {154011} (\bibinfo {year} {2011})}\BibitemShut {NoStop}%
\bibitem [{\citenamefont {Song}\ \emph {et~al.}(2024)\citenamefont {Song}, \citenamefont {Wei}, \citenamefont {Xu}, \citenamefont {Yan}, \citenamefont {Feng}, \citenamefont {Su},\ and\ \citenamefont {Chen}}]{PhysRevA.109.022613}%
  \BibitemOpen
  \bibfield  {author} {\bibinfo {author} {\bibfnamefont {P.}~\bibnamefont {Song}}, \bibinfo {author} {\bibfnamefont {J.}~\bibnamefont {Wei}}, \bibinfo {author} {\bibfnamefont {P.}~\bibnamefont {Xu}}, \bibinfo {author} {\bibfnamefont {L.}~\bibnamefont {Yan}}, \bibinfo {author} {\bibfnamefont {M.}~\bibnamefont {Feng}}, \bibinfo {author} {\bibfnamefont {S.}~\bibnamefont {Su}},\ and\ \bibinfo {author} {\bibfnamefont {G.}~\bibnamefont {Chen}},\ }\bibfield  {title} {\bibinfo {title} {Fast realization of high-fidelity nonadiabatic holonomic quantum gates with a time-optimal-control technique in rydberg atoms},\ }\href {https://doi.org/10.1103/PhysRevA.109.022613} {\bibfield  {journal} {\bibinfo  {journal} {Phys. Rev. A}\ }\textbf {\bibinfo {volume} {109}},\ \bibinfo {pages} {022613} (\bibinfo {year} {2024})}\BibitemShut {NoStop}%
\bibitem [{\citenamefont {Basilewitsch}\ \emph {et~al.}(2024)\citenamefont {Basilewitsch}, \citenamefont {Dlaska},\ and\ \citenamefont {Lechner}}]{PhysRevResearch.6.023026}%
  \BibitemOpen
  \bibfield  {author} {\bibinfo {author} {\bibfnamefont {D.}~\bibnamefont {Basilewitsch}}, \bibinfo {author} {\bibfnamefont {C.}~\bibnamefont {Dlaska}},\ and\ \bibinfo {author} {\bibfnamefont {W.}~\bibnamefont {Lechner}},\ }\bibfield  {title} {\bibinfo {title} {Comparing planar quantum computing platforms at the quantum speed limit},\ }\href {https://doi.org/10.1103/PhysRevResearch.6.023026} {\bibfield  {journal} {\bibinfo  {journal} {Phys. Rev. Res.}\ }\textbf {\bibinfo {volume} {6}},\ \bibinfo {pages} {023026} (\bibinfo {year} {2024})}\BibitemShut {NoStop}%
\bibitem [{\citenamefont {Doultsinos}\ and\ \citenamefont {Petrosyan}(2025)}]{PhysRevResearch.7.023246}%
  \BibitemOpen
  \bibfield  {author} {\bibinfo {author} {\bibfnamefont {G.}~\bibnamefont {Doultsinos}}\ and\ \bibinfo {author} {\bibfnamefont {D.}~\bibnamefont {Petrosyan}},\ }\bibfield  {title} {\bibinfo {title} {Quantum gates between distant atoms mediated by a rydberg excitation antiferromagnet},\ }\href {https://doi.org/10.1103/PhysRevResearch.7.023246} {\bibfield  {journal} {\bibinfo  {journal} {Phys. Rev. Res.}\ }\textbf {\bibinfo {volume} {7}},\ \bibinfo {pages} {023246} (\bibinfo {year} {2025})}\BibitemShut {NoStop}%
\bibitem [{\citenamefont {Fu}\ \emph {et~al.}(2022)\citenamefont {Fu}, \citenamefont {Xu}, \citenamefont {Sun}, \citenamefont {Liu}, \citenamefont {He}, \citenamefont {Li}, \citenamefont {Liu}, \citenamefont {Li}, \citenamefont {Wang}, \citenamefont {Liu},\ and\ \citenamefont {Zhan}}]{PhysRevA.105.042430}%
  \BibitemOpen
  \bibfield  {author} {\bibinfo {author} {\bibfnamefont {Z.}~\bibnamefont {Fu}}, \bibinfo {author} {\bibfnamefont {P.}~\bibnamefont {Xu}}, \bibinfo {author} {\bibfnamefont {Y.}~\bibnamefont {Sun}}, \bibinfo {author} {\bibfnamefont {Y.}~\bibnamefont {Liu}}, \bibinfo {author} {\bibfnamefont {X.}~\bibnamefont {He}}, \bibinfo {author} {\bibfnamefont {X.}~\bibnamefont {Li}}, \bibinfo {author} {\bibfnamefont {M.}~\bibnamefont {Liu}}, \bibinfo {author} {\bibfnamefont {R.}~\bibnamefont {Li}}, \bibinfo {author} {\bibfnamefont {J.}~\bibnamefont {Wang}}, \bibinfo {author} {\bibfnamefont {L.}~\bibnamefont {Liu}},\ and\ \bibinfo {author} {\bibfnamefont {M.}~\bibnamefont {Zhan}},\ }\bibfield  {title} {\bibinfo {title} {High-fidelity entanglement of neutral atoms via a rydberg-mediated single-modulated-pulse controlled-phase gate},\ }\href {https://doi.org/10.1103/PhysRevA.105.042430} {\bibfield  {journal} {\bibinfo  {journal} {Phys. Rev. A}\ }\textbf {\bibinfo {volume} {105}},\ \bibinfo {pages} {042430} (\bibinfo {year}
  {2022})}\BibitemShut {NoStop}%
\bibitem [{\citenamefont {Liu}\ \emph {et~al.}(2019)\citenamefont {Liu}, \citenamefont {Song}, \citenamefont {Xue}, \citenamefont {Wang},\ and\ \citenamefont {Yung}}]{PhysRevLett.123.100501}%
  \BibitemOpen
  \bibfield  {author} {\bibinfo {author} {\bibfnamefont {B.}~\bibnamefont {Liu}}, \bibinfo {author} {\bibfnamefont {X.}~\bibnamefont {Song}}, \bibinfo {author} {\bibfnamefont {Z.}~\bibnamefont {Xue}}, \bibinfo {author} {\bibfnamefont {X.}~\bibnamefont {Wang}},\ and\ \bibinfo {author} {\bibfnamefont {M.}~\bibnamefont {Yung}},\ }\bibfield  {title} {\bibinfo {title} {Plug-and-play approach to nonadiabatic geometric quantum gates},\ }\href {https://doi.org/10.1103/PhysRevLett.123.100501} {\bibfield  {journal} {\bibinfo  {journal} {Phys. Rev. Lett.}\ }\textbf {\bibinfo {volume} {123}},\ \bibinfo {pages} {100501} (\bibinfo {year} {2019})}\BibitemShut {NoStop}%
\bibitem [{\citenamefont {Jiang}\ \emph {et~al.}(2023)\citenamefont {Jiang}, \citenamefont {Scott}, \citenamefont {Friesen},\ and\ \citenamefont {Saffman}}]{PhysRevA.107.042611}%
  \BibitemOpen
  \bibfield  {author} {\bibinfo {author} {\bibfnamefont {X.}~\bibnamefont {Jiang}}, \bibinfo {author} {\bibfnamefont {J.}~\bibnamefont {Scott}}, \bibinfo {author} {\bibfnamefont {M.}~\bibnamefont {Friesen}},\ and\ \bibinfo {author} {\bibfnamefont {M.}~\bibnamefont {Saffman}},\ }\bibfield  {title} {\bibinfo {title} {Sensitivity of quantum gate fidelity to laser phase and intensity noise},\ }\href {https://doi.org/10.1103/PhysRevA.107.042611} {\bibfield  {journal} {\bibinfo  {journal} {Phys. Rev. A}\ }\textbf {\bibinfo {volume} {107}},\ \bibinfo {pages} {042611} (\bibinfo {year} {2023})}\BibitemShut {NoStop}%
\bibitem [{\citenamefont {de~L\'es\'eleuc}\ \emph {et~al.}(2018)\citenamefont {de~L\'es\'eleuc}, \citenamefont {Barredo}, \citenamefont {Lienhard}, \citenamefont {Browaeys},\ and\ \citenamefont {Lahaye}}]{PhysRevA.97.053803}%
  \BibitemOpen
  \bibfield  {author} {\bibinfo {author} {\bibfnamefont {S.}~\bibnamefont {de~L\'es\'eleuc}}, \bibinfo {author} {\bibfnamefont {D.}~\bibnamefont {Barredo}}, \bibinfo {author} {\bibfnamefont {V.}~\bibnamefont {Lienhard}}, \bibinfo {author} {\bibfnamefont {A.}~\bibnamefont {Browaeys}},\ and\ \bibinfo {author} {\bibfnamefont {T.}~\bibnamefont {Lahaye}},\ }\bibfield  {title} {\bibinfo {title} {Analysis of imperfections in the coherent optical excitation of single atoms to rydberg states},\ }\href {https://doi.org/10.1103/PhysRevA.97.053803} {\bibfield  {journal} {\bibinfo  {journal} {Phys. Rev. A}\ }\textbf {\bibinfo {volume} {97}},\ \bibinfo {pages} {053803} (\bibinfo {year} {2018})}\BibitemShut {NoStop}%
\bibitem [{\citenamefont {Evered}\ \emph {et~al.}(2023)\citenamefont {Evered}, \citenamefont {Bluvstein}, \citenamefont {Kalinowski}, \citenamefont {Ebadi}, \citenamefont {Manovitz}, \citenamefont {Zhou}, \citenamefont {Li}, \citenamefont {Geim}, \citenamefont {Wang}, \citenamefont {Maskara}, \citenamefont {Levine}, \citenamefont {Semeghini}, \citenamefont {Greiner}, \citenamefont {Vuleti{\'{c}}},\ and\ \citenamefont {Lukin}}]{Evered2023}%
  \BibitemOpen
  \bibfield  {author} {\bibinfo {author} {\bibfnamefont {S.}~\bibnamefont {Evered}}, \bibinfo {author} {\bibfnamefont {D.}~\bibnamefont {Bluvstein}}, \bibinfo {author} {\bibfnamefont {M.}~\bibnamefont {Kalinowski}}, \bibinfo {author} {\bibfnamefont {S.}~\bibnamefont {Ebadi}}, \bibinfo {author} {\bibfnamefont {T.}~\bibnamefont {Manovitz}}, \bibinfo {author} {\bibfnamefont {H.}~\bibnamefont {Zhou}}, \bibinfo {author} {\bibfnamefont {S.}~\bibnamefont {Li}}, \bibinfo {author} {\bibfnamefont {A.}~\bibnamefont {Geim}}, \bibinfo {author} {\bibfnamefont {T.}~\bibnamefont {Wang}}, \bibinfo {author} {\bibfnamefont {N.}~\bibnamefont {Maskara}}, \bibinfo {author} {\bibfnamefont {H.}~\bibnamefont {Levine}}, \bibinfo {author} {\bibfnamefont {G.}~\bibnamefont {Semeghini}}, \bibinfo {author} {\bibfnamefont {M.}~\bibnamefont {Greiner}}, \bibinfo {author} {\bibfnamefont {V.}~\bibnamefont {Vuleti{\'{c}}}},\ and\ \bibinfo {author} {\bibfnamefont {M.}~\bibnamefont {Lukin}},\ }\bibfield  {title} {\bibinfo {title} {High-fidelity
  parallel entangling gates on a neutral-atom quantum computer},\ }\href {https://doi.org/10.1038/s41586-023-06481-y} {\bibfield  {journal} {\bibinfo  {journal} {Nature}\ }\textbf {\bibinfo {volume} {622}},\ \bibinfo {pages} {268} (\bibinfo {year} {2023})}\BibitemShut {NoStop}%
\bibitem [{\citenamefont {Wang}\ \emph {et~al.}(2025)\citenamefont {Wang}, \citenamefont {Xu}, \citenamefont {Li}, \citenamefont {Vuleti\ifmmode~\acute{c}\else \'{c}\fi{}},\ and\ \citenamefont {Cappellaro}}]{PhysRevApplied.23.024072}%
  \BibitemOpen
  \bibfield  {author} {\bibinfo {author} {\bibfnamefont {G.}~\bibnamefont {Wang}}, \bibinfo {author} {\bibfnamefont {W.}~\bibnamefont {Xu}}, \bibinfo {author} {\bibfnamefont {C.}~\bibnamefont {Li}}, \bibinfo {author} {\bibfnamefont {V.}~\bibnamefont {Vuleti\ifmmode~\acute{c}\else \'{c}\fi{}}},\ and\ \bibinfo {author} {\bibfnamefont {P.}~\bibnamefont {Cappellaro}},\ }\bibfield  {title} {\bibinfo {title} {Individual-atom control in an array through phase modulation},\ }\href {https://doi.org/10.1103/PhysRevApplied.23.024072} {\bibfield  {journal} {\bibinfo  {journal} {Phys. Rev. Appl.}\ }\textbf {\bibinfo {volume} {23}},\ \bibinfo {pages} {024072} (\bibinfo {year} {2025})}\BibitemShut {NoStop}%
\bibitem [{\citenamefont {McDonnell}\ \emph {et~al.}(2022)\citenamefont {McDonnell}, \citenamefont {Keary},\ and\ \citenamefont {Pritchard}}]{PhysRevLett.129.200501}%
  \BibitemOpen
  \bibfield  {author} {\bibinfo {author} {\bibfnamefont {K.}~\bibnamefont {McDonnell}}, \bibinfo {author} {\bibfnamefont {L.}~\bibnamefont {Keary}},\ and\ \bibinfo {author} {\bibfnamefont {J.}~\bibnamefont {Pritchard}},\ }\bibfield  {title} {\bibinfo {title} {Demonstration of a quantum gate using electromagnetically induced transparency},\ }\href {https://doi.org/10.1103/PhysRevLett.129.200501} {\bibfield  {journal} {\bibinfo  {journal} {Phys. Rev. Lett.}\ }\textbf {\bibinfo {volume} {129}},\ \bibinfo {pages} {200501} (\bibinfo {year} {2022})}\BibitemShut {NoStop}%
\bibitem [{\citenamefont {Tsai}\ \emph {et~al.}(2025)\citenamefont {Tsai}, \citenamefont {Sun}, \citenamefont {Shaw}, \citenamefont {Finkelstein},\ and\ \citenamefont {Endres}}]{PRXQuantum.6.010331}%
  \BibitemOpen
  \bibfield  {author} {\bibinfo {author} {\bibfnamefont {R.}~\bibnamefont {Tsai}}, \bibinfo {author} {\bibfnamefont {X.}~\bibnamefont {Sun}}, \bibinfo {author} {\bibfnamefont {A.}~\bibnamefont {Shaw}}, \bibinfo {author} {\bibfnamefont {R.}~\bibnamefont {Finkelstein}},\ and\ \bibinfo {author} {\bibfnamefont {M.}~\bibnamefont {Endres}},\ }\bibfield  {title} {\bibinfo {title} {Benchmarking and fidelity response theory of high-fidelity rydberg entangling gates},\ }\href {https://doi.org/10.1103/PRXQuantum.6.010331} {\bibfield  {journal} {\bibinfo  {journal} {PRX Quantum}\ }\textbf {\bibinfo {volume} {6}},\ \bibinfo {pages} {010331} (\bibinfo {year} {2025})}\BibitemShut {NoStop}%
\bibitem [{\citenamefont {Ga{\"e}tan}\ \emph {et~al.}(2009)\citenamefont {Ga{\"e}tan}, \citenamefont {Miroshnychenko}, \citenamefont {Wilk}, \citenamefont {Chotia}, \citenamefont {Viteau}, \citenamefont {Comparat}, \citenamefont {Pillet}, \citenamefont {Browaeys},\ and\ \citenamefont {Grangier}}]{Gaetan2009}%
  \BibitemOpen
  \bibfield  {author} {\bibinfo {author} {\bibfnamefont {A.}~\bibnamefont {Ga{\"e}tan}}, \bibinfo {author} {\bibfnamefont {Y.}~\bibnamefont {Miroshnychenko}}, \bibinfo {author} {\bibfnamefont {T.}~\bibnamefont {Wilk}}, \bibinfo {author} {\bibfnamefont {A.}~\bibnamefont {Chotia}}, \bibinfo {author} {\bibfnamefont {M.}~\bibnamefont {Viteau}}, \bibinfo {author} {\bibfnamefont {D.}~\bibnamefont {Comparat}}, \bibinfo {author} {\bibfnamefont {P.}~\bibnamefont {Pillet}}, \bibinfo {author} {\bibfnamefont {A.}~\bibnamefont {Browaeys}},\ and\ \bibinfo {author} {\bibfnamefont {P.}~\bibnamefont {Grangier}},\ }\bibfield  {title} {\bibinfo {title} {Observation of collective excitation of two individual atoms in the rydberg blockade regime},\ }\href {https://doi.org/10.1038/nphys1183} {\bibfield  {journal} {\bibinfo  {journal} {Nature Physics}\ }\textbf {\bibinfo {volume} {5}},\ \bibinfo {pages} {115} (\bibinfo {year} {2009})}\BibitemShut {NoStop}%
\bibitem [{\citenamefont {Zeng}\ \emph {et~al.}(2017)\citenamefont {Zeng}, \citenamefont {Xu}, \citenamefont {He}, \citenamefont {Liu}, \citenamefont {Liu}, \citenamefont {Wang}, \citenamefont {Papoular}, \citenamefont {Shlyapnikov},\ and\ \citenamefont {Zhan}}]{PhysRevLett.119.160502}%
  \BibitemOpen
  \bibfield  {author} {\bibinfo {author} {\bibfnamefont {Y.}~\bibnamefont {Zeng}}, \bibinfo {author} {\bibfnamefont {P.}~\bibnamefont {Xu}}, \bibinfo {author} {\bibfnamefont {X.}~\bibnamefont {He}}, \bibinfo {author} {\bibfnamefont {Y.}~\bibnamefont {Liu}}, \bibinfo {author} {\bibfnamefont {M.}~\bibnamefont {Liu}}, \bibinfo {author} {\bibfnamefont {J.}~\bibnamefont {Wang}}, \bibinfo {author} {\bibfnamefont {D.}~\bibnamefont {Papoular}}, \bibinfo {author} {\bibfnamefont {G.}~\bibnamefont {Shlyapnikov}},\ and\ \bibinfo {author} {\bibfnamefont {M.}~\bibnamefont {Zhan}},\ }\bibfield  {title} {\bibinfo {title} {Entangling two individual atoms of different isotopes via rydberg blockade},\ }\href {https://doi.org/10.1103/PhysRevLett.119.160502} {\bibfield  {journal} {\bibinfo  {journal} {Phys. Rev. Lett.}\ }\textbf {\bibinfo {volume} {119}},\ \bibinfo {pages} {160502} (\bibinfo {year} {2017})}\BibitemShut {NoStop}%
\bibitem [{\citenamefont {Reetz-Lamour}\ \emph {et~al.}(2008{\natexlab{a}})\citenamefont {Reetz-Lamour}, \citenamefont {Amthor}, \citenamefont {Deiglmayr},\ and\ \citenamefont {Weidem\"uller}}]{PhysRevLett.100.253001}%
  \BibitemOpen
  \bibfield  {author} {\bibinfo {author} {\bibfnamefont {M.}~\bibnamefont {Reetz-Lamour}}, \bibinfo {author} {\bibfnamefont {T.}~\bibnamefont {Amthor}}, \bibinfo {author} {\bibfnamefont {J.}~\bibnamefont {Deiglmayr}},\ and\ \bibinfo {author} {\bibfnamefont {M.}~\bibnamefont {Weidem\"uller}},\ }\bibfield  {title} {\bibinfo {title} {Rabi oscillations and excitation trapping in the coherent excitation of a mesoscopic frozen rydberg gas},\ }\href {https://doi.org/10.1103/PhysRevLett.100.253001} {\bibfield  {journal} {\bibinfo  {journal} {Phys. Rev. Lett.}\ }\textbf {\bibinfo {volume} {100}},\ \bibinfo {pages} {253001} (\bibinfo {year} {2008}{\natexlab{a}})}\BibitemShut {NoStop}%
\bibitem [{\citenamefont {Baluktsian}\ \emph {et~al.}(2013)\citenamefont {Baluktsian}, \citenamefont {Huber}, \citenamefont {L\"ow},\ and\ \citenamefont {Pfau}}]{PhysRevLett.110.123001}%
  \BibitemOpen
  \bibfield  {author} {\bibinfo {author} {\bibfnamefont {T.}~\bibnamefont {Baluktsian}}, \bibinfo {author} {\bibfnamefont {B.}~\bibnamefont {Huber}}, \bibinfo {author} {\bibfnamefont {R.}~\bibnamefont {L\"ow}},\ and\ \bibinfo {author} {\bibfnamefont {T.}~\bibnamefont {Pfau}},\ }\bibfield  {title} {\bibinfo {title} {Evidence for strong van der waals type rydberg-rydberg interaction in a thermal vapor},\ }\href {https://doi.org/10.1103/PhysRevLett.110.123001} {\bibfield  {journal} {\bibinfo  {journal} {Phys. Rev. Lett.}\ }\textbf {\bibinfo {volume} {110}},\ \bibinfo {pages} {123001} (\bibinfo {year} {2013})}\BibitemShut {NoStop}%
\bibitem [{\citenamefont {Barredo}\ \emph {et~al.}(2015)\citenamefont {Barredo}, \citenamefont {Labuhn}, \citenamefont {Ravets}, \citenamefont {Lahaye}, \citenamefont {Browaeys},\ and\ \citenamefont {Adams}}]{PhysRevLett.114.113002}%
  \BibitemOpen
  \bibfield  {author} {\bibinfo {author} {\bibfnamefont {D.}~\bibnamefont {Barredo}}, \bibinfo {author} {\bibfnamefont {H.}~\bibnamefont {Labuhn}}, \bibinfo {author} {\bibfnamefont {S.}~\bibnamefont {Ravets}}, \bibinfo {author} {\bibfnamefont {T.}~\bibnamefont {Lahaye}}, \bibinfo {author} {\bibfnamefont {A.}~\bibnamefont {Browaeys}},\ and\ \bibinfo {author} {\bibfnamefont {C.}~\bibnamefont {Adams}},\ }\bibfield  {title} {\bibinfo {title} {Coherent excitation transfer in a spin chain of three rydberg atoms},\ }\href {https://doi.org/10.1103/PhysRevLett.114.113002} {\bibfield  {journal} {\bibinfo  {journal} {Phys. Rev. Lett.}\ }\textbf {\bibinfo {volume} {114}},\ \bibinfo {pages} {113002} (\bibinfo {year} {2015})}\BibitemShut {NoStop}%
\bibitem [{\citenamefont {Baßler}\ \emph {et~al.}(2023)\citenamefont {Baßler}, \citenamefont {Zipper}, \citenamefont {Cedzich}, \citenamefont {Heinrich}, \citenamefont {Huber}, \citenamefont {Johanning},\ and\ \citenamefont {Kliesch}}]{Baler2023}%
  \BibitemOpen
  \bibfield  {author} {\bibinfo {author} {\bibfnamefont {P.}~\bibnamefont {Baßler}}, \bibinfo {author} {\bibfnamefont {M.}~\bibnamefont {Zipper}}, \bibinfo {author} {\bibfnamefont {C.}~\bibnamefont {Cedzich}}, \bibinfo {author} {\bibfnamefont {M.}~\bibnamefont {Heinrich}}, \bibinfo {author} {\bibfnamefont {P.}~\bibnamefont {Huber}}, \bibinfo {author} {\bibfnamefont {M.}~\bibnamefont {Johanning}},\ and\ \bibinfo {author} {\bibfnamefont {M.}~\bibnamefont {Kliesch}},\ }\bibfield  {title} {\bibinfo {title} {Synthesis of and compilation with time-optimal multi-qubit gates},\ }\href {https://doi.org/10.22331/q-2023-04-20-984} {\bibfield  {journal} {\bibinfo  {journal} {Quantum}\ }\textbf {\bibinfo {volume} {7}},\ \bibinfo {pages} {984} (\bibinfo {year} {2023})}\BibitemShut {NoStop}%
\bibitem [{\citenamefont {Li}\ \emph {et~al.}(2023)\citenamefont {Li}, \citenamefont {Qian},\ and\ \citenamefont {Zhang}}]{Li_2023}%
  \BibitemOpen
  \bibfield  {author} {\bibinfo {author} {\bibfnamefont {R.}~\bibnamefont {Li}}, \bibinfo {author} {\bibfnamefont {J.}~\bibnamefont {Qian}},\ and\ \bibinfo {author} {\bibfnamefont {W.}~\bibnamefont {Zhang}},\ }\bibfield  {title} {\bibinfo {title} {Proposal for practical rydberg quantum gates using a native two-photon excitation},\ }\href {https://doi.org/10.1088/2058-9565/ace0d5} {\bibfield  {journal} {\bibinfo  {journal} {Quantum Science and Technology}\ }\textbf {\bibinfo {volume} {8}},\ \bibinfo {pages} {035032} (\bibinfo {year} {2023})}\BibitemShut {NoStop}%
\bibitem [{\citenamefont {Levine}\ \emph {et~al.}(2019)\citenamefont {Levine}, \citenamefont {Keesling}, \citenamefont {Semeghini}, \citenamefont {Omran}, \citenamefont {Wang}, \citenamefont {Ebadi}, \citenamefont {Bernien}, \citenamefont {Greiner}, \citenamefont {Vuleti\ifmmode~\acute{c}\else \'{c}\fi{}}, \citenamefont {Pichler},\ and\ \citenamefont {Lukin}}]{Levine2019}%
  \BibitemOpen
  \bibfield  {author} {\bibinfo {author} {\bibfnamefont {H.}~\bibnamefont {Levine}}, \bibinfo {author} {\bibfnamefont {A.}~\bibnamefont {Keesling}}, \bibinfo {author} {\bibfnamefont {G.}~\bibnamefont {Semeghini}}, \bibinfo {author} {\bibfnamefont {A.}~\bibnamefont {Omran}}, \bibinfo {author} {\bibfnamefont {T.~T.}\ \bibnamefont {Wang}}, \bibinfo {author} {\bibfnamefont {S.}~\bibnamefont {Ebadi}}, \bibinfo {author} {\bibfnamefont {H.}~\bibnamefont {Bernien}}, \bibinfo {author} {\bibfnamefont {M.}~\bibnamefont {Greiner}}, \bibinfo {author} {\bibfnamefont {V.}~\bibnamefont {Vuleti\ifmmode~\acute{c}\else \'{c}\fi{}}}, \bibinfo {author} {\bibfnamefont {H.}~\bibnamefont {Pichler}},\ and\ \bibinfo {author} {\bibfnamefont {M.~D.}\ \bibnamefont {Lukin}},\ }\bibfield  {title} {\bibinfo {title} {Parallel implementation of high-fidelity multiqubit gates with neutral atoms},\ }\href {https://doi.org/10.1103/PhysRevLett.123.170503} {\bibfield  {journal} {\bibinfo  {journal} {Phys. Rev. Lett.}\ }\textbf {\bibinfo {volume}
  {123}},\ \bibinfo {pages} {170503} (\bibinfo {year} {2019})}\BibitemShut {NoStop}%
\bibitem [{\citenamefont {Ming}\ \emph {et~al.}(2025)\citenamefont {Ming}, \citenamefont {Fu},\ and\ \citenamefont {Du}}]{tmr4-gtnl}%
  \BibitemOpen
  \bibfield  {author} {\bibinfo {author} {\bibfnamefont {Y.}~\bibnamefont {Ming}}, \bibinfo {author} {\bibfnamefont {Z.}~\bibnamefont {Fu}},\ and\ \bibinfo {author} {\bibfnamefont {Y.}~\bibnamefont {Du}},\ }\bibfield  {title} {\bibinfo {title} {Geometric gates in atomic arrays without rydberg blockade},\ }\href {https://doi.org/10.1103/tmr4-gtnl} {\bibfield  {journal} {\bibinfo  {journal} {Phys. Rev. A}\ }\textbf {\bibinfo {volume} {112}},\ \bibinfo {pages} {042609} (\bibinfo {year} {2025})}\BibitemShut {NoStop}%
\bibitem [{\citenamefont {Graham}\ \emph {et~al.}(2019)\citenamefont {Graham}, \citenamefont {Kwon}, \citenamefont {Grinkemeyer}, \citenamefont {Marra}, \citenamefont {Jiang}, \citenamefont {Lichtman}, \citenamefont {Sun}, \citenamefont {Ebert},\ and\ \citenamefont {Saffman}}]{PhysRevLett.123.230501}%
  \BibitemOpen
  \bibfield  {author} {\bibinfo {author} {\bibfnamefont {T.}~\bibnamefont {Graham}}, \bibinfo {author} {\bibfnamefont {M.}~\bibnamefont {Kwon}}, \bibinfo {author} {\bibfnamefont {B.}~\bibnamefont {Grinkemeyer}}, \bibinfo {author} {\bibfnamefont {Z.}~\bibnamefont {Marra}}, \bibinfo {author} {\bibfnamefont {X.}~\bibnamefont {Jiang}}, \bibinfo {author} {\bibfnamefont {M.}~\bibnamefont {Lichtman}}, \bibinfo {author} {\bibfnamefont {Y.}~\bibnamefont {Sun}}, \bibinfo {author} {\bibfnamefont {M.}~\bibnamefont {Ebert}},\ and\ \bibinfo {author} {\bibfnamefont {M.}~\bibnamefont {Saffman}},\ }\bibfield  {title} {\bibinfo {title} {Rydberg-mediated entanglement in a two-dimensional neutral atom qubit array},\ }\href {https://doi.org/10.1103/PhysRevLett.123.230501} {\bibfield  {journal} {\bibinfo  {journal} {Phys. Rev. Lett.}\ }\textbf {\bibinfo {volume} {123}},\ \bibinfo {pages} {230501} (\bibinfo {year} {2019})}\BibitemShut {NoStop}%
\bibitem [{\citenamefont {Kazemi}\ \emph {et~al.}(2025)\citenamefont {Kazemi}, \citenamefont {Schuler}, \citenamefont {Ertler},\ and\ \citenamefont {Lechner}}]{56qk-rmsz}%
  \BibitemOpen
  \bibfield  {author} {\bibinfo {author} {\bibfnamefont {J.}~\bibnamefont {Kazemi}}, \bibinfo {author} {\bibfnamefont {M.}~\bibnamefont {Schuler}}, \bibinfo {author} {\bibfnamefont {C.}~\bibnamefont {Ertler}},\ and\ \bibinfo {author} {\bibfnamefont {W.}~\bibnamefont {Lechner}},\ }\bibfield  {title} {\bibinfo {title} {Multiqubit parity gates for rydberg atoms in various configurations},\ }\href {https://doi.org/10.1103/56qk-rmsz} {\bibfield  {journal} {\bibinfo  {journal} {Phys. Rev. Res.}\ }\textbf {\bibinfo {volume} {7}},\ \bibinfo {pages} {033269} (\bibinfo {year} {2025})}\BibitemShut {NoStop}%
\bibitem [{\citenamefont {Xu}\ \emph {et~al.}(2021)\citenamefont {Xu}, \citenamefont {Venkatramani}, \citenamefont {Cant\'u}, \citenamefont {\ifmmode~\check{S}\else \v{S}\fi{}umarac}, \citenamefont {Kl\"usener}, \citenamefont {Lukin},\ and\ \citenamefont {Vuleti\ifmmode~\acute{c}\else \'{c}\fi{}}}]{PhysRevLett.127.050501}%
  \BibitemOpen
  \bibfield  {author} {\bibinfo {author} {\bibfnamefont {W.}~\bibnamefont {Xu}}, \bibinfo {author} {\bibfnamefont {A.}~\bibnamefont {Venkatramani}}, \bibinfo {author} {\bibfnamefont {S.}~\bibnamefont {Cant\'u}}, \bibinfo {author} {\bibfnamefont {T.}~\bibnamefont {\ifmmode~\check{S}\else \v{S}\fi{}umarac}}, \bibinfo {author} {\bibfnamefont {V.}~\bibnamefont {Kl\"usener}}, \bibinfo {author} {\bibfnamefont {M.}~\bibnamefont {Lukin}},\ and\ \bibinfo {author} {\bibfnamefont {V.}~\bibnamefont {Vuleti\ifmmode~\acute{c}\else \'{c}\fi{}}},\ }\bibfield  {title} {\bibinfo {title} {Fast preparation and detection of a rydberg qubit using atomic ensembles},\ }\href {https://doi.org/10.1103/PhysRevLett.127.050501} {\bibfield  {journal} {\bibinfo  {journal} {Phys. Rev. Lett.}\ }\textbf {\bibinfo {volume} {127}},\ \bibinfo {pages} {050501} (\bibinfo {year} {2021})}\BibitemShut {NoStop}%
\bibitem [{\citenamefont {Li}\ \emph {et~al.}(2022{\natexlab{a}})\citenamefont {Li}, \citenamefont {Li}, \citenamefont {Yu}, \citenamefont {Qian},\ and\ \citenamefont {Zhang}}]{PhysRevApplied.17.024014}%
  \BibitemOpen
  \bibfield  {author} {\bibinfo {author} {\bibfnamefont {R.}~\bibnamefont {Li}}, \bibinfo {author} {\bibfnamefont {S.}~\bibnamefont {Li}}, \bibinfo {author} {\bibfnamefont {D.}~\bibnamefont {Yu}}, \bibinfo {author} {\bibfnamefont {J.}~\bibnamefont {Qian}},\ and\ \bibinfo {author} {\bibfnamefont {W.}~\bibnamefont {Zhang}},\ }\bibfield  {title} {\bibinfo {title} {Optimal model for fewer-qubit cnot gates with rydberg atoms},\ }\href {https://doi.org/10.1103/PhysRevApplied.17.024014} {\bibfield  {journal} {\bibinfo  {journal} {Phys. Rev. Applied}\ }\textbf {\bibinfo {volume} {17}},\ \bibinfo {pages} {024014} (\bibinfo {year} {2022}{\natexlab{a}})}\BibitemShut {NoStop}%
\bibitem [{\citenamefont {Katoch}\ \emph {et~al.}(2020)\citenamefont {Katoch}, \citenamefont {Chauhan},\ and\ \citenamefont {Kumar}}]{Katoch2020}%
  \BibitemOpen
  \bibfield  {author} {\bibinfo {author} {\bibfnamefont {S.}~\bibnamefont {Katoch}}, \bibinfo {author} {\bibfnamefont {S.~S.}\ \bibnamefont {Chauhan}},\ and\ \bibinfo {author} {\bibfnamefont {V.}~\bibnamefont {Kumar}},\ }\bibfield  {title} {\bibinfo {title} {A review on genetic algorithm: past, present, and future},\ }\href {https://doi.org/10.1007/s11042-020-10139-6} {\bibfield  {journal} {\bibinfo  {journal} {Multimed Tools Appl}\ }\textbf {\bibinfo {volume} {80}},\ \bibinfo {pages} {8091} (\bibinfo {year} {2020})}\BibitemShut {NoStop}%
\bibitem [{\citenamefont {Sun}\ \emph {et~al.}(2020)\citenamefont {Sun}, \citenamefont {Xu}, \citenamefont {Chen},\ and\ \citenamefont {Liu}}]{PhysRevApplied.13.024059}%
  \BibitemOpen
  \bibfield  {author} {\bibinfo {author} {\bibfnamefont {Y.}~\bibnamefont {Sun}}, \bibinfo {author} {\bibfnamefont {P.}~\bibnamefont {Xu}}, \bibinfo {author} {\bibfnamefont {P.}~\bibnamefont {Chen}},\ and\ \bibinfo {author} {\bibfnamefont {L.}~\bibnamefont {Liu}},\ }\bibfield  {title} {\bibinfo {title} {Controlled phase gate protocol for neutral atoms via off-resonant modulated driving},\ }\href {https://doi.org/10.1103/PhysRevApplied.13.024059} {\bibfield  {journal} {\bibinfo  {journal} {Phys. Rev. Appl.}\ }\textbf {\bibinfo {volume} {13}},\ \bibinfo {pages} {024059} (\bibinfo {year} {2020})}\BibitemShut {NoStop}%
\bibitem [{\citenamefont {Mitra}\ \emph {et~al.}(2023)\citenamefont {Mitra}, \citenamefont {Omanakuttan}, \citenamefont {Martin}, \citenamefont {Biedermann},\ and\ \citenamefont {Deutsch}}]{PhysRevA.107.062609}%
  \BibitemOpen
  \bibfield  {author} {\bibinfo {author} {\bibfnamefont {A.}~\bibnamefont {Mitra}}, \bibinfo {author} {\bibfnamefont {S.}~\bibnamefont {Omanakuttan}}, \bibinfo {author} {\bibfnamefont {M.}~\bibnamefont {Martin}}, \bibinfo {author} {\bibfnamefont {G.}~\bibnamefont {Biedermann}},\ and\ \bibinfo {author} {\bibfnamefont {I.}~\bibnamefont {Deutsch}},\ }\bibfield  {title} {\bibinfo {title} {Neutral-atom entanglement using adiabatic rydberg dressing},\ }\href {https://doi.org/10.1103/PhysRevA.107.062609} {\bibfield  {journal} {\bibinfo  {journal} {Phys. Rev. A}\ }\textbf {\bibinfo {volume} {107}},\ \bibinfo {pages} {062609} (\bibinfo {year} {2023})}\BibitemShut {NoStop}%
\bibitem [{\citenamefont {Sch\"{a}fer}\ \emph {et~al.}(2018)\citenamefont {Sch\"{a}fer}, \citenamefont {Ballance}, \citenamefont {Thirumalai}, \citenamefont {Stephenson}, \citenamefont {Ballance}, \citenamefont {Steane},\ and\ \citenamefont {Lucas}}]{Schfer2018}%
  \BibitemOpen
  \bibfield  {author} {\bibinfo {author} {\bibfnamefont {V.}~\bibnamefont {Sch\"{a}fer}}, \bibinfo {author} {\bibfnamefont {C.}~\bibnamefont {Ballance}}, \bibinfo {author} {\bibfnamefont {K.}~\bibnamefont {Thirumalai}}, \bibinfo {author} {\bibfnamefont {L.}~\bibnamefont {Stephenson}}, \bibinfo {author} {\bibfnamefont {T.}~\bibnamefont {Ballance}}, \bibinfo {author} {\bibfnamefont {A.}~\bibnamefont {Steane}},\ and\ \bibinfo {author} {\bibfnamefont {D.}~\bibnamefont {Lucas}},\ }\bibfield  {title} {\bibinfo {title} {Fast quantum logic gates with trapped-ion qubits},\ }\href {https://doi.org/10.1038/nature25737} {\bibfield  {journal} {\bibinfo  {journal} {Nature}\ }\textbf {\bibinfo {volume} {555}},\ \bibinfo {pages} {75} (\bibinfo {year} {2018})}\BibitemShut {NoStop}%
\bibitem [{\citenamefont {Thapliyal}\ \emph {et~al.}(2025)\citenamefont {Thapliyal}, \citenamefont {Das},\ and\ \citenamefont {Wasan}}]{4pzb-9nlg}%
  \BibitemOpen
  \bibfield  {author} {\bibinfo {author} {\bibfnamefont {D.}~\bibnamefont {Thapliyal}}, \bibinfo {author} {\bibfnamefont {I.~K.}\ \bibnamefont {Das}},\ and\ \bibinfo {author} {\bibfnamefont {A.}~\bibnamefont {Wasan}},\ }\bibfield  {title} {\bibinfo {title} {Controlled-$z$ gate fidelity in neutral-atom arrays with finite blockade and near-degenerate rydberg pair states},\ }\href {https://doi.org/10.1103/4pzb-9nlg} {\bibfield  {journal} {\bibinfo  {journal} {Phys. Rev. A}\ }\textbf {\bibinfo {volume} {112}},\ \bibinfo {pages} {052606} (\bibinfo {year} {2025})}\BibitemShut {NoStop}%
\bibitem [{\citenamefont {Petrosyan}\ \emph {et~al.}(2017)\citenamefont {Petrosyan}, \citenamefont {Motzoi}, \citenamefont {Saffman},\ and\ \citenamefont {M\o{}lmer}}]{PhysRevA.96.042306}%
  \BibitemOpen
  \bibfield  {author} {\bibinfo {author} {\bibfnamefont {D.}~\bibnamefont {Petrosyan}}, \bibinfo {author} {\bibfnamefont {F.}~\bibnamefont {Motzoi}}, \bibinfo {author} {\bibfnamefont {M.}~\bibnamefont {Saffman}},\ and\ \bibinfo {author} {\bibfnamefont {K.}~\bibnamefont {M\o{}lmer}},\ }\bibfield  {title} {\bibinfo {title} {High-fidelity rydberg quantum gate via a two-atom dark state},\ }\href {https://doi.org/10.1103/PhysRevA.96.042306} {\bibfield  {journal} {\bibinfo  {journal} {Phys. Rev. A}\ }\textbf {\bibinfo {volume} {96}},\ \bibinfo {pages} {042306} (\bibinfo {year} {2017})}\BibitemShut {NoStop}%
\bibitem [{\citenamefont {Xue}\ \emph {et~al.}(2024)\citenamefont {Xue}, \citenamefont {Xu}, \citenamefont {Li},\ and\ \citenamefont {Li}}]{PhysRevA.110.032619}%
  \BibitemOpen
  \bibfield  {author} {\bibinfo {author} {\bibfnamefont {M.}~\bibnamefont {Xue}}, \bibinfo {author} {\bibfnamefont {S.}~\bibnamefont {Xu}}, \bibinfo {author} {\bibfnamefont {X.}~\bibnamefont {Li}},\ and\ \bibinfo {author} {\bibfnamefont {X.}~\bibnamefont {Li}},\ }\bibfield  {title} {\bibinfo {title} {High-fidelity and robust controlled-$z$ gates implemented with rydberg atoms via echoing rapid adiabatic passage},\ }\href {https://doi.org/10.1103/PhysRevA.110.032619} {\bibfield  {journal} {\bibinfo  {journal} {Phys. Rev. A}\ }\textbf {\bibinfo {volume} {110}},\ \bibinfo {pages} {032619} (\bibinfo {year} {2024})}\BibitemShut {NoStop}%
\bibitem [{\citenamefont {Beterov}\ \emph {et~al.}(2020)\citenamefont {Beterov}, \citenamefont {Tretyakov}, \citenamefont {Entin}, \citenamefont {Yakshina}, \citenamefont {Ryabtsev}, \citenamefont {Saffman},\ and\ \citenamefont {Bergamini}}]{Beterov2020}%
  \BibitemOpen
  \bibfield  {author} {\bibinfo {author} {\bibfnamefont {I.}~\bibnamefont {Beterov}}, \bibinfo {author} {\bibfnamefont {D.}~\bibnamefont {Tretyakov}}, \bibinfo {author} {\bibfnamefont {V.}~\bibnamefont {Entin}}, \bibinfo {author} {\bibfnamefont {E.}~\bibnamefont {Yakshina}}, \bibinfo {author} {\bibfnamefont {I.}~\bibnamefont {Ryabtsev}}, \bibinfo {author} {\bibfnamefont {M.}~\bibnamefont {Saffman}},\ and\ \bibinfo {author} {\bibfnamefont {S.}~\bibnamefont {Bergamini}},\ }\bibfield  {title} {\bibinfo {title} {Application of adiabatic passage in rydberg atomic ensembles for quantum information processing},\ }\href {https://doi.org/10.1088/1361-6455/ab8719} {\bibfield  {journal} {\bibinfo  {journal} {Journal of Physics B: Atomic, Molecular and Optical Physics}\ }\textbf {\bibinfo {volume} {53}},\ \bibinfo {pages} {182001} (\bibinfo {year} {2020})}\BibitemShut {NoStop}%
\bibitem [{\citenamefont {Pagano}\ \emph {et~al.}(2022)\citenamefont {Pagano}, \citenamefont {Weber}, \citenamefont {Jaschke}, \citenamefont {Pfau}, \citenamefont {Meinert}, \citenamefont {Montangero},\ and\ \citenamefont {B\"uchler}}]{PhysRevResearch.4.033019}%
  \BibitemOpen
  \bibfield  {author} {\bibinfo {author} {\bibfnamefont {A.}~\bibnamefont {Pagano}}, \bibinfo {author} {\bibfnamefont {S.}~\bibnamefont {Weber}}, \bibinfo {author} {\bibfnamefont {D.}~\bibnamefont {Jaschke}}, \bibinfo {author} {\bibfnamefont {T.}~\bibnamefont {Pfau}}, \bibinfo {author} {\bibfnamefont {F.}~\bibnamefont {Meinert}}, \bibinfo {author} {\bibfnamefont {S.}~\bibnamefont {Montangero}},\ and\ \bibinfo {author} {\bibfnamefont {H.}~\bibnamefont {B\"uchler}},\ }\bibfield  {title} {\bibinfo {title} {Error budgeting for a controlled-phase gate with strontium-88 rydberg atoms},\ }\href {https://doi.org/10.1103/PhysRevResearch.4.033019} {\bibfield  {journal} {\bibinfo  {journal} {Phys. Rev. Research}\ }\textbf {\bibinfo {volume} {4}},\ \bibinfo {pages} {033019} (\bibinfo {year} {2022})}\BibitemShut {NoStop}%
\bibitem [{\citenamefont {Zhang}\ \emph {et~al.}(2012)\citenamefont {Zhang}, \citenamefont {Gill}, \citenamefont {Isenhower}, \citenamefont {Walker},\ and\ \citenamefont {Saffman}}]{PhysRevA.85.042310}%
  \BibitemOpen
  \bibfield  {author} {\bibinfo {author} {\bibfnamefont {X.}~\bibnamefont {Zhang}}, \bibinfo {author} {\bibfnamefont {A.}~\bibnamefont {Gill}}, \bibinfo {author} {\bibfnamefont {L.}~\bibnamefont {Isenhower}}, \bibinfo {author} {\bibfnamefont {T.}~\bibnamefont {Walker}},\ and\ \bibinfo {author} {\bibfnamefont {M.}~\bibnamefont {Saffman}},\ }\bibfield  {title} {\bibinfo {title} {Fidelity of a rydberg-blockade quantum gate from simulated quantum process tomography},\ }\href {https://doi.org/10.1103/PhysRevA.85.042310} {\bibfield  {journal} {\bibinfo  {journal} {Phys. Rev. A}\ }\textbf {\bibinfo {volume} {85}},\ \bibinfo {pages} {042310} (\bibinfo {year} {2012})}\BibitemShut {NoStop}%
\bibitem [{\citenamefont {Tamura}\ \emph {et~al.}(2020)\citenamefont {Tamura}, \citenamefont {Yamakoshi},\ and\ \citenamefont {Nakagawa}}]{PhysRevA.101.043421}%
  \BibitemOpen
  \bibfield  {author} {\bibinfo {author} {\bibfnamefont {H.}~\bibnamefont {Tamura}}, \bibinfo {author} {\bibfnamefont {T.}~\bibnamefont {Yamakoshi}},\ and\ \bibinfo {author} {\bibfnamefont {K.}~\bibnamefont {Nakagawa}},\ }\bibfield  {title} {\bibinfo {title} {Analysis of coherent dynamics of a rydberg-atom quantum simulator},\ }\href {https://doi.org/10.1103/PhysRevA.101.043421} {\bibfield  {journal} {\bibinfo  {journal} {Phys. Rev. A}\ }\textbf {\bibinfo {volume} {101}},\ \bibinfo {pages} {043421} (\bibinfo {year} {2020})}\BibitemShut {NoStop}%
\bibitem [{\citenamefont {Guo}\ \emph {et~al.}(2020)\citenamefont {Guo}, \citenamefont {Yan}, \citenamefont {Zhang}, \citenamefont {Su},\ and\ \citenamefont {Li}}]{PhysRevA.102.042607}%
  \BibitemOpen
  \bibfield  {author} {\bibinfo {author} {\bibfnamefont {C.}~\bibnamefont {Guo}}, \bibinfo {author} {\bibfnamefont {L.}~\bibnamefont {Yan}}, \bibinfo {author} {\bibfnamefont {S.}~\bibnamefont {Zhang}}, \bibinfo {author} {\bibfnamefont {S.}~\bibnamefont {Su}},\ and\ \bibinfo {author} {\bibfnamefont {W.}~\bibnamefont {Li}},\ }\bibfield  {title} {\bibinfo {title} {Optimized geometric quantum computation with a mesoscopic ensemble of rydberg atoms},\ }\href {https://doi.org/10.1103/PhysRevA.102.042607} {\bibfield  {journal} {\bibinfo  {journal} {Phys. Rev. A}\ }\textbf {\bibinfo {volume} {102}},\ \bibinfo {pages} {042607} (\bibinfo {year} {2020})}\BibitemShut {NoStop}%
\bibitem [{\citenamefont {Xiao}\ \emph {et~al.}(2024)\citenamefont {Xiao}, \citenamefont {Kang}, \citenamefont {Zheng}, \citenamefont {Song}, \citenamefont {Chen},\ and\ \citenamefont {Xia}}]{PhysRevA.109.062610}%
  \BibitemOpen
  \bibfield  {author} {\bibinfo {author} {\bibfnamefont {Y.}~\bibnamefont {Xiao}}, \bibinfo {author} {\bibfnamefont {Y.}~\bibnamefont {Kang}}, \bibinfo {author} {\bibfnamefont {R.}~\bibnamefont {Zheng}}, \bibinfo {author} {\bibfnamefont {J.}~\bibnamefont {Song}}, \bibinfo {author} {\bibfnamefont {Y.}~\bibnamefont {Chen}},\ and\ \bibinfo {author} {\bibfnamefont {Y.}~\bibnamefont {Xia}},\ }\bibfield  {title} {\bibinfo {title} {Effective nonadiabatic holonomic swap gate with rydberg atoms using invariant-based reverse engineering},\ }\href {https://doi.org/10.1103/PhysRevA.109.062610} {\bibfield  {journal} {\bibinfo  {journal} {Phys. Rev. A}\ }\textbf {\bibinfo {volume} {109}},\ \bibinfo {pages} {062610} (\bibinfo {year} {2024})}\BibitemShut {NoStop}%
\bibitem [{\citenamefont {Jin}\ and\ \citenamefont {Jing}(2024)}]{PhysRevA.109.012619}%
  \BibitemOpen
  \bibfield  {author} {\bibinfo {author} {\bibfnamefont {Z.}~\bibnamefont {Jin}}\ and\ \bibinfo {author} {\bibfnamefont {J.}~\bibnamefont {Jing}},\ }\bibfield  {title} {\bibinfo {title} {Geometric quantum gates via dark paths in rydberg atoms},\ }\href {https://doi.org/10.1103/PhysRevA.109.012619} {\bibfield  {journal} {\bibinfo  {journal} {Phys. Rev. A}\ }\textbf {\bibinfo {volume} {109}},\ \bibinfo {pages} {012619} (\bibinfo {year} {2024})}\BibitemShut {NoStop}%
\bibitem [{\citenamefont {Saffman}\ and\ \citenamefont {Walker}(2005)}]{PhysRevA.72.022347}%
  \BibitemOpen
  \bibfield  {author} {\bibinfo {author} {\bibfnamefont {M.}~\bibnamefont {Saffman}}\ and\ \bibinfo {author} {\bibfnamefont {T.}~\bibnamefont {Walker}},\ }\bibfield  {title} {\bibinfo {title} {Analysis of a quantum logic device based on dipole-dipole interactions of optically trapped rydberg atoms},\ }\href {https://doi.org/10.1103/PhysRevA.72.022347} {\bibfield  {journal} {\bibinfo  {journal} {Phys. Rev. A}\ }\textbf {\bibinfo {volume} {72}},\ \bibinfo {pages} {022347} (\bibinfo {year} {2005})}\BibitemShut {NoStop}%
\bibitem [{\citenamefont {Maller}\ \emph {et~al.}(2015)\citenamefont {Maller}, \citenamefont {Lichtman}, \citenamefont {Xia}, \citenamefont {Sun}, \citenamefont {Piotrowicz}, \citenamefont {Carr}, \citenamefont {Isenhower},\ and\ \citenamefont {Saffman}}]{PhysRevA.92.022336}%
  \BibitemOpen
  \bibfield  {author} {\bibinfo {author} {\bibfnamefont {K.}~\bibnamefont {Maller}}, \bibinfo {author} {\bibfnamefont {M.}~\bibnamefont {Lichtman}}, \bibinfo {author} {\bibfnamefont {T.}~\bibnamefont {Xia}}, \bibinfo {author} {\bibfnamefont {Y.}~\bibnamefont {Sun}}, \bibinfo {author} {\bibfnamefont {M.}~\bibnamefont {Piotrowicz}}, \bibinfo {author} {\bibfnamefont {A.}~\bibnamefont {Carr}}, \bibinfo {author} {\bibfnamefont {L.}~\bibnamefont {Isenhower}},\ and\ \bibinfo {author} {\bibfnamefont {M.}~\bibnamefont {Saffman}},\ }\bibfield  {title} {\bibinfo {title} {Rydberg-blockade controlled-not gate and entanglement in a two-dimensional array of neutral-atom qubits},\ }\href {https://doi.org/10.1103/PhysRevA.92.022336} {\bibfield  {journal} {\bibinfo  {journal} {Phys. Rev. A}\ }\textbf {\bibinfo {volume} {92}},\ \bibinfo {pages} {022336} (\bibinfo {year} {2015})}\BibitemShut {NoStop}%
\bibitem [{\citenamefont {Liang}\ \emph {et~al.}(2022)\citenamefont {Liang}, \citenamefont {Shen}, \citenamefont {Chen},\ and\ \citenamefont {Xue}}]{PhysRevApplied.17.034015}%
  \BibitemOpen
  \bibfield  {author} {\bibinfo {author} {\bibfnamefont {Y.}~\bibnamefont {Liang}}, \bibinfo {author} {\bibfnamefont {P.}~\bibnamefont {Shen}}, \bibinfo {author} {\bibfnamefont {T.}~\bibnamefont {Chen}},\ and\ \bibinfo {author} {\bibfnamefont {Z.}~\bibnamefont {Xue}},\ }\bibfield  {title} {\bibinfo {title} {Composite short-path nonadiabatic holonomic quantum gates},\ }\href {https://doi.org/10.1103/PhysRevApplied.17.034015} {\bibfield  {journal} {\bibinfo  {journal} {Phys. Rev. Appl.}\ }\textbf {\bibinfo {volume} {17}},\ \bibinfo {pages} {034015} (\bibinfo {year} {2022})}\BibitemShut {NoStop}%
\bibitem [{\citenamefont {Lee}\ \emph {et~al.}(2019)\citenamefont {Lee}, \citenamefont {Kim}, \citenamefont {Jo}, \citenamefont {Song},\ and\ \citenamefont {Ahn}}]{PhysRevA.99.043404}%
  \BibitemOpen
  \bibfield  {author} {\bibinfo {author} {\bibfnamefont {W.}~\bibnamefont {Lee}}, \bibinfo {author} {\bibfnamefont {M.}~\bibnamefont {Kim}}, \bibinfo {author} {\bibfnamefont {H.}~\bibnamefont {Jo}}, \bibinfo {author} {\bibfnamefont {Y.}~\bibnamefont {Song}},\ and\ \bibinfo {author} {\bibfnamefont {J.}~\bibnamefont {Ahn}},\ }\bibfield  {title} {\bibinfo {title} {Coherent and dissipative dynamics of entangled few-body systems of rydberg atoms},\ }\href {https://doi.org/10.1103/PhysRevA.99.043404} {\bibfield  {journal} {\bibinfo  {journal} {Phys. Rev. A}\ }\textbf {\bibinfo {volume} {99}},\ \bibinfo {pages} {043404} (\bibinfo {year} {2019})}\BibitemShut {NoStop}%
\bibitem [{\citenamefont {Li}\ \emph {et~al.}(2022{\natexlab{b}})\citenamefont {Li}, \citenamefont {Shao},\ and\ \citenamefont {Li}}]{PhysRevApplied.18.044042}%
  \BibitemOpen
  \bibfield  {author} {\bibinfo {author} {\bibfnamefont {X.}~\bibnamefont {Li}}, \bibinfo {author} {\bibfnamefont {X.}~\bibnamefont {Shao}},\ and\ \bibinfo {author} {\bibfnamefont {W.}~\bibnamefont {Li}},\ }\bibfield  {title} {\bibinfo {title} {Single temporal-pulse-modulated parameterized controlled-phase gate for rydberg atoms},\ }\href {https://doi.org/10.1103/PhysRevApplied.18.044042} {\bibfield  {journal} {\bibinfo  {journal} {Phys. Rev. Appl.}\ }\textbf {\bibinfo {volume} {18}},\ \bibinfo {pages} {044042} (\bibinfo {year} {2022}{\natexlab{b}})}\BibitemShut {NoStop}%
\bibitem [{\citenamefont {Reetz-Lamour}\ \emph {et~al.}(2008{\natexlab{b}})\citenamefont {Reetz-Lamour}, \citenamefont {Deiglmayr}, \citenamefont {Amthor},\ and\ \citenamefont {Weidem\"{u}ller}}]{ReetzLamour2008}%
  \BibitemOpen
  \bibfield  {author} {\bibinfo {author} {\bibfnamefont {M.}~\bibnamefont {Reetz-Lamour}}, \bibinfo {author} {\bibfnamefont {J.}~\bibnamefont {Deiglmayr}}, \bibinfo {author} {\bibfnamefont {T.}~\bibnamefont {Amthor}},\ and\ \bibinfo {author} {\bibfnamefont {M.}~\bibnamefont {Weidem\"{u}ller}},\ }\bibfield  {title} {\bibinfo {title} {Rabi oscillations between ground and rydberg states and van der waals blockade in a mesoscopic frozen rydberg gas},\ }\href {https://doi.org/10.1088/1367-2630/10/4/045026} {\bibfield  {journal} {\bibinfo  {journal} {New Journal of Physics}\ }\textbf {\bibinfo {volume} {10}},\ \bibinfo {pages} {045026} (\bibinfo {year} {2008}{\natexlab{b}})}\BibitemShut {NoStop}%
\bibitem [{\citenamefont {Šibalić}\ \emph {et~al.}(2017)\citenamefont {Šibalić}, \citenamefont {Pritchard}, \citenamefont {Adams},\ and\ \citenamefont {Weatherill}}]{SIBALIC2017319}%
  \BibitemOpen
  \bibfield  {author} {\bibinfo {author} {\bibfnamefont {N.}~\bibnamefont {Šibalić}}, \bibinfo {author} {\bibfnamefont {J.}~\bibnamefont {Pritchard}}, \bibinfo {author} {\bibfnamefont {C.}~\bibnamefont {Adams}},\ and\ \bibinfo {author} {\bibfnamefont {K.}~\bibnamefont {Weatherill}},\ }\bibfield  {title} {\bibinfo {title} {Arc: An open-source library for calculating properties of alkali rydberg atoms},\ }\href {https://doi.org/https://doi.org/10.1016/j.cpc.2017.06.015} {\bibfield  {journal} {\bibinfo  {journal} {Computer Physics Communications}\ }\textbf {\bibinfo {volume} {220}},\ \bibinfo {pages} {319} (\bibinfo {year} {2017})}\BibitemShut {NoStop}%
\bibitem [{\citenamefont {Wood}\ and\ \citenamefont {Gambetta}(2018)}]{PhysRevA.97.032306}%
  \BibitemOpen
  \bibfield  {author} {\bibinfo {author} {\bibfnamefont {C.~J.}\ \bibnamefont {Wood}}\ and\ \bibinfo {author} {\bibfnamefont {J.~M.}\ \bibnamefont {Gambetta}},\ }\bibfield  {title} {\bibinfo {title} {Quantification and characterization of leakage errors},\ }\href {https://doi.org/10.1103/PhysRevA.97.032306} {\bibfield  {journal} {\bibinfo  {journal} {Phys. Rev. A}\ }\textbf {\bibinfo {volume} {97}},\ \bibinfo {pages} {032306} (\bibinfo {year} {2018})}\BibitemShut {NoStop}%
\bibitem [{\citenamefont {Chang}\ \emph {et~al.}(2023)\citenamefont {Chang}, \citenamefont {Wang}, \citenamefont {Jen},\ and\ \citenamefont {Chen}}]{Chang2023}%
  \BibitemOpen
  \bibfield  {author} {\bibinfo {author} {\bibfnamefont {T.}~\bibnamefont {Chang}}, \bibinfo {author} {\bibfnamefont {T.}~\bibnamefont {Wang}}, \bibinfo {author} {\bibfnamefont {H.}~\bibnamefont {Jen}},\ and\ \bibinfo {author} {\bibfnamefont {Y.}~\bibnamefont {Chen}},\ }\bibfield  {title} {\bibinfo {title} {High-fidelity rydberg controlled-z gates with optimized pulses},\ }\href {https://doi.org/10.1088/1367-2630/ad0fa9} {\bibfield  {journal} {\bibinfo  {journal} {New Journal of Physics}\ }\textbf {\bibinfo {volume} {25}},\ \bibinfo {pages} {123007} (\bibinfo {year} {2023})}\BibitemShut {NoStop}%
\bibitem [{\citenamefont {Tang}\ \emph {et~al.}(2022)\citenamefont {Tang}, \citenamefont {Yang}, \citenamefont {Li},\ and\ \citenamefont {Shao}}]{Tang2022}%
  \BibitemOpen
  \bibfield  {author} {\bibinfo {author} {\bibfnamefont {S.}~\bibnamefont {Tang}}, \bibinfo {author} {\bibfnamefont {C.}~\bibnamefont {Yang}}, \bibinfo {author} {\bibfnamefont {D.}~\bibnamefont {Li}},\ and\ \bibinfo {author} {\bibfnamefont {X.}~\bibnamefont {Shao}},\ }\bibfield  {title} {\bibinfo {title} {Implementation of quantum algorithms via fast three-rydberg-atom ccz gates},\ }\href {https://doi.org/10.3390/e24101371} {\bibfield  {journal} {\bibinfo  {journal} {Entropy}\ }\textbf {\bibinfo {volume} {24}},\ \bibinfo {pages} {1371} (\bibinfo {year} {2022})}\BibitemShut {NoStop}%
\bibitem [{\citenamefont {Ebadi}\ \emph {et~al.}(2022)\citenamefont {Ebadi}, \citenamefont {Keesling}, \citenamefont {Cain}, \citenamefont {Wang}, \citenamefont {Levine}, \citenamefont {Bluvstein}, \citenamefont {Semeghini}, \citenamefont {Omran}, \citenamefont {Liu}, \citenamefont {Samajdar}, \citenamefont {Luo}, \citenamefont {Nash}, \citenamefont {Gao}, \citenamefont {Barak}, \citenamefont {Farhi}, \citenamefont {Sachdev}, \citenamefont {Gemelke}, \citenamefont {Zhou}, \citenamefont {Choi}, \citenamefont {Pichler}, \citenamefont {Wang}, \citenamefont {Greiner}, \citenamefont {Vuletić},\ and\ \citenamefont {Lukin}}]{Ebadi2022}%
  \BibitemOpen
  \bibfield  {author} {\bibinfo {author} {\bibfnamefont {S.}~\bibnamefont {Ebadi}}, \bibinfo {author} {\bibfnamefont {A.}~\bibnamefont {Keesling}}, \bibinfo {author} {\bibfnamefont {M.}~\bibnamefont {Cain}}, \bibinfo {author} {\bibfnamefont {T.}~\bibnamefont {Wang}}, \bibinfo {author} {\bibfnamefont {H.}~\bibnamefont {Levine}}, \bibinfo {author} {\bibfnamefont {D.}~\bibnamefont {Bluvstein}}, \bibinfo {author} {\bibfnamefont {G.}~\bibnamefont {Semeghini}}, \bibinfo {author} {\bibfnamefont {A.}~\bibnamefont {Omran}}, \bibinfo {author} {\bibfnamefont {J.}~\bibnamefont {Liu}}, \bibinfo {author} {\bibfnamefont {R.}~\bibnamefont {Samajdar}}, \bibinfo {author} {\bibfnamefont {X.}~\bibnamefont {Luo}}, \bibinfo {author} {\bibfnamefont {B.}~\bibnamefont {Nash}}, \bibinfo {author} {\bibfnamefont {X.}~\bibnamefont {Gao}}, \bibinfo {author} {\bibfnamefont {B.}~\bibnamefont {Barak}}, \bibinfo {author} {\bibfnamefont {E.}~\bibnamefont {Farhi}}, \bibinfo {author} {\bibfnamefont {S.}~\bibnamefont {Sachdev}}, \bibinfo {author}
  {\bibfnamefont {N.}~\bibnamefont {Gemelke}}, \bibinfo {author} {\bibfnamefont {L.}~\bibnamefont {Zhou}}, \bibinfo {author} {\bibfnamefont {S.}~\bibnamefont {Choi}}, \bibinfo {author} {\bibfnamefont {H.}~\bibnamefont {Pichler}}, \bibinfo {author} {\bibfnamefont {S.}~\bibnamefont {Wang}}, \bibinfo {author} {\bibfnamefont {M.}~\bibnamefont {Greiner}}, \bibinfo {author} {\bibfnamefont {V.}~\bibnamefont {Vuletić}},\ and\ \bibinfo {author} {\bibfnamefont {M.}~\bibnamefont {Lukin}},\ }\bibfield  {title} {\bibinfo {title} {Quantum optimization of maximum independent set using rydberg atom arrays},\ }\href {https://doi.org/10.1126/science.abo6587} {\bibfield  {journal} {\bibinfo  {journal} {Science}\ }\textbf {\bibinfo {volume} {376}},\ \bibinfo {pages} {1209–1215} (\bibinfo {year} {2022})}\BibitemShut {NoStop}%
\bibitem [{\citenamefont {Cesa}\ and\ \citenamefont {Pichler}(2023)}]{PhysRevLett.131.170601}%
  \BibitemOpen
  \bibfield  {author} {\bibinfo {author} {\bibfnamefont {F.}~\bibnamefont {Cesa}}\ and\ \bibinfo {author} {\bibfnamefont {H.}~\bibnamefont {Pichler}},\ }\bibfield  {title} {\bibinfo {title} {Universal quantum computation in globally driven rydberg atom arrays},\ }\href {https://doi.org/10.1103/PhysRevLett.131.170601} {\bibfield  {journal} {\bibinfo  {journal} {Phys. Rev. Lett.}\ }\textbf {\bibinfo {volume} {131}},\ \bibinfo {pages} {170601} (\bibinfo {year} {2023})}\BibitemShut {NoStop}%
\end{thebibliography}

\end{document}